\documentclass[5p,final,authoryear,times]{elsarticle}
\usepackage[titletoc,toc,title]{appendix}
\journal{NeuroImage}
\usepackage{hyperref}
\hypersetup{colorlinks = true,linkcolor = blue, anchorcolor = blue, citecolor = blue, pdfstartpage=1, filecolor = red,urlcolor = black}
\usepackage{graphicx}
\usepackage{amsmath}
\usepackage{amssymb}
\usepackage{amsfonts}
\usepackage{mathrsfs}
\usepackage{times}
\usepackage{soul}
\usepackage{xcolor}
\usepackage{xspace}
\usepackage{sidecap}    
\usepackage[english]{babel}

\def\cite{\citep}

\newcommand{\non}{\nonumber}
\newcommand{\be}{\begin{equation}}
\newcommand{\ee}{\end{equation}}
\newcommand{\bea}{\begin{eqnarray}}
\newcommand{\eea}{\end{eqnarray}}
\newcommand{\eq}[1]{(\ref{#1})}
\renewcommand{\d}{{\rm d}}

\newcommand{\q}{{\bf q}}
\newcommand{\g}{{\bf \hat g}}

\newcommand{\n}{{\bf \hat n}}

\renewcommand{\P}{{\cal P}}

\newcommand{\lp}{\left(}
\newcommand{\rp}{\right)}
\newcommand{\lb}{\left[}
\newcommand{\rb}{\right]}
\newcommand{\la}{\left<}
\newcommand{\ra}{\right>}

\newcommand{\ts}[1]{\textstyle{#1}}
\newcommand{\ds}[1]{\displaystyle{#1}}

\newcommand{\Y}{{\cal Y}}
\renewcommand{\O}{{\cal O}}
\newcommand{\K}{{\cal K}}
\newcommand{\D}{{\cal D}}
\newcommand{\N}{{\cal N}}

\newcommand{\sgn}{\,{\rm sgn}\,}
\def\Re{\mbox{\,Re\,}}
\def\Im{\mbox{\,Im\,}}
\newcommand{\units}[1]{\,{\mathrm{#1}}}

\newcommand{\Da}{{D_{a}}}
\newcommand{\Depar}{{D_{e}^{\parallel}}}
\newcommand{\Deperp}{{D_{e}^{\perp}}}
\newcommand{\dDe}{{\Delta_{e}}}
\newcommand{\Dbar}{{\bar D}}
\newcommand{\deperp}{{d_{e}^{\perp}}}
\newcommand{\dde}{{\delta_{e}}}
\newcommand{\lmax}{l_{\rm max}}

\newcommand{\dm}{{\Delta m}}

\newcommand{\rem}[1]{\!\xspace}
\newcommand{\del}[1]{\!\xspace}

\newcommand{\new}{\!\xspace}
\newcommand{\newr}{\!\xspace}

\newcommand{\keep}{\color{black}}

\usepackage[fulladjust]{marginnote}
\newcommand{\mpar}[1]{\!\xspace} \newcommand{\rmpar}[1]{\!\xspace}
\newcommand{\mparr}[1]{\!\xspace}

\begin{document}
\begin{frontmatter}

\title{{\bf Rotationally-invariant mapping of scalar and orientational metrics of neuronal microstructure with diffusion MRI}}
\author[add1]{Dmitry S. Novikov\corref{cor1}}
\ead{dima@alum.mit.edu}
\author[add1]{Jelle Veraart}
\author[add1,add2]{Ileana O. Jelescu} 
\author[add1]{Els Fieremans}
\address[add1]{Bernard and Irene Schwartz Center for Biomedical Imaging, Department of Radiology,
New York University School of Medicine, New York, NY, USA}
\address[add2]{Centre d'Imagerie Biom\'edicale, \'Ecole Polytechnique F\'ed\'erale de Lausanne, Lausanne, Switzerland}
\cortext[cor1]{Corresponding author}
\date{\today}

\begin{abstract}
\noindent 
We develop a general analytical and numerical framework for estimating intra- and extra-neurite water fractions and diffusion coefficients, as well as neurite orientational dispersion, in each imaging voxel. By employing a set of rotational invariants and their expansion in the powers of diffusion weighting, we analytically uncover the nontrivial topology of the parameter estimation landscape, showing that multiple branches of parameters describe the measurement almost equally well, with only one of them corresponding to the biophysical reality.
A comprehensive acquisition shows that the branch choice \new varies across the brain. \keep 
Our framework reveals hidden degeneracies in MRI parameter estimation for neuronal tissue, provides microstructural and orientational maps in the whole brain without constraints or priors, and connects modern biophysical modeling with clinical MRI. 
\end{abstract}


\end{frontmatter}

\section{Introduction and overview of results}

\begin{SCfigure*}[][h!!]
\centering
\includegraphics[width=5.1in]{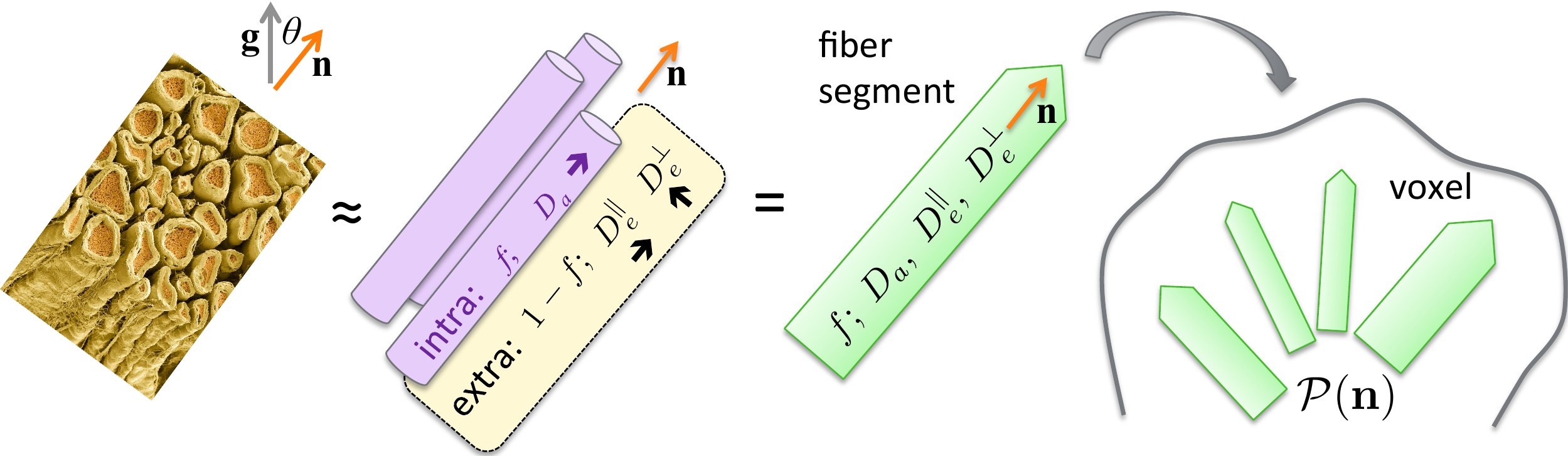}\hspace{1mm}
\caption{
{\bf Standard Model of diffusion in neuronal tissue.} In the long time limit, elementary fiber segments (fascicles), consisting of intra- and extra-neurite compartments, 
are described by at least 4 independent parameters: $f$, $\Da$, $\Depar$ and $\Deperp$. Within a macroscopic imaging voxel, such segments contribute to the directional dMRI signal according to their ODF $\P(\n)$. Due to its rich orientational content, the total number of parameters (\ref{Np}) characterizing a voxel is about 30 -- 50.
}
\label{fig:SM}
\end{SCfigure*}

Brownian motion of water molecules is strongly hindered by neurite walls \cite{beaulieu2002}. 
Serendipitously, this sensitivity to tissue microstructure can be probed with NMR for diffusion times 
$t\sim 1 - 1000\,$ms, 
\mpar{R4.3}
corresponding to a diffusion length (rms molecular displacement) $\ell(t) \sim 1 - 100\,\mu$m commensurate with cell dimensions. 
The resulting diffusion MRI (dMRI) signal, acquired over a macroscopic imaging voxel, is an indirect but powerful probe into neuronal structure at the scale $\ell(t)$, 2--3 orders of magnitude below the MRI  resolution.

The dMRI signal is generally anisotropic \cite{Basser1994,beaulieu2002,jones-book}, non-Gaussian \cite{LeBihan,kroenke2004,assaf2004,Kiselev2007,jespersen2007,jespersen2010,wmdki,ferizi2015,highb,jensen2016}, and 
time-dependent \cite{Stanisz1997,assaf-NAA-1998,mesopnas,fieremans2016}.  
Description of this complex process simplifies at long  times $t\sim 100\,$ms, used in clinical dMRI, when
$\ell(t) \sim 10\,\mu$m exceeds typical  neurite diameters $a \lesssim 1\,{\rm \mu m}$. 
In this regime, diffusion approaches its Gaussian limit separately in the intra- and extra-neurite spaces, Fig.~\ref{fig:SM}. Biologically distinct hindrances lead to coarse-grained \cite{review-nbm} diffusion coefficients inside ($\Da$) and outside ($\Depar$ and $\Deperp$) neurites within an elementary fiber fascicle; 
transverse diffusion $\sim a^2/t \ll \Da$  inside  neurites then becomes negligible \cite{kroenke2004}. 
The  dMRI signal (voxel-averaged diffusion propagator) is an ensemble average over contributions of individual fascicles within a voxel.

Here we investigate in detail the general picture of anisotropic Gaussian compartments, Fig.~\ref{fig:SM}, 
as an overarching ``Standard Model", such that previously used biophysical models \cite{kroenke2004,jespersen2007,jespersen2010,assaf2004,KM,wmdki,noddi,sotiropoulos2012,ferizi2015,jelescu2016}
follow as special cases. 
To set up the model, we first represent the dMRI signal,
parameterized by diffusion weighting $b=q^2 t$ 
and measured in the unit direction $\g$, 
\new \mpar{R2.3, R3}
as a convolution%
\footnote{\label{foot:dn}
\new The convolution is on a unit sphere $|\n|=1$ \cite{Healy1998}. 
We normalize $\d\n \equiv {\sin\theta \d\theta\, \d\phi \over 4\pi}$ 
such that $\int\!\d\n \cdot 1  \equiv 1$, while $\K|_{b=0} = S_0 \equiv S|_{b=0}$. 
}  
\be \label{S}
S_{\g}(b) = \int_{|\n|=1}\!\! \d\n\, \P(\n) \, \K(b, \g\cdot\n)
\ee
\mpar{$S(0)$ incl in $\K$}
between the fiber orientation distribution function (ODF) $\P(\n)$ 
normalized to $\int\d\n\, \P(\n)\equiv 1$, and the response kernel $\K$
from a perfectly aligned fiber segment (fascicle) pointing in the direction $\n$. 
The kernel $\K(b, \g\cdot\n)$ depends on the relative angle $\theta$, $\cos\theta \equiv \g\cdot\n$.
The general representation (\ref{S}) gave rise to a number of methods for deconvolving  
the fiber ODF from the dMRI signal for a given $|\q|=q$ shell in $q$-space, using different empirical forms of the kernel  
\cite{Tournier2004,Anderson2005,Tournier2007,DellAcqua2007,Jian2007,Kaden2007,White2009}. 

Our work focuses on quantifying the microstructural origins of the signal, Fig.~\ref{fig:SM}, which dictates the kernel's functional form  
\be \label{K}
\K(b,\xi) = S_0  \lb f e^{-b\Da \xi^2} + (1-f) e^{-b\Deperp-b(\Depar-\Deperp)\xi^{2}} \rb, 
\ee
\keep
with $\xi = \g\!\cdot\!\n$, to be a sum of the exponential (in $b$) contributions from intra- and extra-neurite spaces, 
modeled by axially-symmetric Gaussian compartments. 
\new 
Here we neglect the myelin water compartment  due to its short $T_2$ time \cite{Mackay1994} as compared to clinical NMR echo time $T_E$. We emphasize that the fractions $f$ and $1-f$ are the relative signal fractions, and not the absolute volume fractions, 
\mpar{R2.5, R4.1}
due to generally different $T_2$ values for the intra- and extra-neurite compartments \cite{Dortch2013}, and due to neglecting myelin water. \keep 
Further compartments, such as  isotropic cerebrospinal fluid (CSF), can in principle be added to  kernel \eq{K}; 
here we will study in depth the two-compartment kernel \eq{K}, and will later comment on its generalizations.  
\new
We also note that a major limitation of the kernel (\ref{K}) is sharing the scalar parameter values among different fiber tracts passing through a voxel, noted by \cite{Assaf2005,DeSantis2016}, which prompted assigning different (albeit constant) fiber responses to different \mparr{R3.1}
tracts \newr to deconvolve the ODF \cite{Sherbondy2010,Tournier2011,Girard2017}. 
\keep

The {\it scalar parameters} $f$, $\Da$, $\Depar$ and $\Deperp$, and the {\it tensor parameters} (the spherical harmonics coefficients of the ODF $\P(\n)$), carry distinct biophysical significance. Deconvolving the voxel-wise fiber ODF, instead of relying on the empirical directions from the signal \eq{S}, in principle provides a more adequate starting point for  fiber tractography, an essential tool for mapping structural brain connectivity and for presurgical planning \cite{Behrens2007,Descoteaux2009,Farquharson2013,Sotiropoulos2013a,Wilkins2015}. \mpar{R2.3, R3}

The scalar parameters of the kernel (\ref{K}) make dMRI measurements specific to $\mu$m-level manifestations of disease processes, such as demyelination  \cite{axlossdemyel,jelescu-cpz-2016} ($\Deperp$), axonal loss \cite{axlossdemyel} ($f$),  beading \cite{budde-frank2010} ($\Da$), inflammation and oedema 
\new \mpar{R2.6, R4.2}
(potentially, mostly $\Da$ for cytotoxic and mostly $\Depar, \Deperp$ for vasogenic oedema \cite{unterberg2004}). 
\keep
Since the precise nature and pathological changes in microarchitecture of restrictions leading to $\Da$, $\Depar$ and $\Deperp$ values are unknown, ideally, to become specific to pathology, one needs to estimate $f$, $\Da$, $\Depar$ and $\Deperp$ separately, without constraints or priors. 

Here we establish the mathematical structure of the general parameter estimation problem \eq{S}--\eq{K}, reveal its hidden degeneracies, \mpar{R2.1: shortened,  fwd refs to Eqs removed}
and elucidate the information content of the dMRI signal, in order to design a parameter estimation algorithm. The outline of our main steps and results is as follows: 
\begin{itemize}

\item {\bf Scalar-tensor factorization.} We factorize the model \eq{S}--\eq{K} in the spherical haromics (SH) basis, and employ {rotational invariants}, to separate the estimation of scalar parameters $x \equiv \{ f, \Da, \Depar, \Deperp\}$ from the ODF SH coefficients, Sec.~\ref{sec:fact}. 

\item{\bf Radial-angular connection and parameter count.} 
\mpar{R3}
We derive, \new using Taylor expansion of the model \eq{S}--\eq{K}, 
the connection between the {radial} $q$-space sensitivity and the {angular} resolution for the ODF expanded 
in the SH basis, and thus identify  
the physically preferred role the SH basis plays in dMRI acquisitions, Sec.~\ref{sec:moments}--\ref{sec:par-model}. 
\keep

\item {\bf Parameter landscape. } We  show that the estimation of the scalar parameters in the space of rotational invariants 
is generally {\it degenerate}, and uncover the nontrivial topology of the parameter landscape, Fig.~\ref{fig:landscape} in Sec.~\ref{sec:landscape}. 

\item {\bf LEMONADE} (Linearly Estimated Moments provide Orientations of Neurites And their Diffusivities Exactly). We analytically study this topology by expanding the model \eq{S} in the powers of $b$, 
deriving and exactly solving the  system of equations 
relating the signal's moments to the model parameters, Sec.~\ref{sec:lemo}.  

\mpar{R4.4: \del{at least}}
\item {\bf Fit degeneracies as LEMONADE branches.} We establish the two types of degeneracies, Sec.~\ref{sec:landscape}--\ref{sec:bimodality}. The {\it discrete degeneracy} amounts to the two sets (``branches") $x_\pm$ of parameters that exactly satisfy the low-$b$ expansion of model \eq{S}--\eq{K}, and fit realistic dMRI data equally well. 
Furthermore, these branches are not just isolated minima, but  narrow trenches in the parameter landscape, Fig.~\ref{fig:landscape}; 
up to $\O(b^2)$, each point on a trench satisfies the model \eq{S}--\eq{K} {\it exactly}. 
This {\it continuous degeneracy} rationalizes poor precision in parameter estimation, tying it to  inherently low information content of clinical acquisitions. 

\item {\bf Branch selection.} Out of the two parameter branches, 
only one corresponds to the biophysical reality, and the other  should be discarded. Branch selection is nontrivial, Sec.~\ref{sec:branch}, 
as it depends on the ground truth values, 
and is generally brain region-specific, as we show in Figs.~\ref{fig:maps-all}-\ref{fig:bars} 
based on 21-shell human dMRI with $b\leq 10\,\mathrm{ms/\mu m^2}$. 

\mparr{R3.2}
\item {\bf Algorithm.}  
\newr
We designed a nonlinear estimation algorithm that is initialized using the selected branch. \keep
We  produce parameter maps (Fig.~\ref{fig:maps-all}), \new their histograms (Fig.~\ref{fig:branch}) and ROI bar plots (Fig.~\ref{fig:bars}), \keep
as well as fiber ODFs and tracts reconstructed via locally estimated kernels \eq{K}, Fig.~\ref{fig:odf}.

\item
Finally, in Sec.~\ref{sec:discussion} 
we discuss noise propagation, ways to improve precision  by using ``orthogonal" acquisitions, 
and critically assess the previously used constraints, Fig.~\ref{fig:constraints}.  
\end{itemize}

\section{Theory}
\label{sec:th}
\mpar{{\it Theory} Sec introduced; restructured, split into sub-sections}

\subsection{Scalar-tensor factorization and rotational invariants}
\label{sec:fact}
\bigskip

\new Signal \eq{S} \keep factorizes in the SH basis \new for any kernel $\K$
\cite{Healy1998,Tournier2004,Anderson2005,jespersen2007,Jian2007,DellAcqua2007},  \keep
\be \label{fact}
S_{lm}(b, x)  = p_{lm} \, K_l(b, x) \,.
\ee
Here $S_{lm}$ and $p_{lm}$ are the SH coefficients of the signal $S$ and of the ODF
\be \label{P=Y}
\P(\n) \approx 1 + \sum_{l = 2, 4, \dots}^{\lmax} \sum_{m=-l}^l p_{lm} Y_{lm}(\n) \,,
\ee
up to order $\lmax$ which practically depends on the dMRI sampling and signal-to-noise ratio (SNR), as it will be discussed below. 
The functions $K_l(b,x)$ are projections of our model kernel \eq{K} onto the Legendre polynomials (Eq.~\eq{Kl} in \ref{app:fact}) for a given set $x = \{ f, \Da, \Depar, \Deperp\}$ of scalar parameters.

To factor out the dependence on the choice of the physical basis in three-dimensional space (via $m=-l\dots l$), recall that any rotation corresponds to an orthogonal transformation on the $(2l+1)$-dimensional vectors $S_{lm}$ and $p_{lm}$, belonging to the irreducible representation of the SO(3) group of rotations of weight $l$. Hence, the 2-norms $\| S_l \|^2 = \sum_{m=-l}^l |S_{lm}|^2$ and $\| p_l \|^2 = \sum_{m=-l}^l |p_{lm}|^2$ do not depend on the choice of the physical basis. 
Introducing basis-independent {\it rotational invariants} 
\be \label{rotinv}
S_l  \equiv \| S_l \| /\N_l \,, \quad p_l  \equiv \| p_l \| /\N_l \,, \quad \N_l = \sqrt{4\pi(2l+1)}
\ee
of the signal and ODF  [cf. \cite{kazhdan2003} for using such approach for image matching in computer vision, 
\new \cite{Mirzaalian2016} for dMRI data harmonization, and \keep
\rmpar{R1.10}
\del{beautiful} \cite{baydiff} for the related machine-learning dMRI parameter estimation framework], 
the ODF SHs $p_{lm}$ are factored out: 
\be \label{S=pK}
S_l (b,x) = p_l \, K_l(b,x) \,, \quad l = 0, \ 2, \ \dots \,.
\ee
Here, $p_0\equiv 1$ (ODF normalization); the remaining ODF invariants, one for each $l$, characterize its anisotropy, 
with the normalization factor $\N_l$ chosen so that $0\leq p_l \leq 1$ (cf. \ref{app:pl}).

It now appears logical to first estimate the scalar parameters $x$, together with just a few basis-independent $p_l $, $l = 2, \dots$,
from the greatly reduced system of equations \eq{S=pK}, one for each $l$.
A standard way to solve any such nonlinear system is to minimize the corresponding 
``energy''  function 
\be \label{F}
F^2(x,\{p_l\}) \equiv \frac{1}{\big(1+\frac{L}2\big) N_b} \sum_{l=0,2,\dots}^{L} \sum_{j=1}^{N_b} \, w_{lj}\, [S_l(b_j,x) - p_l K_l(b_j,x)]^2 
\ee
with respect to $x$ and a few $p_l $, $l=0,2,\dots,L$. 
\rmpar{R3}
\new
We will refer to Eq.~(\ref{F}) as the rotationally invariant (RotInv) nonlinear objective function for minimization. 
\keep
\mpar{R3.3}
Here $b_j$ are the radii of $N_b$ shells in $q$-space for a \new uniform \keep spherical sampling; 
all units for diffusion coefficients and for $1/b$ are $\mathrm{\mu m^2/ms}$  hereon.  
\new The weights $w_{lj} = 1/\sigma_{lj}^2$ are chosen for the optimal precision, 
in terms of the effective noise variances $\sigma_{lj}^2 \sim  (2l+1)/N_j$ in each shell, after estimating $2l+1$ independent parameters $p_{lm}$ over the  $N_j$ diffusion directions in the shell $b=b_j$.\keep

The estimated scalar parameters $x$ will allow us to reconstruct the kernel components $K_l(b,x)$, 
and to subsequently evaluate all  ODF coefficients $p_{lm}$
using Eq.~\eq{fact}, based on the linearly estimated $S_{lm}$ from the measured  signal --- much like in a 
\mpar{R2.2, R3}
\new spherical deconvolution approach \cite{Tournier2004,Tournier2007}, 
albeit with the voxel-wise, rather than globally estimated kernel (\ref{K}), cf. \cite{Anderson2005,Barmpoutis2009,Schultz2013}. 
\keep

While the above rotationally invariant framework looks conceptually simple and completely general [cf.  \cite{Anderson2005,jespersen2007,baydiff}], 
the rest of this paper will be devoted to uncovering and resolving hidden degeneracies of parameter estimation problem \eq{S}
because of the kernel \eq{K} specific to multi-compartmental diffusion in neuronal tissue.  
Understanding these degeneracies is essential for proper initialization and/or constraining any kind of solution for nonlinear system \eq{S=pK}, 
e.g. via minimization of Eq.~\eq{F}, or by \new machine learning \keep methods \cite{baydiff}. 
This is especially relevant because it has been known all too well that direct nonlinear fitting of model \eq{S}--\eq{K} to realistic noisy data is unreliable. Hence, parameter estimation from clinical acquisitions has so far reverted to making severe restrictions on 
\mparr{R3.4}
the ODF shape: either assuming a highly aligned bundle \citep[white matter tract integrity (WMTI)]{KM,wmdki}, or a special Gaussian-like ODF shape characterized by one \citep[NODDI]{noddi} or two \cite{noddi-bingham} parameters, in order to reduce the dimensionality of the problem 
\mpar{\del{ from $\sim 30-50$ down to $\sim 3$.}} 
\new by an order of magnitude.\keep

Furthermore, the problem (\ref{F}) has multiple degenerate minima, as we will investigate in detail below in Sec.~\ref{sec:results}, cf. Fig.~\ref{fig:landscape}. This echoes our previous recent study, where, 
even assuming the simplest, 1-parameter ODF shape \cite{noddi}, unconstrained nonlinear fitting  reveals multiple biophysically plausible minima in the (4+1) - dimensional parameter space, and flat directions along them \cite{jelescu2016}.
Fixing all diffusion coefficients in Eq.~\eq{K} and ODF shape $\P(\n)$ in Eq.~\eq{S}, as done in NODDI \cite{noddi,noddi-bingham}, introduces {\it a priori} unknown bias \cite{jelescu2016} for the remaining few parameters, and thereby leads to the loss of specificity --- the main motivation for employing microstructural modeling. 
\mpar{R1.1}
\new
For example, diffusivity changes were recently found  \cite{khan2016ni}  in regions with no detectable neurite density changes; these details would otherwise have been lost or attributed to other parameters in an analysis using fixed diffusivities. 
\keep

\new
To understand analytically the information content of our acquisitions,  in the rest of this Section we analyze the parameter estimation for the model (\ref{S})--(\ref{K})  {\it perturbatively in} $b$. 
This approach is based on the Taylor expansion of the signal \eq{S} 
\be\label{taylor}
{S_\g(b) \over S_0}  = 1 - b M^{(2)}_{i_1 i_2} \, g_{i_1} g_{i_2} + \frac{b^{2}}{2!} \, M^{(4)}_{i_1 \dots i_4} \, g_{i_1} \dots g_{i_4} - \dots
\ee
in the fully symmetric moment tensors $M^{(l)}_{i_1\dots i_l}$. (Einstein's convention of summation over repeated indices is assumed hereon.)  
The expansion in moments has a one-to-one correspondence with the cumulant expansion of $\ln S_\g(b)$ \cite{Kiselev2010_diff_book}, 
and is also equivalent to Taylor-expanding the above nonlinear equations (\ref{fact}), (\ref{S=pK}), and (\ref{F}), matching the moments of the signal and of the model at each order in $b$. 

The perturbative approach, which we develop term-by-term in $b$, is useful in the three ways. 
First, it enables us to elucidate the role of the order $\lmax$ in Eq.~(\ref{P=Y}), via the radial-angular connection in the $q$-space, Sec.~\ref{sec:moments}, and to count the number $N_p(\lmax)$ of model parameters, which depends on the maximal even power $\lmax$ of the diffusion weighting $q^{\lmax} \sim b^{\lmax/2}$ to which an acquisition is sensitive, at a given signal-to-noise ratio (SNR), Sec.~\ref{sec:par-model}. Second, it helps us to develop intuition by analytically studying the problem's degeneracies, Sec.~\ref{sec:lemo}, which will qualitatively persist in the subsequent full non-perturbative numerical treatment. Third, it will help initialize the full, non-perturbative numerical parameter estimation, Eq.~(\ref{F}) above. 

\keep

\subsection{Radial-angular connection in the $q$-space}
\label{sec:moments}
\bigskip

Expanding the model (\ref{S})--(\ref{K}) in the powers of $b$, one can readily see that 
moments $M^{(l)}_{i_1\dots i_l}$ are proportional  to angular averages $\langle n_{i_1}\dots n_{i_l} \rangle$ over the ODF $\P(\n)$. 
This follows from expanding the exponential terms containing $\xi = n_i g_i$ in  kernel (\ref{K}), so that subsequent terms have the form $b \langle n_i n_j \rangle g_i g_j$, 
$b^2 \langle n_{i_1}\dots n_{i_4}\rangle \\ g_{i_1}\dots g_{i_4}$, etc (cf. \ref{app:lemo} for explicit formulas up to $l=6$). 

It is crucial that the maximal even order $l$ of the product  $\langle n_{i_1}\dots n_{i_l} \rangle$ always appears with
the corresponding power $q^l \sim b^{l/2}$ of the diffusion weighting. This observation underpins the \new perturbative \keep {\it radial-angular connection} in the $q$-space: Practical sensitivity to the maximal order $q^{\lmax}$ automatically sets the sensitivity to the maximal order $\langle n_{i_1}\dots n_{i_{\lmax}} \rangle$ of the ODF average, and, therefore, to the maximal order $\lmax$ in its SH expansion (\ref{P=Y}).
This key relation between the expansion (\ref{taylor}) of the signal \eq{S} and of the ODF \eq{P=Y} rests on the fact that  linear combinations of symmetric tensors $n_{i_1}\dots n_{i_l}$ form the SH set $Y_{lm}(\n)$ (cf. Eq.~(\ref{Y=Yn}) in \ref{app:lemo}). Therefore,  linear combinations of the ODF averages $\langle n_{i_1}\dots n_{i_l} \rangle$ correspond to the ODF SH coefficients $p_{lm}$.   

The physical coupling between products $q_{i_1}\dots q_{i_l}$ in $q$-space imaging \cite{Callaghan1988} and the ODF averages $\langle n_{i_1}\dots n_{i_l} \rangle$  emphasizes the preferred role the SH basis plays in dMRI. 
Essentially, the $q$-space measurement directly couples to the SH basis coefficients. 
Since we do not know the functional form of the ODF shape {\it a priori}, we cannot claim that we are able to determine ODF sharper than what is permitted by the practical sensitivity to the corresponding order 
$q^{\lmax}$. 

Hence, if our acquisition is only sensitive to, e.g. $\O(q^4\sim b^2)$ (equivalent to diffusion kurtosis imaging, or DKI \cite{DKI}), parametrizing the ODF with $p_{lm}$ with $l>4$ would amount to ODF overfitting. 
\mpar{R3.1}
\new Conversely, apparent sensitivity to $p_{lm}$ up to order, e.g., $\lmax=6$, 
is equivalent to the presence of notable $\O(b^3)$ contributions for a given $b$-shell (at a given SNR level). 
At finite $b$, the corresponding $K_l(b)$ values numerically connect the radial and angular sensitivities in the $q$-space. We will turn to this point in Discussion, Sec.~\ref{sec:paramvals}. 
\keep

\subsection{Parameter count: Signal via its moments, term-by-term}
\label{sec:par-mom}
\bigskip

Let us count the number of parameters in  expansion \eq{taylor} as a function of the maximal  order $l_{\rm max}$. 
A term $M^{(l)}_{i_1\dots i_l}$ of rank $l$ is a fully symmetric tensor, which can be represented in terms of symmetric trace-free (STF) tensors of rank $l$, $l-2$, $\dots$, $2$, $0$. Each set of  STF-$l$ tensors realizes a $2l+1$-dimensional irreducible representation of the SO(3) group of rotations, equivalent \cite{thorne} to the set of $2l+1$ SH $Y_{lm}$ (cf. \ref{app:lemo}). 
Hence, the total number of independent components in the rank-$l$ moment (or cumulant) is 
$n_c(l) = \sum_{l' =0, 2, \dots}^{l} \! (2l'+1) = \frac12 (l+1)(l+2)$.
Truncating the  series \eq{taylor} at $l = l_{\rm max}$ means that  
we determine all components of $M^{(l)}_{i_1\dots i_l}$ for $l = 0, 2, \dots, l_{\rm max}$, 
with the total number of independent, i.e. ``informative", components
\be \label{Nc}
N_c(l_{\rm max}) = \sum_{l= 2, 4, \dots}^{l_{\rm max}} \!\!\! n_c(l) 
= \frac{1}{12}\, l_{\rm max}^3 + \frac{5}8 \, l_{\rm max}^2 + \frac{17}{12} \, l_{\rm max}
\ee
\mparr{R3.5}
corresponding to $N_c = 6, \ 21, \ 49, \ 94 \dots$ for $l_{\rm max}=2, \ 4,\ 6, \ 8 \dots$.
(We did not include the unweighted $S_0$ in our counting.) 
Eq.~\eq{Nc} returns familiar numbers of DTI, DKI, etc components, which can be practically estimated via polynomial regression
of $\ln S_\g(b)$. 

\subsection{Parameter count: Model (\ref{S})--(\ref{K}), term-by-term}
\label{sec:par-model}
\bigskip

\mpar{R2,R3}
\new
So far, we counted the parameters of the {\it representation} (\ref{taylor}) of the signal in terms of its moments. This representation is completely general --- it only respects the time-reversal symmetry of the problem (only even orders $l$ are included).  
We now count the number of parameters defining the {\it biophysical model} \eq{S}--\eq{K} with ODF (\ref{P=Y}) up to $\lmax$. 
\keep
The $N_s = 4$ {\it scalar} parameters from kernel \eq{K} in the absence of CSF (or $N_s = 5$ if the CSF compartment is added), 
are complemented by $\lmax(\lmax+3)/2$ {\it tensor} parameters $p_{lm}$, obtained as $\sum_{l=2,4,\dots}^{\lmax} (2l+1) \equiv n_c(\lmax)-1$ (since \new $p_{00}\equiv \sqrt{4\pi}$ \keep \mpar{R2.4}
is set by the ODF normalization), yielding
\be \label{Np}
N_p(\lmax) = N_s + \lmax(\lmax+3)/2 \,.
\ee
Hence, $N_p = 9, \ 18, \ 31, \ 48, \ \dots$ for  $\lmax = 2, \ 4, \ 6, \ 8, \ \dots$ already for the two-compartment kernel \eq{K}, without including $S_0$ and CSF fraction in the count. 

Equation (\ref{Np}) reveals that the model complexity grows fast, as $\lmax^2$ due to the ODF, if we are to account for the rich orientational content of the realistic fiber ODFs in the brain. For the achievable  $\lmax \sim 4-8$, the dMRI signal in principle ``contains'' a few dozen parameters, none of which are known  {\it a priori}.

A superficial reason for poor fit quality is the high dimensionality (\ref{Np}) of the parameter space. However, comparing $N_c(l_{\rm max})$ with the corresponding number of model parameters $N_p(l_{\rm max})$, it naively looks like the series \eq{taylor} is overdetermined, $N_c \geq N_p$, already for $l_{\rm max} \geq 4$ --- meaning that the acquisition looks sufficiently informative starting from the DKI level. 

Unfortunately, this simplistic parameter count does not present the full picture, as {\it the information content is not evenly distributed among all the $N_c(\lmax)$ components}. 
It turns out that for $l_{\rm max}=4$ (DKI), there are not enough relations for  scalar model parameters $x$, while too many for  tensor  parameters $p_{lm}$. All model parameters can be determined from the series \eq{taylor} {\it only} starting from 
$l_{\rm max} \geq 6$, as we will now explain.

\newpage
\subsection{Perturbative approach to model (\ref{S})--(\ref{K}): LEMONADE}
\label{sec:lemo}
\bigskip

\new 
In \ref{app:lemo}, we derive the LEMONADE system of equations (\ref{lemo-m}), relating the linear combinations of the signal's moments to the model parameters. The rotationally invariant form of the LEMONADE system, one of the central results of this work, is as follows: 
\keep
\begin{subequations}\label{lemo}
\bea 
\label{D00}
\mkern-36mu
M^{(2),0} &=&   f \Da + (1-f) (3\Deperp + \dDe)   \\ 
\label{D20}
\mkern-36mu
\frac{M^{(2),2}}{p_{2}} &=&   f\Da + (1-f) \dDe   \\ 
\label{M00}
\mkern-36mu
M^{(4),0} &=& f\Da^{2} \! + \! (1-f) \lb \! 5 \Deperp^{2} + \frac{10}3 \Deperp \dDe + \dDe^{2}\rb \qquad \\
\label{M20}
\mkern-36mu
\frac{M^{(4),2}}{p_{2}} &=&   f\Da^{2} + (1-f) \lb \frac{7}3 \Deperp \dDe + \dDe^{2}\rb  \qquad \\
\label{L00} 
\mkern-36mu
M^{(6),0} &=& f\Da^{3} \\ \nonumber 
&  + & (1-f) \lb 7\Deperp^{2}(\Deperp + \dDe) + \frac{21}5 \Deperp \dDe^{2}  + \dDe^{3}\rb \qquad \\ 
\label{L20} 
\mkern-36mu
\frac{M^{(6),2}}{p_{2}} &=&  f\Da^{3} + (1-f) \lb  \frac{21}5 \Deperp^{2}\dDe + \frac{18}5 \Deperp \dDe^{2}  + \dDe^{3}\rb. \qquad 
\eea
\end{subequations}
Here $\dDe \equiv \Depar - \Deperp$. The system \eq{lemo} involves  rotationally invariant components $M^{(L),l}$ 
[defined in Eq.~\eq{MLl}] of the tensors  $M^{(l)}_{i_1\dots i_l}$, up to  minimal orders $L\leq 6$ and $l\leq 2$, enough to find all  4 scalar kernel parameters $x$, and ODF anisotropy invariant $p_2$. 

\new
The first four LEMONADE equations, corresponding  to $\lmax=4$, contain 5 unknowns. (Introducing an extra equation $M^{(4),4} \propto p_4$ does not help, as it introduces yet an extra unknown ODF invariant $p_4$.) This is why the $\lmax=4$ acquisition, equivalent to relying on the signal's curvature, or kurtosis, is fundamentally insufficient. This is an important fundamental limitation for the parameter estimation in the Standard Model. 
\keep 

The full system's 6 equations are in principle enough to determine 5 model parameters; even with added CSF compartment with its fraction and an isotropic $D_{\rm CSF}=3\,\mathrm{\mu m^2/ms}$, one could in principle still determine the 6 parameters from the appropriately modified system \eq{lemo}. 
\new
At the $\lmax = 6$ level, one could introduce even more  equations: $M^{(6),4} \propto p_4$, or $M^{(6),6} \propto p_6$, at the expense of  extra unknown ODF invariants $p_4$ and $p_6$. Practically, they would rely on the components $M^{(L),l}$ with high $l$, which are less accurately estimated. Tying $p_l$ with different $l$ to each other could be a way to increase precision, but at the expense of constraining the ODF shape to a particular functional form. Therefore, the LEMONADE system (\ref{lemo}) is minimally complex yet still fully unconstrained. 
\keep
LEMONADE is solved exactly in \ref{app:lemo-sol}, and its output is subsequently used to initialize the subsequent nonlinear minimization, Eq.~(\ref{F}) in Sec.~\ref{sec:fact} above.

\new
Crucially, LEMONADE, and hence the original nonlinear problem (\ref{F}), has multiple biophysically plausible solutions. 
The choice of the physical solution, to initialize problem (\ref{F}), is nontrivial, Sec.~\ref{sec:landscape}; in Sec.~\ref{sec:branch} below we will rationalize our recipe for making such selection voxel-wise. In the {\it Methods}, Sec.~\ref{sec:impl} and Fig.~\ref{fig:maps-all}, we summarize all RotInv computational steps. 
\keep

\section{Methods}

\subsection{In vivo dMRI}
\label{sec:invivo}
\bigskip

Three healthy volunteers underwent imaging on a Siemens Prisma 3T whole-body MRI scanner. 
The study was approved by the local Institutional Review Board, and informed consent was obtained and documented from all participants. 
The MRI scanner was equipped with a $80 \,\mathrm{mT/m}$ gradient system and a 64-channel receiver head coil. The body coil was used for transmission. A monopolar diffusion-weighted EPI  sequence was used to acquire the dMRI data. Diffusion weighting was applied along 64 isotropically distributed gradient directions for each of the 21 $b$-values  equidistantly distributed in the range $[0:0.5:10]\, \mathrm{ms/\mu m^2}$. Following imaging parameters were kept constant throughout the data acquisition sequence: $\mathrm{TR/TE}: 4000/105\,\mathrm{ms}$,  matrix: $80 \times 80$, NEX: 1, in-plane resolution:  $3 \times 3 \,\mathrm{mm^2}$, slice thickness: $3\,\mathrm{mm}$, slices: 38, parallel imaging: GRAPPA with acceleration factor 2, reconstructed using the adaptive combine  algorithm to ensure Rician data distribution, multiband acceleration with factor 2, and no partial Fourier. Total scan time was just below 2 hours per person.

\subsection{Image processing}
\bigskip

MP-PCA noise estimation and denoising method \cite{MPnoise,veraart2016-noise} allowed us to preserve only the significant principal components and to strongly reduce the noise in the data and to estimate spatially varying noise map. The positive signal bias, inherent to low-SNR magnitude MR data, was removed by using the method of moments \cite{koay}, where the denoised signal was used as a proxy for the Rician expectation value. 
\mpar{R2.3, R3.5}
Denoised and Rice-floor-corrected images were subsequently corrected for Gibbs ringing \cite{kellner2016}, geometric eddy current distortions and subject motion \cite{Andersson2016}.

\subsection{Estimating moments from dMRI signal}
\bigskip

Expansions in moments and in cumulants are mathematically equivalent; the combinatorial relation between moments and cumulants (``linked cluster expansion") was established in statistical physics \cite{mayer1941}, and is reviewed in \cite{Kiselev2010_diff_book}.  
While expanding in moments (\ref{taylor}) is the most optimal for the analytical treatment (contributions from different tissue compartments add up), we observe that estimation from the dMRI signal is more accurate from the cumulant series. 
Hence, for  accuracy, we first estimate the cumulant tensors $C^{(l)}$
\be \label{cumexp}
\ln {S_\g(b) \over S_0} =  - b C^{(2)}_{i_1 i_2} g_{i_1} g_{i_2} + b^2 C^{(4)}_{i_1 \dots i_4} g_{i_1}\dots g_{i_4} - \dots 
\ee
and the unweighted $S_0$, via $b$-matrix pseudoinversion applied to $\ln S_\g(b)$, with voxel-specific weights \cite{veraart2013}  up to $l_{\rm max} = 6$.
We then convert $C^{(l)}_{i_1 \dots i_l}$ to the moments $M^{(l)}_{i_1 \dots i_l}$ as derived in \ref{app:mom-cum}.
For  unbiased estimation, we use only  shells within $0\leq b \leq 2.5$, where the cumulant series is expected to converge.

\subsection{Implementation and computation time} 
\label{sec:impl}
\bigskip

\mpar{R3.4 Secs reordered}
Processing steps (Fig.~\ref{fig:maps-all}) were implemented in {\it Matlab} (MathWorks, Natick, MA, USA), according to Eqs.~\eq{cumexp}, \eq{MLlm}, \eq{MLl}, and \eq{Dbar}--\eq{fp2} 
using standard library functions, and the Levenberg-Marquardt algorithm for subsequent nonlinear minimization of Eq.~\eq{F} initialized by LEMONADE output $x$ and $p_2$. 
For the whole brain (34383 voxels within the WM+GM mask for subject 1,  \new at our relatively coarse resolution) \keep on a desktop iMac (4 cores), it took under 2 min for estimating the cumulants using the $b$-matrix pseudoinversion with the voxel-specific weights \cite{veraart2013}, together with recalculating the moments \eq{MLlm} from the cumulants, \ref{app:mom-cum} (only the range $b\leq 2.5$ was used for unbiased estimation); 1.5 min for LEMONADE calculation (both branches); and $\sim 5$ min for  nonlinear  fitting, Eq.~(\ref{F}),
including all shells 
\del{(both branches for all voxels, for $b\leq 10$ and $b\leq 2.5$ respectively)}, 
using the corresponding LEMONADE solution (Sec.~\ref{sec:branch}) as fit initialization. Nonlinear fitting achieves  speedup because of the initial values being already quite close to the minima of $F$; we also precomputed corresponding integrals \eq{Kl} and their first derivatives in a broad range.

\section{Results}
\label{sec:results}

\subsection{Topology of parameter landscape}
\label{sec:landscape}
\bigskip

\begin{figure}[b!]
{\bf a}\includegraphics[width=1.6in]{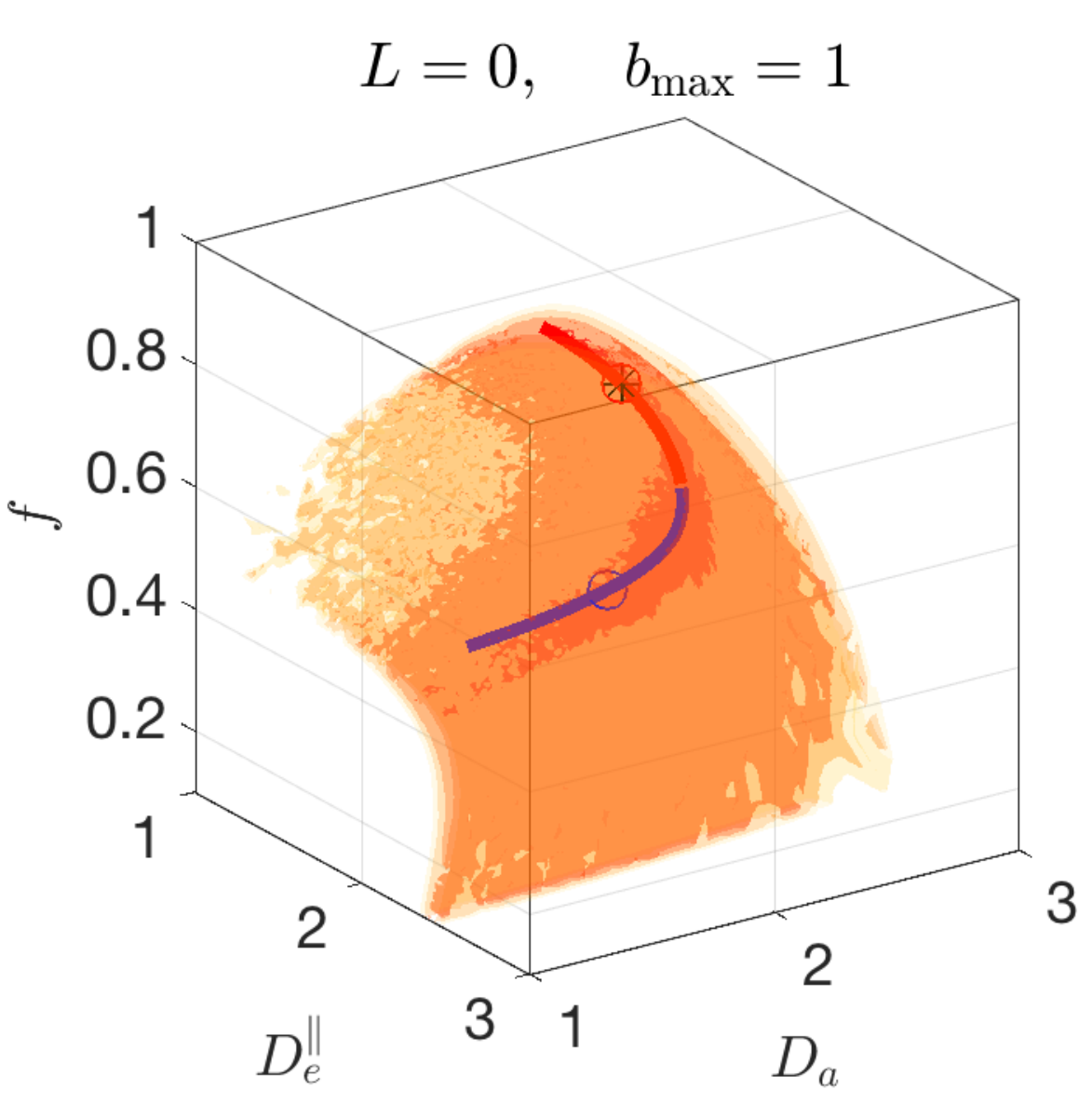}{\bf b}\includegraphics[width=1.6in]{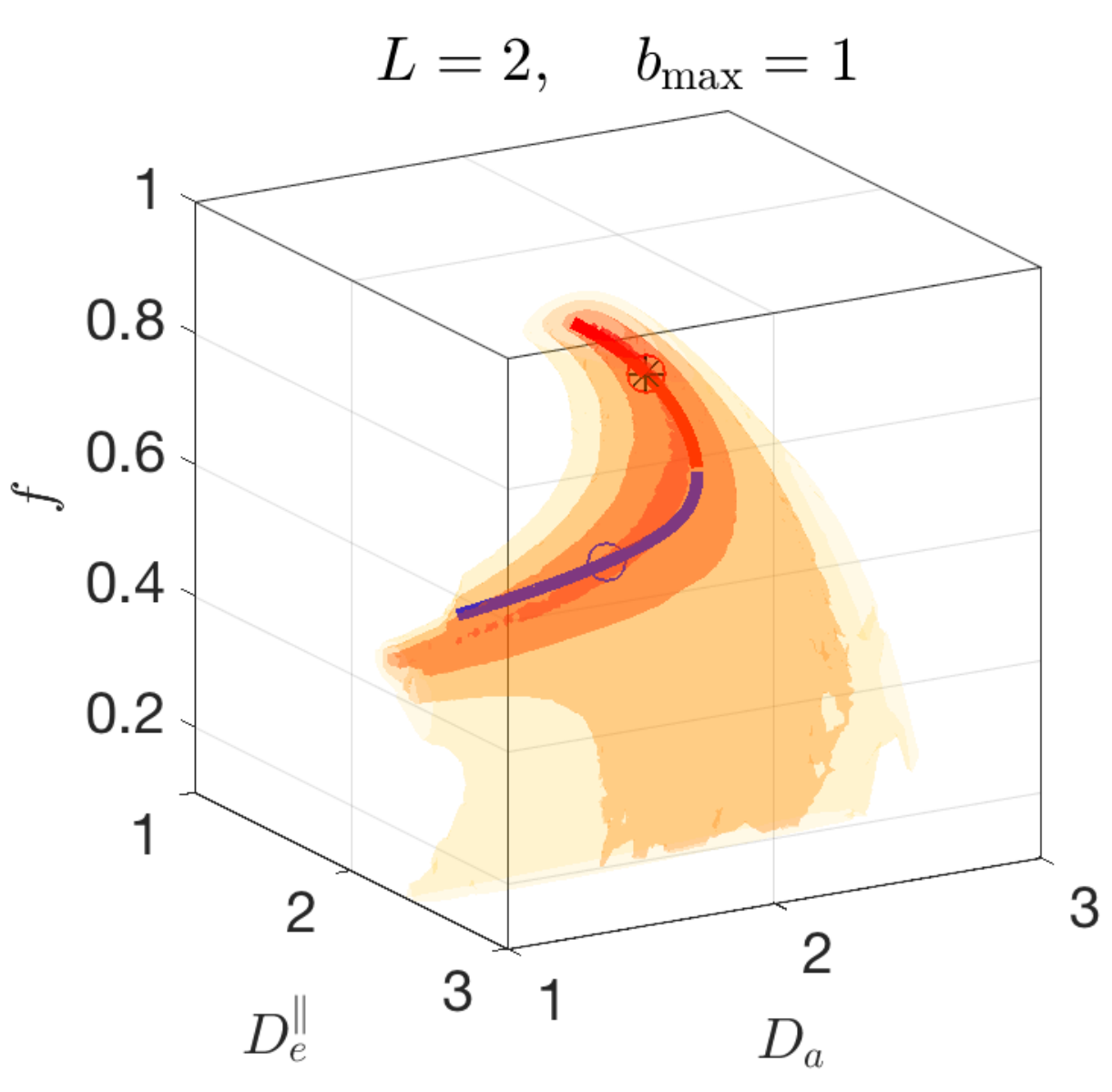}\\
{\bf c}\includegraphics[width=1.6in]{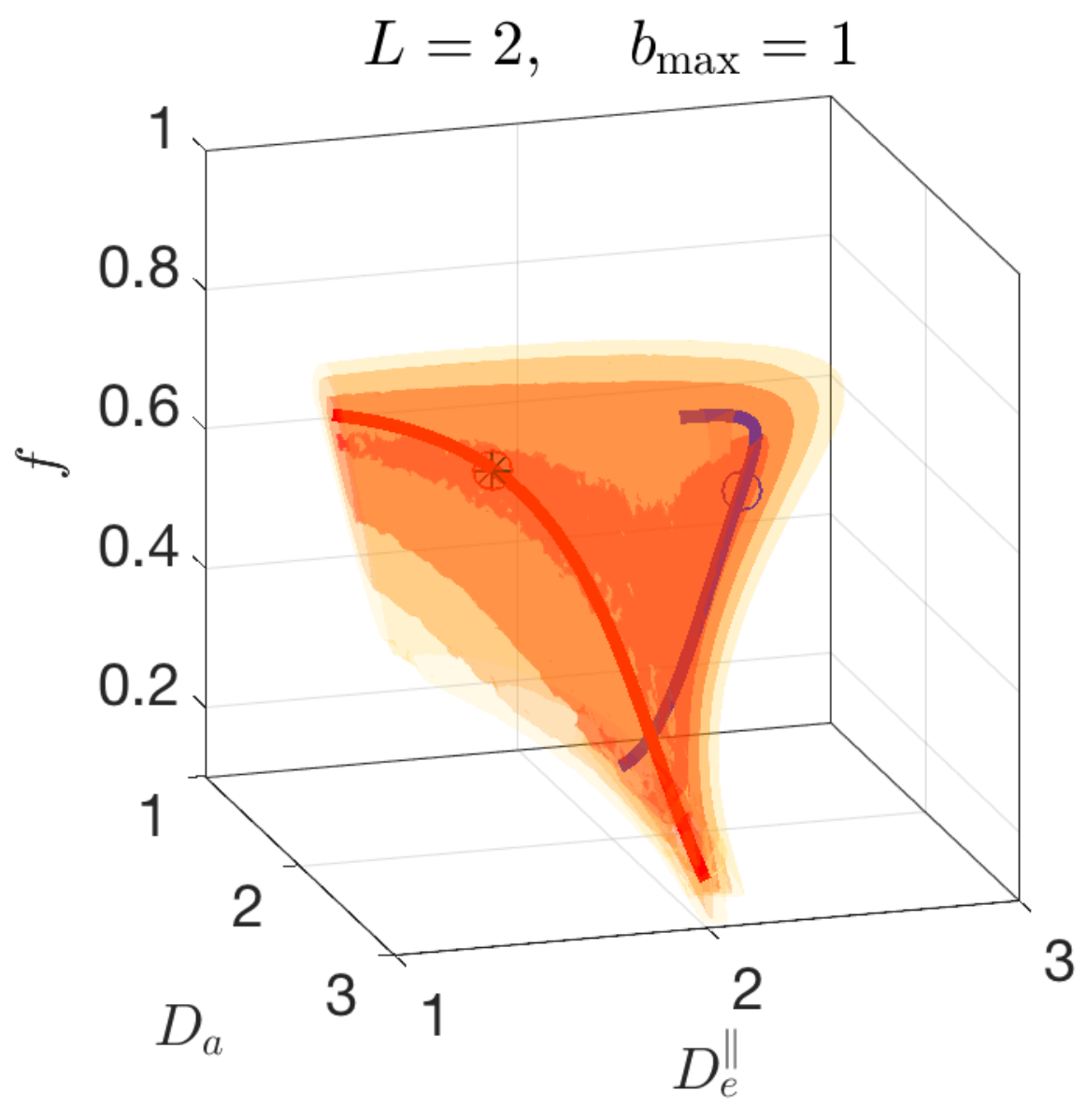}{\bf d}\includegraphics[width=1.6in]{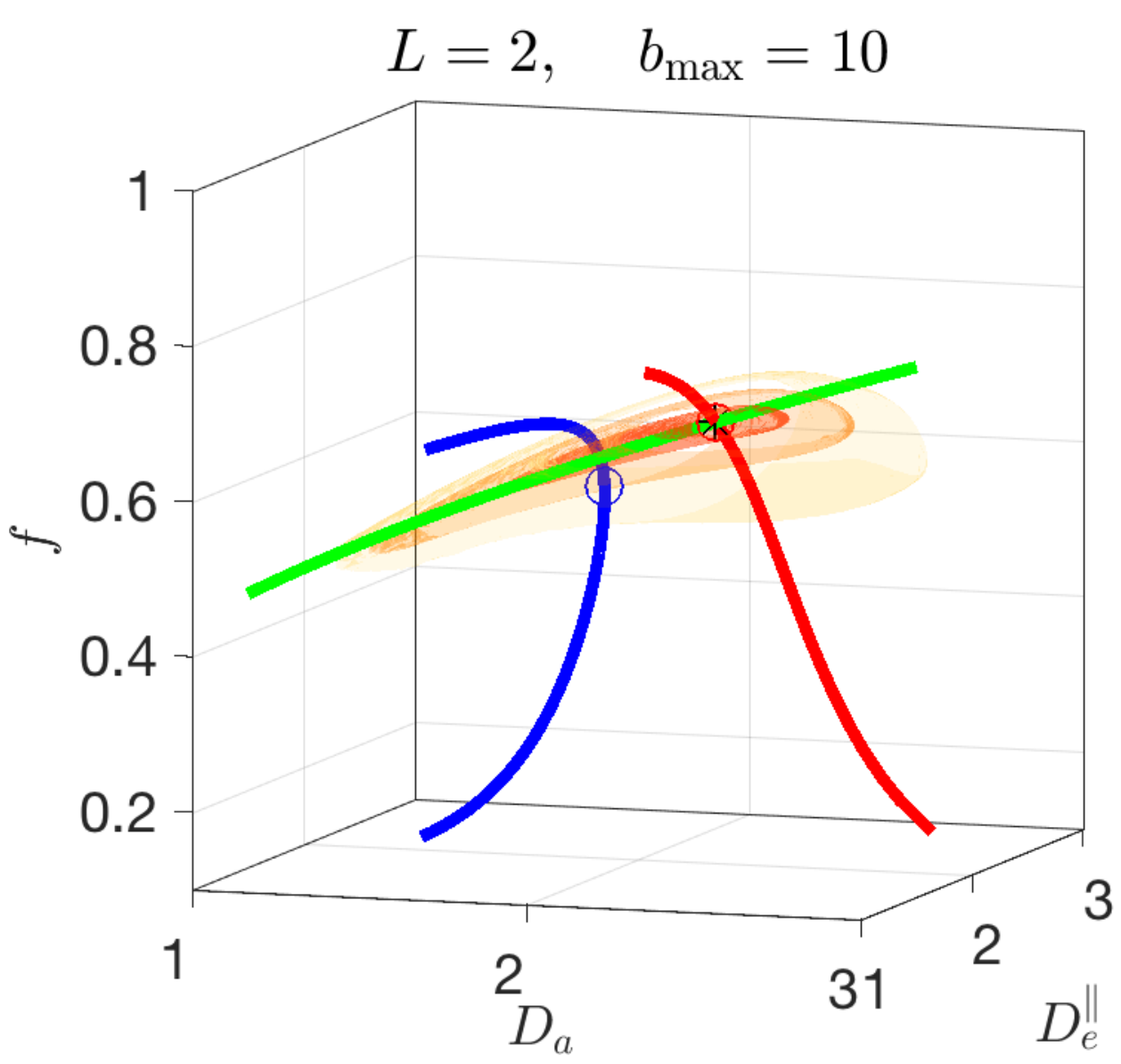}%
\caption{
{\bf Degeneracies in scalar parameter estimation} depending on the maximal  invariant $L$ used, Eq.~\eq{F}. 
Contour surfaces are darker for lower values of $F$. 
Low-``energy" landscape of nonlinear RotInv problem \eq{F} for system \eq{S=pK} is a 2-dimensional surface for 
$l=0$ invariant ({\bf a}), and  two 1-dimensional trenches with $l=0, \, 2$ invariants included ({\bf b, c}). 
The trenches can either match to form a single 1-dimensional manifold ({\bf b}), or be disjoint ({\bf c, d}), depending on the ground truth values (see Supplementary Figs.~\ref{fig:landscape08}--\ref{fig:landscape04}). 
Circles denote minima in both trenches, with the true minimum marked by $*$. 
Exact branches $\zeta=\pm$ of system \eq{lemo} (Sec.~\ref{sec:lemo})
are drawn in red and blue, reproducing the trenches up to $\O(b^2)$.  
Including large $b$ data further limits the landscape to the surface $f/\sqrt{\Da}=\mbox{const}$ arising solely from intra-neurite space \cite{jensen2016,highb} (green section in {\bf d}), helping, but not fully curing all degeneracies (see text and Figs.~\ref{fig:landscape08}--\ref{fig:landscape04}).  
}
\label{fig:landscape}
\end{figure}

\begin{figure*}[th!!]
\centering
\includegraphics[width=4.8in]{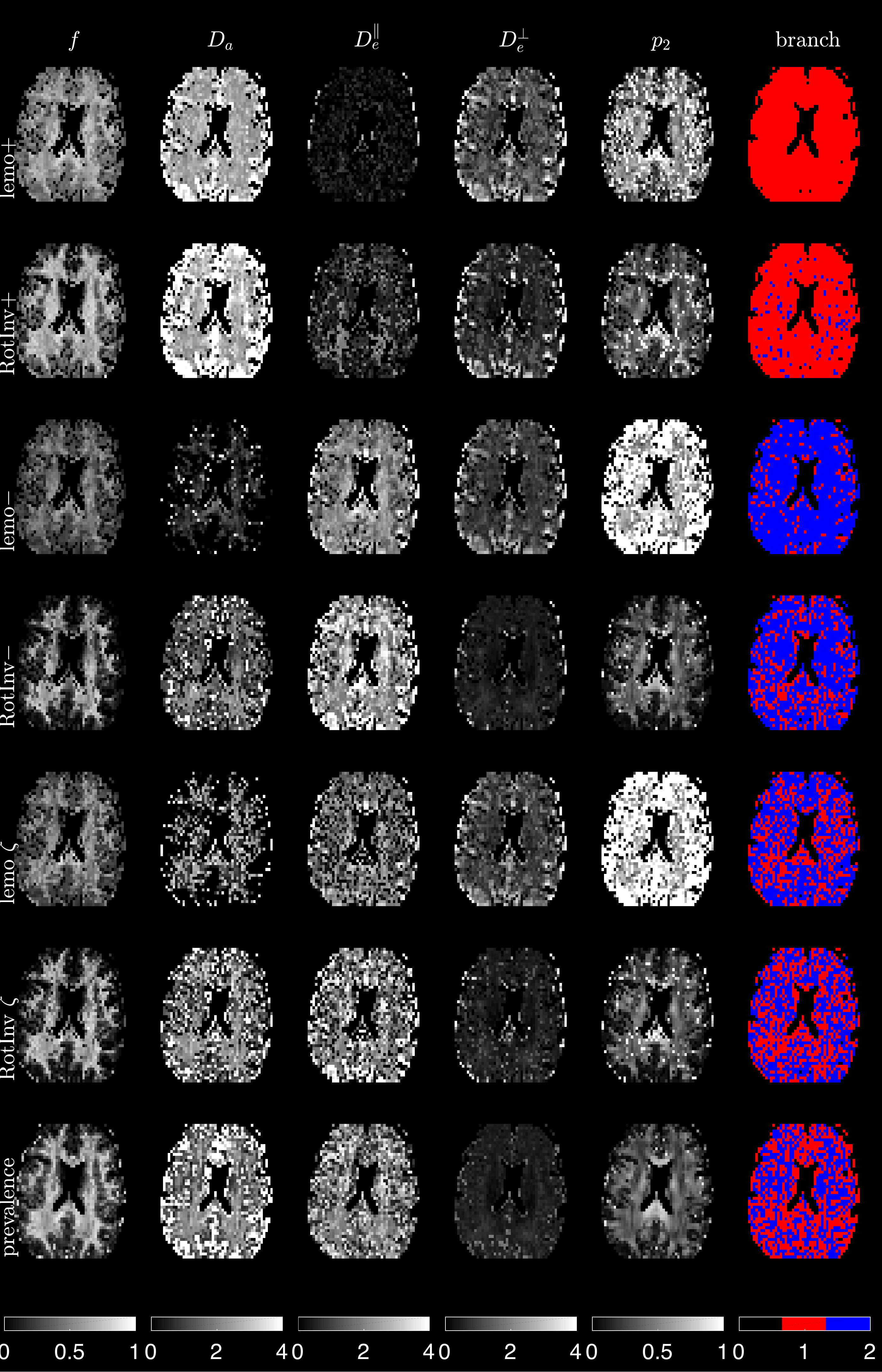}
\caption{
{\bf Parametric maps for both branches (LEMONADE and nonlinear RotInv estimation), and prevalence maps.} 
Exemplary maps for subject 1, mid-brain axial slice. 
\newr
The {\it branch} maps (last column) are calculated using Eq.~\eq{branch} based on the voxel-wise estimated parameters, with $\zeta=\pm$ corresponding to red/blue. \keep    
{\bf lemo$\pm$:} outputs of LEMONADE $\zeta =\pm$ branches, respectively (solution of Eqs.~\eq{lemo}), 
using only the  shells within $0\leq b \leq 2.5$. 
Note that $f_+> f_-$, as well as $\Da_+ > \Depar_+$ and $\Da_- < \Depar_-$, practically consistent with Eq.~\eq{branch}. 
For $\zeta = +$ branch, the output $\Depar < \Deperp$ is likely to be a result of the bias of moments estimation (a similar bias was observed in numerical simulations), since it is biophysically more plausible that $\Depar \gtrsim \Deperp$. 
\newr In a few voxels, the branch index $\zeta$ flips due to noise and partial volume effects affecting solution of Eqs.~(\ref{L00})--(\ref{L20}). \keep
{\bf RotInv$\pm$:}  Full nonlinear estimation outputs of gradient-descent minimization, Eq.~\eq{F}, using all $b$ shells, initialized via the corresponding {\it lemo$\pm$} maps. We observe the same qualitative features as in the LEMONADE maps, while numerically bringing $\Depar$ and $\Da$ closer. 
Importantly, the branch index $\zeta$ is fairly stable (cf. also histograms in Fig.~\ref{fig:branch}) --- for the vast majority of voxels, the nonlinear fitting of the full problem \eq{F} does not change the branch index $\zeta = \pm$ 
\newr calculated based on the estimated parameters.  \keep 
 \new 
 {\bf lemo $\zeta$:} Selecting the LEMONADE branch $\zeta=\pm$ voxel-wise, based on the proximity of the solutions for the redundant Eqs.~\eq{L00} and \eq{L20}, cf. Sec.~\ref{sec:branch} for details. 
{\bf RotInv $\zeta$:} Full RotInv output based on {\it lemo $\zeta$} from previous row.  
{\bf Prevalence:} Maps generated using the prevalence method (see text), current proxy for the ground truth. 
Note quantitative similarity of the parameter maps between {\it RotInv $\zeta$} and {\it prevalence}.  
Extra-axonal diffusion tensors are practically always prolate, with $\Deperp$ values typically below 1, while $\Da$ and $\Deperp$ around 2. 
\del{The same combination of the $\zeta=\pm$ RotInv maps for WM/GM, but now calculated only based on the $0\leq b \leq 2.5$ measurements, a proxy for a clinically feasible acquisition. While the results are noisier, the overall correspondence with the full acquisition is evident.} 
\del{Note branch map roughly corresponding to WM/GM contrast. 
(However, extra-neurite tensor is not always prolate.)} 
\keep
}
\label{fig:maps-all}
\end{figure*}

\begin{figure*}[t]
{\bf a}\includegraphics[width=7.2in]{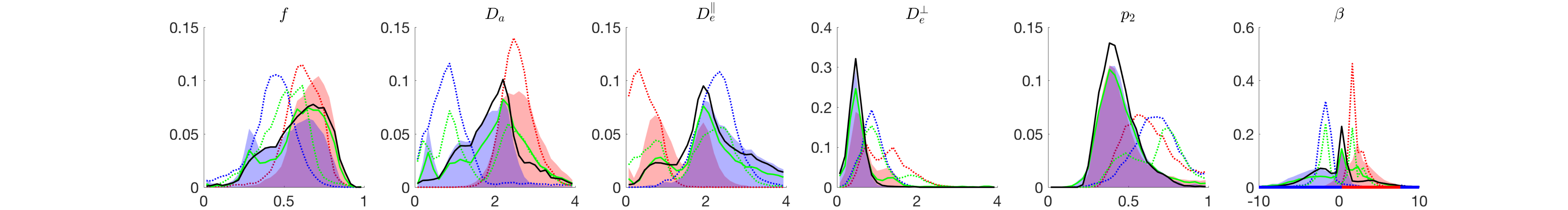}
{\bf b}\includegraphics[width=7.2in]{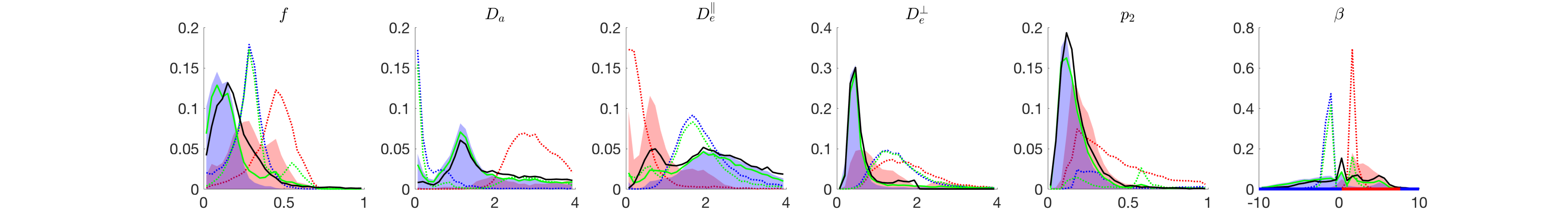}
\caption{
\new
{\bf Degeneracy in parameter estimation with human dMRI data as bimodality in parameter histograms.}
Parameter histograms for WM, {\bf a}, and for GM, {\bf b}.   
Red/blue dashed  histograms  correspond to $\zeta = \pm$ solutions 
$\{f, \Da, \Depar, \Deperp, p_2\}_\pm$ of LEMONADE system \eq{lemo};  filled histograms are obtained via RotInv fitting \eq{F} up to $b\leq 10$, initialized by the $\zeta=\pm$ solutions of system \eq{lemo}.
\newr The last column shows the branch selection parameter $\beta$, Eq.~(\ref{branch}), with the red/blue intervals corresponding to the $\pm$ branches. 
\keep
Nonlinear fitting \eq{F} brings the $\Da$ and $\Depar$ values closer to each other, and reduces $\Deperp$, as in the maps in Fig.~\ref{fig:maps-all}. 
Green dashed and solid histograms correspond to the branch selection method of Sec.~\ref{sec:branch} (cf. Fig.~\ref{fig:maps-all}, {\it lemo $\zeta$}) and the respectively initialized full RotInv output (Fig.~\ref{fig:maps-all}, {\it RotInv $\zeta$}). 
Note the qualitative and quantitative agreement between the voxel-wise branch selection output of RotInv (green) and prevalence method (black) described in Sec.~\ref{sec:branch}.  
\del{which mostly agrees with $\zeta=+$ in WM and with $\zeta= -$ in GM, cf. histograms of the branch ratio $\beta$ falling within red/blue intervals, and Fig.~3 for the branch index maps.} 
\keep
}
\label{fig:branch}
\end{figure*}

\begin{figure}[t]
\includegraphics[width=3.3in]{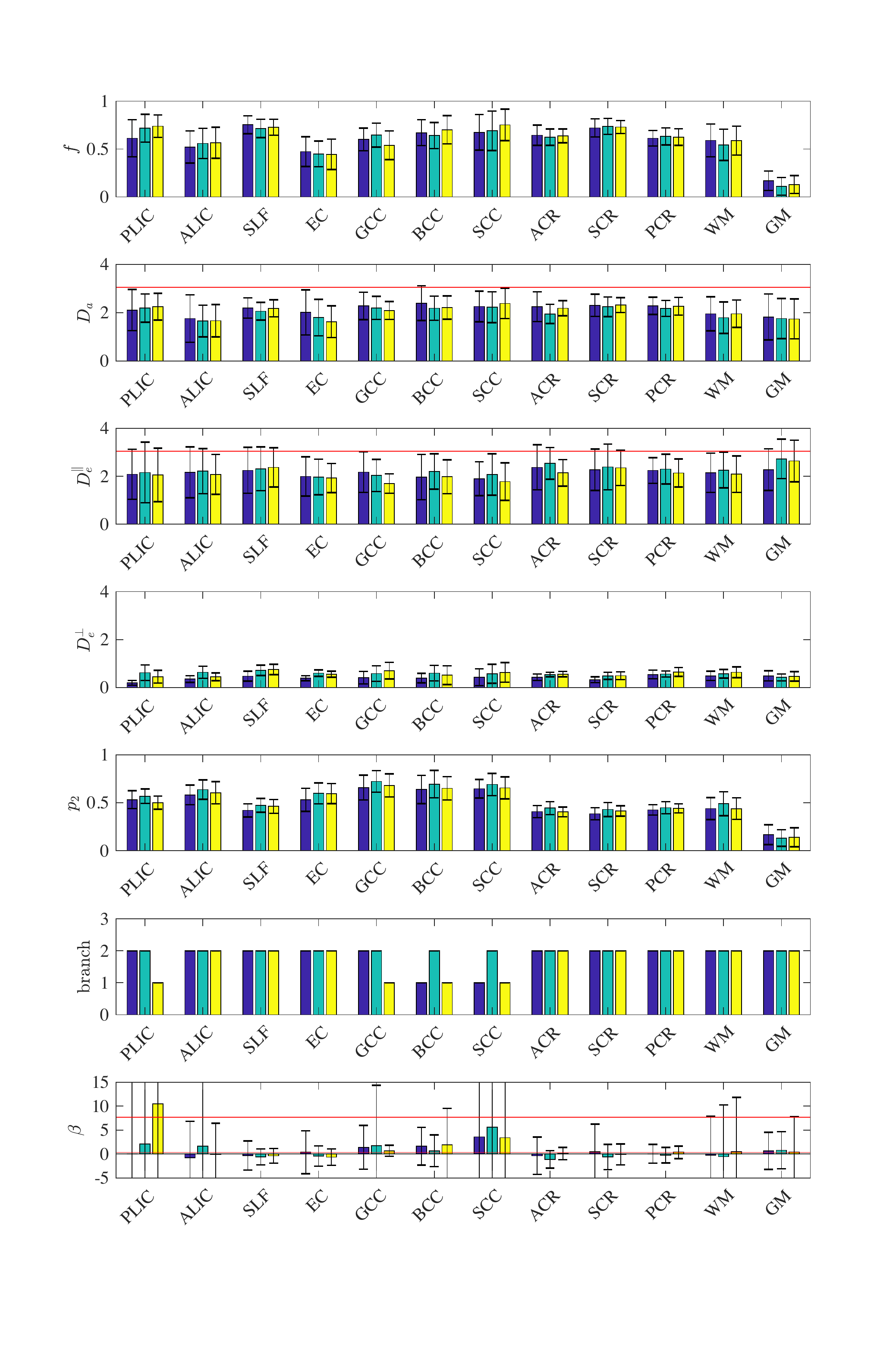}
\caption{
\new 
\rem{New Fig. }
{\bf ROI values.} 
Mean ROI values for the 3 subjects, based on the prevalence method, in 10 standard WM ROIs (JHU atlas), 
as well as in all WM and GM voxels. The maximal water diffusivity value $3\units{\mu m^2/ms}$ for the body temparature is drawn as red lines for $\Da$ and $\Depar$. 
Inter-ROI variability clearly exceeds inter-subject variability for $f$ and $p_2$; the ROI differences in diffusivities are present albeit less pronounced.  
The `branch'  values (1, 2 for $+$, $-$ correspondingly) are calculated based on ROI means according to Eq.~\eq{branch}, while the corresponding $\beta$ values are calculated voxel-wise, and averaged over ROIs. The $\zeta=+$ branch would correspond to $\beta$ falling in-between the two red lines, and $\zeta=-$ otherwise, cf. Eq.~\eq{branch}. Branch assignment based on the fairly noisy estimates of $\beta$ is not robust and needs further validation. 
\keep
}
\label{fig:bars}
\end{figure}

The contour plots of values of RotInv energy function $F$, Eq.~\eq{F}, shown in Fig.~\ref{fig:landscape} (cf.  also Supplementary Figs.~\ref{fig:landscape08}--\ref{fig:landscape04} for more examples), illustrate that the minimization landscape is generally quite {\it flat} in at least 1 dimension, and there exist  {\it multiple minima}. 
We emphasize from the outset that these degeneracies are {\it intrinsic} to the  problem \eq{S}--\eq{K}, and are not introduced by the above RotInv framework or the particular way \eq{F} of solving the system \eq{S=pK}. 
Rather, this framework allows us to uncover their general origin -- namely, the multi-compartmental character of the kernel \eq{K}. 

We now focus on the topology of the low-energy landscape of $F$ in order to understand degeneracies in parameter estimation, 
which is crucial for initializing the search for parameters $x$ within the biophysically correct domain, 
and for speeding up the solution of system \eq{S=pK}. 
As mentioned in {\it Theory} Section above, our analytical method to study this topology, is to approximate the  signal \eq{S} by its low-$b$ expansion \eq{taylor}, whose consecutive terms are equivalent to diffusion tensor imaging (DTI, $\sim b$), diffusion kurtosis imaging (DKI, $\sim b^2$), etc. 
Empirically, it is  known that DKI \cite{DKI} approximates clinical dMRI signal ($b_{\rm max} \sim 1 - 2$) quite well, further justifying studying the series \eq{taylor} up to 
${\cal O}(b^2)$, and perhaps, up to ${\cal O}(b^3)$. 

For low enough $b$ (typically used in the clinic), nonlinear fitting \eq{F} practically corresponds to matching first few moments of the signal \eq{taylor} to those of the model \eq{S}. 
In Sec.~\ref{sec:lemo} above we exactly derived this matching, the LEMONADE system \eq{lemo}, for up to $\O(b^3)$. 
We can now calculate the dimensionality of the low-energy ``attractor" manifolds in Fig.~\ref{fig:landscape} by  simple counting of equations. 

Taking the $l=0$ invariant alone, $S_0 = K_0$, is exactly equivalent to isotropic (or ``powder") signal averaging \cite{jespersen2013,lasic2014},  
\mparr{R3.6}
\newr a result subsequently used 
in the ``spherical mean technique" (SMT) \cite{kaden2016mrm,smt}.  \keep \mpar{R3.6}
Expanding the relation $S_0 = K_0$ up to $\O(b^2)$ yields a 2-dimensional {\it surface}, in accord with the two equations \eq{D00} ($\sim b$) and \eq{M00} ($\sim b^2$) for the 4 scalar parameters $x$, cf. Fig.~\ref{fig:landscape}a (note that isotropic averaging discards nontrivial $p_l$ with $l>0$). \new This rationalizes the empirical need for the $4-2=2$ constraints introduced in the SMT; their validity will be discussed in Sec.~\ref{sec:constraints} below. \keep 
While, theoretically, the $l=0$ technique allows one to factor out the ODF, we can see that practically determining all 4 scalar parameters $x$ from the single invariant $K_0$ requires the sensitivity to the signal's moments up to $M^{(8)}$ (namely, to their full traces), i.e. up to $b^4$ in expansion (\ref{taylor}), which is extremely hard to achieve. Intuitively, it is quite obvious that discarding the orientational content of the signal makes parameter estimation less informative. 

According to the analysis in Sec.~\ref{sec:lemo} above, staying at the same level $\O(b^2)$, and including the $K_2(b)$ invariant, $L=2$, adds one extra parameter $p_2$ describing the sensitivity to the lowest-order ODF anisotropy, and two extra equations, (\ref{D20}) and (\ref{M20}), 
turning the surface into the two narrow 1-dimensional {\it trenches} in parameter space, Fig.~\ref{fig:landscape}b,c (the first 4 equations of the system \eq{lemo} for  5 parameters: $x$ and $p_2$). Having a perfectly flat trench is obvious from the counting of equations; getting \new what looks like \keep {\it two} trenches is nontrivial. 

The trenches $f_\zeta(p_2)$ (Eq.~\eq{fp2} in \ref{app:lemo-sol}), labeled by branch index $\zeta=\pm$, Eq.~(\ref{zeta}), 
\new represent the exactly derived ``flat" one-dimensional continuous manifold that is determined by rotationally invariant combinations 
of signal's moment tensors $M^{(2)}$ and $M^{(4)}$. 
The parameterization of this manifold, derived as the two branches of quadratic equation \eq{qeq}, makes it appear within the physically meaningful domain of parameters as either one trench (the $\zeta=\pm$ parts merge), Fig.~\ref{fig:landscape}a,b, or as two disjoint trenches, Fig.~\ref{fig:landscape}c, depending on the ground truth values. 
\keep 
Physically, the dual branches come from having two tissue compartments, cf. the ``toy model" of \ref{app:toy}. There, we emphasize that both sets of values (obtained after imposing an extra constraint $p_2=p_4=1$ \mparr{R3.9} \newr of a single-fiber case) \keep can look perfectly plausible, and the ``wrong'' set corresponds to swapping  intra- and extra-neurite parameters --- which carries the danger of  misrepresenting parameter changes in pathology. 
Including the $K_4(b)$ invariant adds one extra parameter $p_4$ and only one equation for $M^{(4),4}$ (since $K_{l}(b)|_{b\to 0} \sim b^{l/2}$); hence, further invariants $K_{l>2}$ do not provide extra equations relative to the number of unknowns up to $b^2$. 

As a result, if an acquisition is only sensitive to $\O(b^2)$, due to e.g. $b$-range or SNR limitations (as it is often the case in the clinic), the parameter estimation problem will be in general ``doubly degenerate'', as we empirically observed recently for a particular ODF shape \cite{jelescu2016}: with respect to selecting the trench (discrete degeneracy), and due to the perfect flatness (zero curvature) along either trench (continuous degeneracy). Our exact solution of system \eq{lemo}  establishes that these two kinds of degeneracy are general ``hidden'' features of  problem \eq{S}--\eq{K}, due to every point in each branch exactly matching the $b$ and $b^2$ terms in the Taylor expansion \eq{taylor}. 

Furthermore, as already mentioned in Sec.~\ref{sec:lemo},  simplistic counting of parameters, without separating them into the scalar and ODF parts, can be misleading: the $N_c(4) = 21$ DKI parameters 
are not enough to determine the corresponding $N_p(4)=18$ model parameters, Eq.~\eq{Np}, since the excess parameters over-determine the ODF, whereas the  kernel \eq{K} remains under-determined. (This issue becomes even more severe if the kernel has more than 2 compartments, e.g. by adding the CSF pool.) This means that, unfortunately,  a popular DKI acquisition is not enough to resolve two-compartment model parameters due to a perfect 1-dimensional degeneracy within a chosen trench, unless, e.g., $p_2$ is fixed by the ODF shape: $p_2\to 1$, the aligned fibers assumption \cite{KM,wmdki} (which still then requires making a discrete branch choice). 

We also note that including the $l>0$ invariants in system \eq{S=pK} is only possible for anisotropic ODFs, with $p_l > 0$. 
Physically, it is expected since the less symmetric the system, the more inequivalent ways it enables for probing its properties; this intuition has had far-reaching consequences in quantum theory of excitations of non-spherical nuclei \cite{bohr-mottelson}. For our classical physics purposes, it means that it is overall harder to determine parameters of gray rather than white matter.  
Fortunately, in the brain,  ODF is at least somewhat anisotropic; the anisotropy parameter $p_2$ 
is generally nonzero even in the gray matter, as we can see {\it a posteriori}, cf. 
Figs.~\ref{fig:maps-all}--\ref{fig:bars}.

\subsection{Bimodality of parameter  estimation in human dMRI} 
\label{sec:bimodality}
\bigskip

In Figures \ref{fig:maps-all} and \ref{fig:branch} we demonstrate the double degeneracy of parameter estimation problem \eq{S}--\eq{K}, anticipated from topology of Fig.~\ref{fig:landscape}, 
using a dedicated human dMRI acquisition (cf. {\it Methods}). 
We first solve the LEMONADE system \eq{lemo} (using  subset with $b\leq 2.5$), map all parameters (cf. Fig.~\ref{fig:maps-all} {\it lemo $\pm$}), and plot histograms for its both branches $\{f, \Da, \Depar, \Deperp, p_2\}_{\zeta=\pm}$ (red and blue dotted lines in Fig.~\ref{fig:branch}) in white matter (WM, $\sim 11,000$ voxels), and in gray matter (GM, $\sim 13,000$ voxels), selected using probability masks. 
(WM mask was further thresholded by ${\mathrm{FA}>0.4}$ to minimize partial volume effects.)

We then use pairs of LEMONADE solutions with $\zeta=\pm$ in each voxel to initialize the full gradient-descent nonlinear RotInv minimization \eq{F}, for which all the data with $b\leq 10$ is used (cf. Fig.~\ref{fig:maps-all} {\it RotInv $\pm$}). This leads to the corresponding shaded histograms in  Fig.~\ref{fig:branch}. We can see that the output of full optimization \eq{F} is qualitatively similar to that based on the Taylor expansion \eq{taylor}, with bimodal parameter histograms corresponding to the fundamental degeneracy of the parameter landscape corresponding to the two distinct branches of solutions. 

Our analysis shows that the two branches are qualitatively distinct in the following ways: $f_+ > f_-$; also, usually, $D_a > \Depar$ for $\zeta = +$ and $D_a < \Depar$ for $\zeta = -$ (cf. Eq.~\eq{branch} below).  
Generally, neither solution can be discarded based on parameter values alone, as they often fall within plausible biophysical bounds $0<D\lesssim 3$ and $0<f<1$. (We show values up to $D=4$ to illustrate the role of noise in broadening the histograms.) 
Figure \ref{fig:noiseprop} in Supplementary Section \ref{app:noise} shows that improvement in accuracy gained by nonlinear minimization \eq{F} 
relative to LEMONADE occurs because small errors in estimated moments in the finite range of $b$ translate into greater errors in the LEMONADE solutions, mostly due to errors in estimating the moment tensor $M^{(6)}$.

\subsection{Branch index $\zeta$ in terms of ground truth parameters} 
\bigskip

Our analysis in Fig.~\ref{fig:maps-all} (cf. {\it branch} column) and in Fig.~\ref{fig:branch} highlights the {\it stability of the branch index $\zeta$}: The exact bimodality, following from the topology of minimization landscape at low $b$, still affects parameter estimation even at very high $b$, \new since the $\O(b^3)$ terms and beyond generally result in a local minimum in either trench. \keep Hence, branch assignment is akin to a discrete topological index, characterizing which part of the parameter space a given imaging voxel belongs to, based on its ground truth values.  

In \ref{app:lemo-sol}, Eq.~(\ref{zeta}),  we derive that the choice of the $\zeta =  + $ branch corresponds to the ratio $\beta$ between ground truth compartment diffusivities falling within the interval 
\be \label{branch}
4 - \sqrt{\ts{40\over 3}}  < \beta  < 4 + \sqrt{\ts{40\over 3}} \,, \quad \beta =  \frac{\Da - \Depar}{\Deperp} \,, 	
\ee
and the $\zeta = -$ branch should be chosen otherwise. 

Branch choice is nontrivial because we {\it a priori} do not know the ground truth values $\Da$, $\Depar$ and $\Deperp$ entering Eq.~\eq{branch}; besides, these values may generally vary across the brain, and be altered in pathology. 
Noise may affect estimated diffusivities enough to switch the estimated branch ratio $\beta$ (Figs.~\ref{fig:maps-all}--\ref{fig:bars} and \ref{fig:noiseprop}), which is particularly noisy due to the division by a small $\Deperp$ and its imprecise estimator. 

%

 %

\begin{figure*}[t!!]
\centering
{\bf a}\ \includegraphics[width=2.0in]{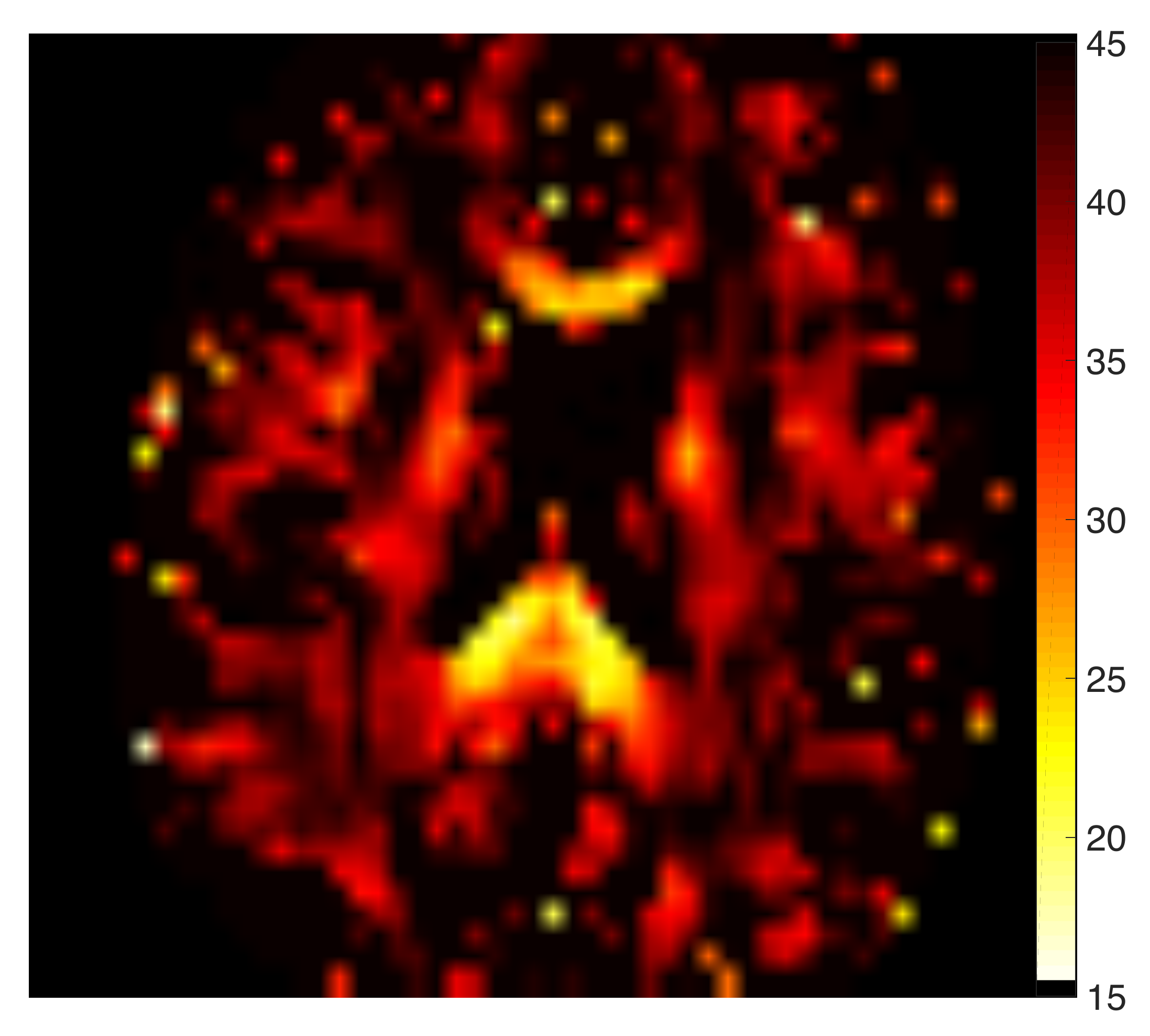}\hspace{8mm}
\quad {\bf b}\ \includegraphics[width=4.0in]{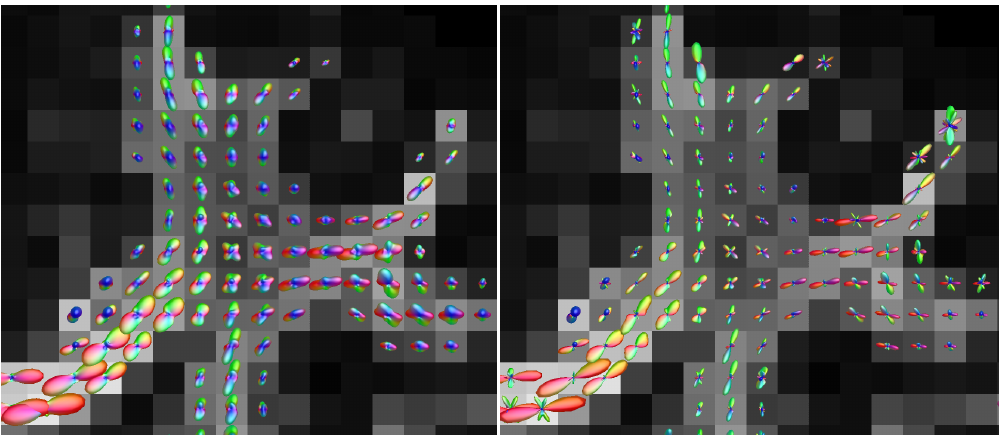}\\
{\bf c}~\includegraphics[width=6.5in]{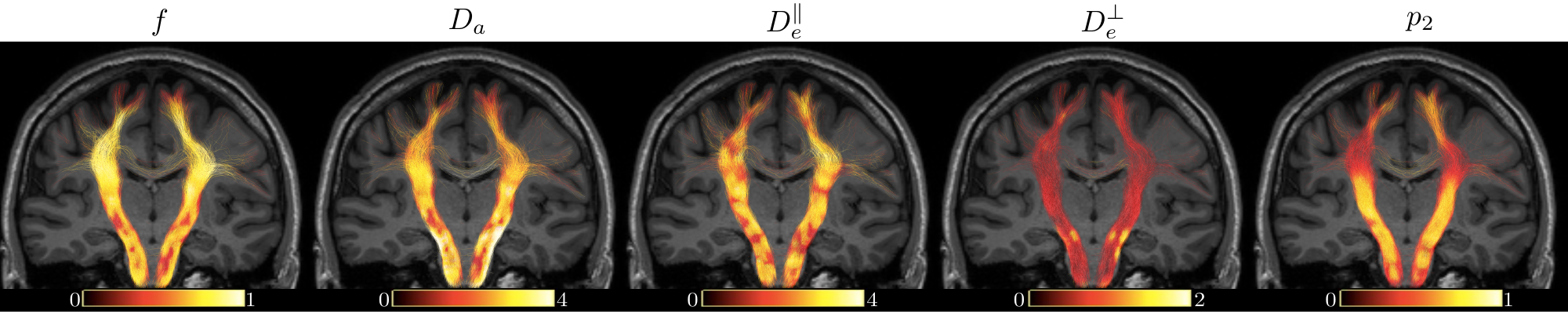}%
\caption{
{\bf ODF dispersion, local kernel-based tractography, and parameter maps.} 
{\bf a,}
Fiber orientation dispersion angle $\theta_{\rm disp}$ from $p_2$ (prevalence method) emphasizes major WM tracts. 
{\bf b,} 
Empirical reference ODF $\P^{(0)}$ (left, see text and \ref{app:K0}),
and the fiber ODF calculated using $p_{lm}$ from Eq.~\eq{fact} (right), for the $b=5$ shell,  with $l\leq l_{\rm max}=6$. 
Note the strong ODF sharpening effect, due to the deconvolution with locally estimated kernels $K_l(b,x)$.
{\bf c,}
Coronal view of a structural MR image (MPRAGE) overlaid with a reconstruction of the corticospinal fiber tract colored according to the value of the respective parametric maps $x$ and $p_2$ (via prevalence method, Fig.~\ref{fig:maps-all}) at each point along the track. Fiber tracts have been reconstructed using improved probabilistic streamline tractography \new \cite{Tournier2010} \keep  by integration over local fiber ODFs calculated using $p_{lm}$ estimated from Eq.~\eq{fact} using the voxel-wise estimated kernel $K_l(b,x)$, for the $b=5$ shell with maximal SH order $\lmax=6$. 
}
\label{fig:odf}
\end{figure*}

\subsection{Branch selection} 
\label{sec:branch}
\bigskip

Sensitivity to $\O(b^3)$ terms and beyond, ideally, will 
determine which branch $\zeta=\pm$ is the correct one, as well as the true minimum (solution) along this branch. 
Practically, however, branch selection  from realistic noisy data turns out to be quite challenging.
Relying on very large $b$ data is also tricky: the scaling \cite{highb,mckinnon2016} $S\sim f/\sqrt{b \Da}$ for $b\Da\gg 1$ 
is similar for both branches, since 
$f_+> f_-$ and $\Da_+ > \Da_-$, compensating each other's effect.
(Note, however, the $1/b$-scaling of the ODF variance suggested in \cite{highb} as a method to uncouple $f$ and $\Da$;
\new the precision of this uncoupling has so far been generally not enough to make a decisive branch choice for single fiber populations in WM.) \keep 
\del{which supports our WM branch choice $\zeta=+$  here.}

For now, to get the best available proxy for the ground truth, and to select the branch, we use our  wide-$b$-range dedicated dMRI acquisition and calculate the scalar parameters independently of the branch location, solely based on the {\it prevalence}.
In order to not bias our outcome on the branch choice, for each voxel, we initialize the problem \eq{F} using 100 random starting points within the biophysically plausible parameter range (such as $0<f, p_2<1$, and $0<D<3$ for all diffusivities), observe that the fit outcomes typically cluster around a few points in the parameter space, and select 
\mparr{R3.10}\newr
the mean of the largest cluster \keep (after excluding the outcomes outside the plausible bounds). 
An increasing $b$-range broadens the basin of attraction of the true minimum in a model calculation \cite{jelescu2016}.  
Supplementary Fig.~\ref{fig:noiseprop} justifies this method using simulations for a range of ground truth values with added noise. 
Certainly, this method may still fail in some voxels, but the overall maps (Fig.~\ref{fig:maps-all} bottom row, and Fig.~\ref{fig:odf}c) look sufficiently smooth and biophysically plausible. \new While we do not have an independent validation method for the prevalence parameters, \keep 
\mpar{new Fig}
we performed the prevalence calculation for all three subjects and observed that the prevalence maps and histograms are 
similar, as are their ROI values (Fig.~\ref{fig:bars}). 

Histograms of branch ratio \eq{branch} in Fig.~\ref{fig:branch}, as well as branch maps in Fig.~\ref{fig:maps-all}, 
\del{suggest that the $\zeta=+$ branch dominates in WM while $\zeta =-$ branch prevails in GM.} 
\new 
unfortunately yield no clear branch selection depending on the ROI. It looks like the left inequality, $4-\sqrt{40/3} < \beta$, is practically most often choosing the branch, but this choice is inconclusive, cf. bottom two rows in Fig.~\ref{fig:bars} --- presumably, due to the ground truth values varying across the brain, and because of the effects of the noise in estimating the three diffusivities. 
(Since typical $\Deperp \lesssim 0.5$, the left inequality may be practically often identified with $\zeta=\sgn(\Da-\Depar)$; however, the exact branch selection involves all three diffusivities.)
\keep
There is a cluster of voxels with $\Da \approx \Depar$ such that $\beta \approx 4-\sqrt{40/3} \approx 0.35$ (cf. also Fig.~\ref{fig:constraints});  
for those,  branches merge since the discriminant ${\cal D}\to 0$, Eq.~\eq{Discr}, and   both estimated parameter sets coincide.

\new
Prevalence calculations take long time. Practically, since we cannot so far see any obvious ``global" branch selection method, we tested {\it choosing the branch voxel-wise}, based on having the two redundant LEMONADE equations \eq{L00} and \eq{L20} for $M^{(6),0}$ and $M^{(6),2}$.
As described at the end of \ref{app:lemo-sol}, their independent solutions for $p_2$ (after excluding all scalar parameters using the first 4 LEMONADE equations), which we label ${p_2}_\zeta^{(6),0}$ and ${p_2}_\zeta^{(6),2}$, produce the corresponding sets $x^{(6),0}_\zeta$ and $x^{(6),2}_\zeta$ of scalar parameters, for both branches. 
Only the correct solution would yield the same set of parameters $x_\zeta\equiv x^{(6),0}_\zeta  = x^{(6),2}_\zeta$  when using either of these two equations. Noise inevitably causes the discrepancy between the solutions $x^{(6),0}_\zeta$ and $x^{(6),2}_\zeta$ even for the correct branch $\zeta$; however, for sufficiently small noise, these solutions in the correct branch will be closer to each other than those for the wrong branch. This branch choice, cf.  \ref{app:lemo-sol}, is reasonable for SNR$\, = 100$ (and not very conclusive for the SNR$\,=33$), Supplementary Fig.~\ref{fig:noiseprop}. Its output, taken as the mean between the sets 
$x^{(6),0}_\zeta$ and $x^{(6),2}_\zeta$ for the chosen $\zeta$  is shown in Fig.~\ref{fig:maps-all} ({\it lemo $\zeta$} row), and Fig.~\ref{fig:branch} (dashed green lines). We then use it to initialize the full nonlinear RotInv problem \eq{F}, which leads to the output marked by {\it RotInv $\zeta$} in Fig.~\ref{fig:maps-all} and by solid green lines in the histograms of Fig.~\ref{fig:branch}. 
We can see that this output generally matches quite well the results of the prevalence method. Furthermore, we observe strong similarity in the ROI-averaged values between both methods, Fig.~\ref{fig:bars} and Supplementary Fig.~\ref{fig:bars-sel}.  
\keep
\del{Maps from a clinically feasible acquisition, $b\leq 2.5$, calculated with branch selection according to $\zeta=+/-$ in WM/GM,  shown in Fig.~3f, are qualitatively similar to the full-$b$-range maps.}

\subsection{ODF and fiber tracking using locally estimated kernels}
\bigskip

Figure \ref{fig:odf}b demonstrates that fiber ODFs (right panel), calculated using factorization relation \eq{fact}, are notably sharper than the ``reference" signal ODFs (left panel).
Here, we calculated the SH coefficients of the reference signal ODF $\P^{(0)}$ via $p^{(0)}_{lm} = (-)^{l/2}S_{\! lm}$, cf. \ref{app:K0}. 
The logic  was to maximally preserve the ``raw" diffusion-weighted signal, and to only 
rotate its $l=2, 6, 10, \dots$ harmonics, so that the directionality of the resulting $\P^{(0)}$ mimics that of the genuine fiber ODF, while the blurring due to the diffusion is preserved.  
We notice that the projections (\ref{Kl}) of the kernel onto the Legendre polynomials have the alternating sign $(-)^{l/2}$, which physically turns a prolate fiber ODF into an oblate signal. In \ref{app:K0} we show that such a procedure corresponds to a (de)convolution with a singular kernel, concentrated along the equator $\xi = \n\cdot\g=0$ of the unit sphere $\n=1$ (relative to the diffusion direction $\g$), akin to the Funk-Radon transform \cite{Tuch2004,Descoteaux2009,jensen2016} --- albeit taking a {\it fractional derivative} 
$\partial_\xi^{1/2}|_{\xi = 0}$ in the direction transverse to the equator, instead of merely averaging over the equatorial points via $\delta(\xi)$, as in the FRT. Such a ``blur-preserving" procedure can serve as a natural reference for the model-based ODF deconvolution methods. 
Note that the coefficients $|K_l(b) |$ generally decrease with $l$. Therefore, dividing by the kernel $K_l(b)$ gives larger weight (relative to $(-)^{l/2}$) to the higher-order spherical harmonics, sharpening the fiber ODF relative to $\P^{(0)}$, Fig.~\ref{fig:odf}b. 
Small spurious ODF peaks are intrinsic to unconstrained SH deconvolution due to low order truncation and thermal noise \cite{Tournier2004}, and might be further mitigated by adopting the constrained approaches \cite{Tournier2007}. 
Here,  we  show all parameters estimated without any constraints.  

Fiber ODFs, rather than empirical signal ODFs, should be in principle a better starting point for any  fiber tracking algorithm, as  physics of diffusion is factored out. Figure \ref{fig:odf}c shows an example of the probabilistic tractography outcome, 
\mpar{R2.3}
\new using iFOD2 algorithm with default settings \cite{Tournier2010} as part of the MRtrix3.0 open source package by \citet[{\tt www.mrtrix.org}]{mrtrix},  \keep relying on ODFs \eq{P=Y} using locally estimated kernels \eq{K} via Eq.~(\ref{fact}), with the scalar parameters and $p_2$ drawn on the streamlines.  
Furthermore, voxel-wise estimated $f, \Da, \Depar, \Deperp$ and $p_{lm}$ can serve as a  starting point for mesoscopic global fiber tracking \cite{reisert2014mesoft} that can provide further regularization of  problem \eq{S}--\eq{K}
by correlating over adjacent voxels.

\section{Discussion}
\label{sec:discussion}

The rotational invariant framework for the overarching model \eq{S} generalizes  previous methods which constrained parameter values or ODF shape, reveals a  nontrivial topology of the fitting landscape, and explains its degeneracies, and associated issues with accuracy and precision in  modern quantitative approaches to dMRI-based neuroimaging.

\subsection{Parameter values}
\label{sec:paramvals}
\bigskip

We observe that scalar parameter values in Fig.~\ref{fig:maps-all} (especially {\it RotInv $\zeta$} and {\it prevalence}) exhibit  WM/GM contrast, with the $T_2$-weighted axonal water fraction $f$ highest in major tracts, approaching $0.7-0.8$. 
The neurite fraction drops significantly in GM, 
\new
down to about 0.2,  in agreement with the recent study based on isotropic diffusion weighting \cite{CODIVIDE}. There could be a number of explanations to this observation.  First, the $T_2$ values 
in extra- and intra-neurite spaces are generally different, which re-weighs the compartment fractions.  
\mpar{R2.5}
In fact, in our recent RotInv-based work \cite{teddi}, where we varied echo time for probing compartment $T_2$ values, we generally found $T_2^a > T_2^e$ for WM, effectively increasing the $T_2$-weighted fraction $f$ in WM; this balance may be different in GM.  
Second, water within cell bodies in GM may effectively add to extra-neurite fraction \cite{chklovskii-optimization2002}. 
Furthermore, such low fraction could be rationalized if some of neurites (e.g. dendrites or glia) happen to be in the \new intermediate or even fast water exchange regime as recently argued by \cite{Yang2017}; \keep in this case, the low $f\sim 0.2$ could be dominated by myelinated axons whose fraction is lower in GM. In that case, model (\ref{S}) for GM should be augmented by incorporating exchange and possibly time-dependence of effective diffusion metrics \cite{mesopnas,fieremans2016}. 
\keep

Scalar parameters $x$ do not abruptly differ between corpus callosum and the crossing regions such as centrum semiovale, cf. Fig.~\ref{fig:odf}c, emphasizing that our approach is able to separate the spatially varying orientational dispersion $\P(\n)$ from the kernel $\K$. 
Conversely, $p_2$ drops significantly in WM crossing regions, Fig.~\ref{fig:odf}c (panel $p_2$), as well as in GM 
(Fig.~\ref{fig:maps-all} and Fig.~\ref{fig:odf}a).
Typical fiber orientation dispersion angle $\theta_{\rm disp}$, calculated as 
\mpar{R3.8}
\new
$\theta_{\rm disp} \equiv \cos^{-1} \sqrt{\langle\cos^2\theta\rangle}$, cf. Eq.~(\ref{p2-odf}) in \ref{app:pl}, 
\keep 
delineates major WM tracts in Fig.~\ref{fig:odf}a; the values $\theta_{\rm disp} \approx 20^\circ$ in genu and splenium agree remarkably well with the range $14^\circ-22^\circ$ observed recently from NAA diffusion and from histology in human corpus callosum \cite{ronen2014}. Even stronger orientational dispersion occurs in other WM regions, stressing the need to account for it \cite{noddi,noddi-bingham,ferizi2015}.

\new
Relation \eq{fact} describes how fast the signal's harmonics $S_{lm}$ decrease as function of $l$.   
As an estimate, for a  HARDI shell of $b=3$, assuming $\Da \approx \Depar \approx 2$ and neglecting $\Deperp$, 
we numerically find 
$(-)^{l/2} K_l \approx \{0.36,\ 0.14, \ 0.055, \ 0.019, \ 0.0055, \ 0.0014 \}$ for $l = 0, 2, \dots, 10$. 
\mpar{R2.5}
This calculation gives an estimate of the highest SH order $l$ to be observed in the signal $S_{lm}$, as function of SNR, assuming that 
$p_{lm}$ are not small, and suggests  $\lmax = 6-8$ as a practical upper bound at typical in vivo SNR values, in agreement with  numerical estimates \cite{White2009} and an observation in voxels with single fiber populations \cite{Tournier2013}. 
Conversely, the same estimate tells that at reasonable SNR, involving the $S_4(b)$ signal invariant to estimate scalar parameters may help only marginally (we did not see practical difference with our MRI data), while higher-order $S_l(b)$ will likely drown in the noise. 
The  decrease of $K_l(b)$ with $l$ and $b$ embodies a non-perturbative analog of the perturbative radial-angular connection (Sec.~\ref{sec:moments}). 
\keep

\subsection{Parameter (in)dependence and possible constraints}
\label{sec:constraints}
\bigskip

\begin{figure}[t!!]
{\bf a}\includegraphics[width=3.5in]{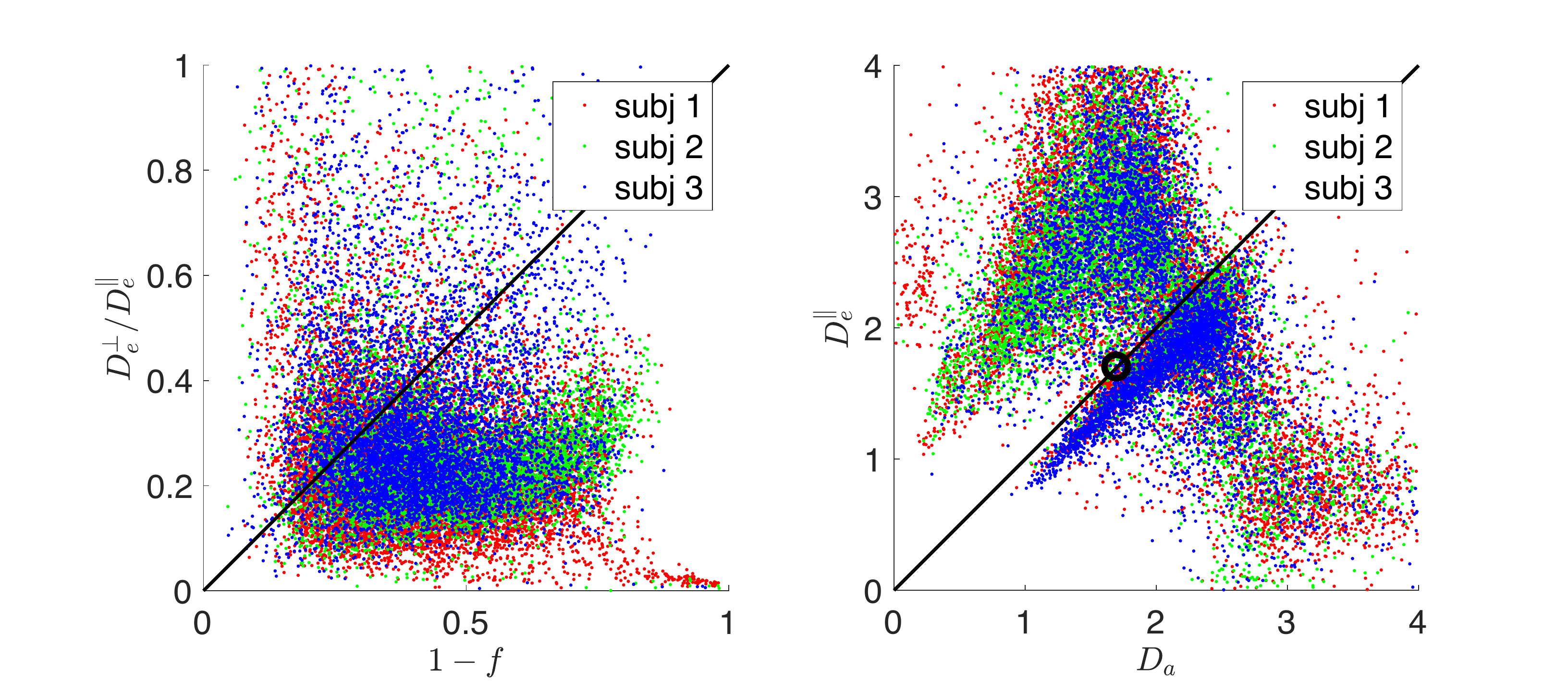}
{\bf b}\includegraphics[width=3.5in]{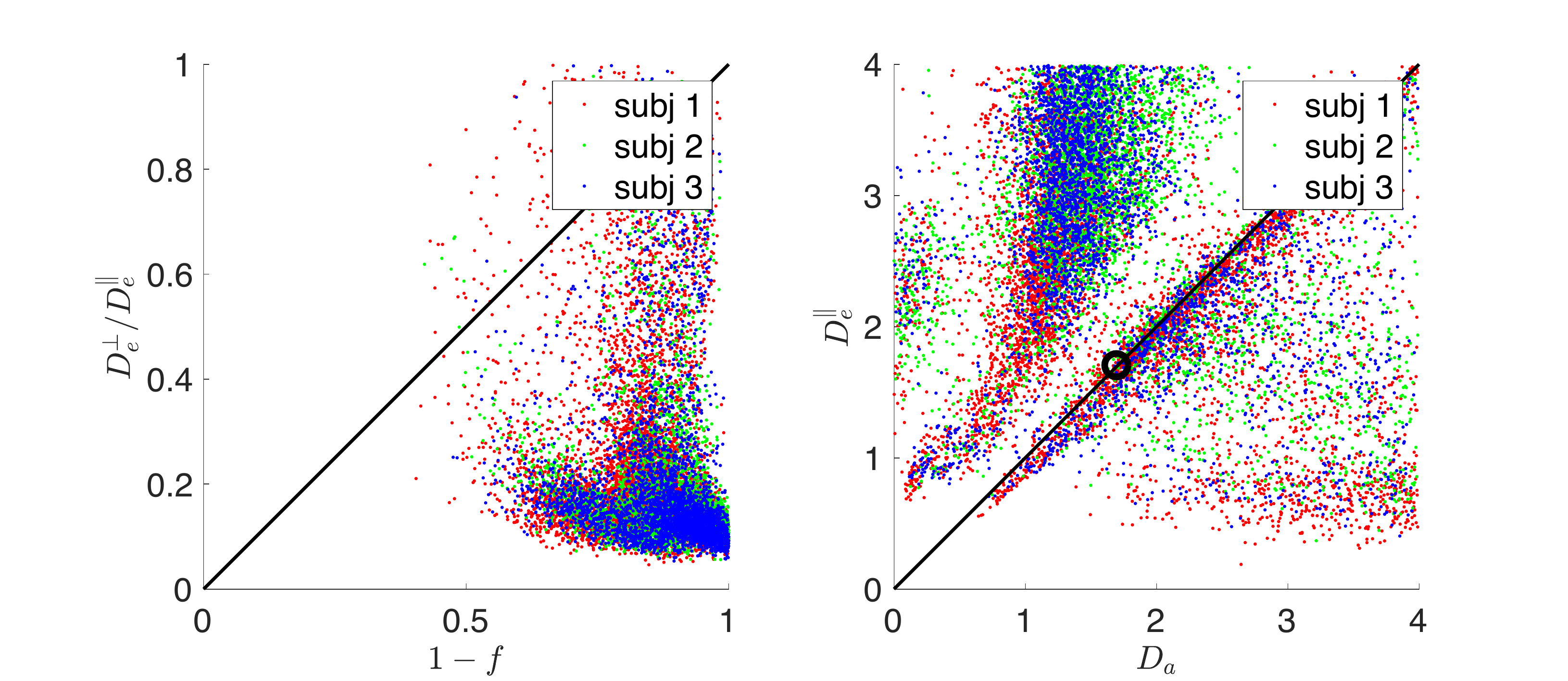}
\caption{
{\bf Are the  scalar model parameters $f$, $\Da$, $\Depar$ and $\Deperp$ independent?} 
To see whether we can rely on relations between the scalar parameters in order to increase precision in their estimation, 
we investigate the validity of widely used constraints \cite{Alexander2010,noddi,noddi-bingham,smt} 
using parameter values estimated from the prevalence method. 
Generally, these constraints fail.  
Colors correspond to 3 subjects in WM ({\bf a}) and GM ({\bf b}). Interestingly,  constraints fail  more dramatically in GM. 
}
\label{fig:constraints}
\end{figure}

Overall, we find that all three diffusivities generally vary across brain. We also observe that $D_a$ is estimated more precisely than $\Depar$ and $\Deperp$ (cf. Fig.~\ref{fig:noiseprop} in Supplementary Sec.~\ref{app:noise}). 
This can be expected due to the extra-neurite space being suppressed by the factor $e^{-b\Deperp}$ relative to the intra-axonal one, cf. Eq.~(\ref{K}).  
Scatter plots in Fig.~\ref{fig:constraints} confirm that $\Da$, $\Depar$ and $\Deperp$ should be generally estimated independently,  since neither of the  widely employed constraints of NODDI and SMT, $\Depar = \Deperp/(1-f)$, $\Da = \Depar$ \cite{Alexander2010,noddi,noddi-bingham,smt} and $\Da = \Depar = 1.7$ \cite{Alexander2010,noddi,noddi-bingham} seem to be universally valid. 
Given the ``shallow" fit landscape, employing unjustified constraints generally results in notable and unpredictable bias in the remaining parameters, which is likely to translate is unknown biases in parameter trends when applied to pathologies. 

\new 
We can see that the tortuosity constraint $\Depar = \Deperp/(1-f)$ (left panels in Fig.~\ref{fig:constraints}) is especially questionable: the tortuosity $\Depar/\Deperp \sim 5$ seems to be very high, as compared to the mean-field estimate \cite{szafer1995} used previously. The fact that the mean-field estimate fails, by $\sim 100\%$, at the biologically relevant fiber packing densities has been previously shown  both analytically and numerically \cite{shrink-remove}. 

\mpar{R3.7}
Whether $\Da = \Depar$ can be adopted as a constraint at least in some regions remains inconclusive. To test this assumption, we solved the LEMONADE system (\ref{lemo}) exactly under the additional assumption $\Da=\Depar$, based on the first 4 equations, equivalent to $1+d_2(p_2) = d_a(p_2)$, cf. Eqs.~(\ref{dimensionless}) in \ref{app:lemo-sol}; the branch selection for the corresponding quadratic equation 
\[
\lb 9M^{(4),0} - 5(M^{(2),0})^2 \rb p_2^2 + 2 \lb 9M^{(4),2} - 7M^{(2),0}M^{(2),2}\rb p_2 = 8 (M^{(2),2})^2
\]
is definitive in this case (always the $+$ branch of the corresponding discriminant, since otherwise $p_2<0$), and all parameters $x$ are given via simple explicit expressions (\ref{dimensionless}) based on the found $p_2$. 
However, due to bias in the moments, as well as likely the inapplicability of the global constraint $\Da=\Depar$, 
the parameter values remain biased so that, for instance, often times $f>1$. This solution also proved notably less optimal for initializing the full RotInv estimation (\ref{F}) (unphysical values in many voxels, and overall lack of agreement with the prevalence outcomes). 
These observations speak against imposing such a global constraint on the diffusivities.    

On the other hand, there is recent evidence that $\Da \approx \Depar$ in rat spinal cord \cite{skinner2017}. 
\new Recent time-dependent diffusion study \cite{Jespersen2017-arxiv} in fixed porcine spinal cord favors the $\zeta=+$ branch. 
However, we note that spinal cord morphology, and effects of tissue fixation, can make these estimate sufficiently different from those in {\it in vivo} human brain.  
\keep 
Overall, it remains an open question whether and by how much $\Da$ and $\Depar$  differ, and by how much each of them changes in different pathologies.  

At this time, we are not aware of  relations between parameters able to constrain the problem without introducing bias. 
Precision improvement can come from  ``orthogonal" measurements \cite{dhital-isodw,szczepankiewicz2015,DIVIDE,CODIVIDE,skinner2017,dhital2017,dhital-intra} 
cutting through the trenches (as discussed below), as well as from searching for solutions within the physical parameter ranges for $x$ 
(e.g. by creating libraries of $K_l(b,x)$), based on better understanding of the ground truth. 
This should be a subject of future work.  

\keep

\subsection{Branch selection and ``orthogonal" measurements}
\bigskip

\del{While results in Fig.~4 favor branch selection $\zeta=\pm$ for WM/GM,}
\new 
Overall, our current branch choice method does not seem optimal, and we suggest that branch selection remains an essential problem for quantifying neuronal microstructure, \keep
 to be ultimately validated using  very strong diffusion gradients (e.g. employing unique  Connectom scanners with gradients up to 300\,mT/m), as well as adding ``orthogonal'' acquisitions such as extra-neurite water suppression by strong gradients \cite{skinner2017} and isotropic diffusion weighting \cite{dhital-isodw,szczepankiewicz2015,DIVIDE,martins2016,CODIVIDE}. 
In particular, isotropic weighting (spherical tensor encoding), yielding 
\be \label{Siso}
S(b)/S_0 = f e^{-b\Da} + (1-f) e^{-b(\Depar+2\Deperp)} \,, 
\ee
seems to produce relations $\Da \approx \Depar + 2\Deperp$ due to an empirically small iso-weighted kurtosis of signal (\ref{Siso}) \cite{dhital-isodw,CODIVIDE,dhital2017}. 
While this can be interpreted as favoring the $\zeta=+$ branch, this relation cannot be used as a global constraint: \citet{szczepankiewicz2015} show it failing in thalamus, apparently consistent with the $\zeta=-$ selection in GM (note however that thalamus is a GM/WM mixture).
It is also interesting to further investigate the branch-merging case of $\Da \approx \Depar$ \cite{skinner2017}.  

\mpar{R1.9}
\new
We also note that branch selection (\ref{branch}) for the unconstrained problem (\ref{S})--(\ref{K}) is qualitatively similar but quantitatively different from that in the WMTI highly-aligned tracts case \cite{KM,wmdki}, cf. the toy model of \ref{app:toy}. 
In Fig.~\ref{fig:maps-all} we see that the {\it lemo}$-$ branch qualitatively corresponds to the standard WMTI choice $\Da < \Depar$, also preferring larger $p_2$ (strongly aligned fibers). 
While qualitatively, the ``wrong branch" in both the full model  (\ref{S})--(\ref{K}) and WMTI \cite{wmdki} corresponds, roughly, to swapping of intra- and extra-neurite parameters, there is no exact correspondence; for instance,  $f$ and $\Deperp$ are also different between the branches.   
The difference between WMTI and the full model comes from the fact that in the toy model (WMTI prototype), the perfectly-aligned fiber constraint $p_2 = p_4 = 1$ has been implemented, together with effectively mixing the LEMONADE equations with moments $M^{(4),2m}$ and $M^{(4),4m}$. Therefore, the branch choice based on $\sgn(\Depar-\Da)$ is sufficiently different from that of Eq.~(\ref{branch}). 
\keep

\subsection{Limitations and generalizations}
\bigskip

Overall, our experience shows that it is quite difficult to estimate  scalar parameters, even with a  highly oversampled acquisition.
The scalar-tensor factorization works in such a way, that while ODF parameters $p_{lm}$ ``lie on a surface",  scalar parameters are quite hidden, so that  directional dMRI acquisition is not very sensitive  to their values. 
Therefore, our unconstrained fit, Eq.~(\ref{F}), yields maps which for some  parameters are quite noisy (Fig.~\ref{fig:maps-all}), with unphysical values masked. 
Noise in voxelwise maps is not a bug but a feature: it  arises from the degeneracies (``trenches") in the landscape, Fig.~\ref{fig:landscape}, inherent to the general model \eq{S}--\eq{K}. 
Similar issue has been empirically noted in the related machine-learning method \cite{baydiff}, where model degeneracies manifest themselves in priors strongly affecting the estimation outcome. 
\mpar{\del{``honest maps"}}
Here, our goal  has been to uncover \new and to understand the uncertainties in the {\it unconstrained maps} \keep achieved with a very extensive directional human dMRI acquisition. 
Effect of noise is somewhat alleviated when looking through semi-transparent-drawn tracts (Fig.~\ref{fig:odf}c), where the anatomical trends in all parameters become particularly obvious. 

\mpar{R2.5, R4.1}
\new Our main limiting assumption has been that all fibers in a given voxel share the same scalar parameters, which may be questionable when anatomically different tracts cross. This can lead to particularly challenging branch selection, where some ``average" values, such as $\Da \approx \Depar$, corresponding to the two branches merging, could be preferred --- if  fiber tracts belonging to different branches cross. Quantification of the role of this limitation may be possible with adding, say, isotropic weighting measurement to an extensive multi-shell protocol. 
\keep

Generalizing, for any number of compartments in the kernel \eq{K}, scalar parameters can be determined from a  set \eq{S=pK} of basis-independent rotational invariants. 
Branch-selection degeneracy of the scalar sector will persist for 3 or more compartments. 
Relating  moments \eq{taylor} to kernel parameters can be used to analyze this degeneracy. If the added compartment(s) are isotropic, the LEMONADE branches will correspond to the anisotropic 2-compartment part of the kernel $\K$, determining the respective higher-dimensional ``low-energy'' manifolds in  parameter space. Methods other than gradient descent (e.g. library-based or machine-learning \cite{baydiff}) can be utilized for solving system \eq{S=pK}; applicability of all such methods hinges on resolving intrinsic degeneracies for driving the estimation towards the biophysically correct parameter domain.

\section{Outlook}

Using SO(3) symmetry and representation theory, we separated the parameter estimation problem for  neuronal  microstructure into scalar and tensor (ODF) sectors. 
Taylor-expansion analysis of the scalar sector reveals a nontrivial topology of the parameter landscape, with the first few moments exactly determining  two narrow  trenches along which the parameters approximate dMRI measurements almost equally well. This  degeneracy is intrinsic to model \eq{S}--\eq{K} with {\it any} ODF, revealing  issues with accuracy and precision in parameter estimation.  

Branch selection criterion \eq{branch}  determines the domain for the physical solution. 
\del{The  choice $\zeta=\pm$ for WM/GM}
\new Our voxel-wise LEMONADE-inspired branch choice \keep remains to be validated by estimating ground truth compartment diffusivity values in animal studies, and by using strong diffusion gradients or alternative acquisition schemes. 

\mparr{R3.13}\newr Notwithstanding \keep the branch selection uncertainty,  the combination of a linearized solution for the moments and the subsequent nonlinear optimization gives rise to an unconstrained algorithm for parametric maps in the whole brain, that does not imply any parameter priors, and yields scalar and ODF maps in the whole brain in about 10 min on a desktop computer. Precision still remains a challenge due to the fundamental degeneracies of the parameter estimation problem. 
Our analysis shows that commonly used constraints between the scalar parameters generally do not hold, and can severely bias the remaining parameters due to the nontrivial topology of the minimization landscape. 

We believe our approach sets the analytical foundation \new and poses further questions necessary \keep for building an unbiased non-invasive clinically feasible mapping of key neuronal microstructure parameters orders of magnitude below MRI resolution, opening a window into architectural, orientational and functional changes in pathology, aging and development, and bridging the gap between biophysical modeling, basic neuroscience, and clinical MRI.

\section*{Acknowledgments}

This work has benefited from fruitful discussions with Marco Reisert, Valerij Kiselev, Bibek Dhital and Elias Kellner.
Research was supported 
by the National Institute of Neurological Disorders and Stroke of the NIH under award number R01NS088040.
Photo credit to Tom Deerinck and Mark Ellisman (National Center for Microscopy and Imaging Research) for the histology image illustrating a fiber fascicle in Fig.~\ref{fig:SM}.

\newpage
\appendix

\section{Scalar-tensor factorization for the full problem \eq{S}--\eq{K}}
\setcounter{figure}{0}
\renewcommand\thefigure{A.\arabic{figure}} 
\renewcommand\theequation{A.\arabic{equation}}  
\label{app:fact}
\mpar{Appendix Sections reshuffled}

Since kernel \eq{K} is axially symmetric, it can be expanded in even-order Legendre polynomials $P_l(\xi)$  (i.e. in the $m=0$ SHs), as function of $\xi\equiv \cos\theta$ (for given scalar parameters $x$):
\bea \label{K=KlPl}
\K(b,\xi) &=& \sum_{l = 0, 2, \dots} (2l+1) K_l(b) P_l(\xi) \,, \\ 
\label{Kl}
K_l(b) &\equiv&  \int_0^1\! \d \xi\, \K(b,\xi) P_l(\xi) \,.
\eea
Applying the SH addition formula 
\be \label{SHaddition}
P_l(\g\cdot \n) = {4\pi\over 2l+1} \sum_{m=-l}^l Y_{lm}(\g)Y^*_{lm}(\n) 
\ee
to $P_l(\xi)$ from kernel (\ref{K=KlPl}) plugged into  model (\ref{S}), 
yields Eq.~\eq{S=pK} of the main text, where SH components $S_{lm}$ (with even $l$ only, due to $\n \to -\n$ symmetry) 
are defined in a standard way (note the definition of $\d\n$ or $\d\g$ in footnote \ref{foot:dn}),
\bea \label{S=SH}
S_\g(b) &\equiv& \sum_{l=0,2,\dots}\sum_{m=-l}^l S_{lm}(b) Y_{lm}(\g) \,, \\  
S_{lm}(b) &=& 4\pi \int_{|\g|=1}\!\d\g \, S_\g(b) Y^*_{lm}(\g) \,. 
\eea
Rotationally invariant functions \eq{Kl} were first analytically calculated and used by  \cite{Anderson2005} for a single-compartment kernel (\ref{K}),  
by \cite{jespersen2007} for the full fitting \eq{fact}, and later by \cite{White2009,baydiff}.

\mpar{R2.4}
\new
We note that the normalization of the measure $\d\n$ (footnote \ref{foot:dn}) 
with the normalization $\int\!\d\n\,\P(\n)\equiv 1$ of the ODF \eq{P=Y} sets $\P \equiv 1$ for the completely isotropic case, 
with the kernel and its $l=0$ invariant normalized to $\K|_{b=0} = K_0|_{b=0} \equiv S_0$. 
\keep

\section{Geometric meaning and normalization of the ODF rotational invariants}
\setcounter{figure}{0}
\renewcommand\thefigure{B.\arabic{figure}} 
\renewcommand\theequation{B.\arabic{equation}}  
\label{app:pl}
\mpar{new Appendix}

\new
\mpar{R3.2}
The normalization factor $\N_l$ in Eq.~(\ref{rotinv}) is chosen such that the maximal value of $p_l$ is unity. 
Indeed, for the maximally anisotropic fiber ODF, with all fibers oriented along ${\bf \hat z}$, the SH coefficients 
\[
p_{lm} = \delta_{m,0} \cdot 4\pi \int_0^1 \d \cos\theta\, Y_{l0}(\theta) \, \delta(\cos\theta-1) \equiv \delta_{m,0} \cdot \N_l \,,
\]
where we used $Y_{l0}(\theta) = \sqrt{2l+1\over 4\pi} P_l(\cos\theta)$ and the Legendre polynomials normalization $P_l(1) \equiv 1$. 
This yields $p_l \equiv 1$ for all $l$ for such a singular ODF.  Less anisotropic ODFs will correspond to $0 \leq p_l < 1$. 
Therefore, defined in this way, the invariants $p_l$ can serve as a purely geometric normalized measure for the ODF anisotropy in each SO(3) sector $l$, with all the diffusion physics factored out. The $p_2$ invariant may be thought of as a purely geometric fiber fractional anisotropy, and so on. 

To better appreciate the geometric meaning of $p_l$, consider first the $l=2$ case. The $m=0$ SH component
\bea \non
\mkern-36mu
p_{20} &=& 4\pi \int\! \d\n \, \P(\n) \, Y_{20}(\n) = \N_2 \int\!\d\n\, \P(\n) \, P_2(\cos\theta)  
\\
&\equiv& \N_2 \cdot {3\langle \cos^2\theta\rangle -1 \over 2} \,,
\label{p20-odf}
\eea
where $\langle \dots \rangle$ stands for taking the mean over the ODF, as before. Hence, $p_{20}/\N_2$ is a measure of the angular ODF dispersion $\langle \cos^2\theta \rangle$. 
\mpar{R3.8}
Consider now the single-fiber voxels, where one could approximately assume axial symmetry around the principal fiber direction; in the basis where the $ {\bf \hat z}$ axis is along the fiber direction, 
the corresponding $|p_{2m}| \ll p_{20}$, and $\theta$ is the angle of the ODF orientation dispersion.  
Therefore, $p_{20}/\N_2 \approx \| p_2 \|/\N_2 \equiv p_2$ is a measure of the fiber orientation dispersion:
\be \label{p2-odf}
\langle \cos^2\theta \rangle  \simeq {2p_2 + 1\over 3} \,.
\ee
This intuition has been applied to define the axially-symmetric liquid crystal order parameter \cite{emsley1985}, and more recently to characterize single-fiber populations in dMRI \cite{lasic2014} in the fiber reference frame.  
Written via the rotational invariant $p_2$, this relation can be applied irrespective of the choice of the basis, applies to fibers without axial symmetry, and also underscores the benefit of normalizing the invariant $p_2$ such that $0 \leq p_2 \leq 1$. 

Likewise, $p_{l0} = \N_l \cdot \langle P_l (\cos\theta) \rangle$; in the fiber basis, with $\| p_l \| \simeq p_{l0}$ dominated by $p_{l0}$ when the ODF is approximately axially symmetric, we obtain 
\be \label{pl-odf}
p_l \simeq \langle P_l (\cos\theta) \rangle \,, \quad 0\leq p_l \leq 1 \,.
\ee
This relation allows one to recursively express any (even) moment $\langle \cos^L \theta \rangle$ via the invariants 
$p_l$, $l =  2\dots L$, calculated in any basis. Odd-$l$  moments  vanish by inversion symmetry.   
\keep

\section{Scalar-tensor factorization for the moments: LEMONADE}
\setcounter{figure}{0}
\renewcommand\thefigure{C.\arabic{figure}} 
\renewcommand\theequation{C.\arabic{equation}}  
\label{app:lemo}

Let us first give the outline of the LEMONADE derivation: 

\begin{enumerate}
\item
We relate all  tensor components $M^{(L)}_{i_1\dots i_L}$ to the scalar parameters 
$x$ and to the ODF SH components $p_{lm}$, for $L\leq \lmax = 6$, Eqs.~(\ref{Msym}). 
These $N_c(6) = 49$ redundant  relations for the $N_p(6) = 31$ parameters embody the perturbative parameter count of Secs.~\ref{sec:par-mom} and \ref{sec:par-model}. 
However, this large number of nonlinear equations makes brute-force investigation difficult. 

\item
This problem factorizes in SH basis, $M^{(L)}_{i_1\dots i_L} \to M^{(L),lm}$, Eq.~(\ref{MLlm}). 
Here the index $L$ corresponds to the order $b^{L/2}$, and index $l\leq L$ labels  irreducible representations of SO(3). 
This transformation is the mathematical manifestation of the products $n_{i_1}\dots n_{i_l}$ forming the set of SH, mentioned in Sec.~\ref{sec:moments} above. As a result, it is now the set of 
transformed tensor components $M^{(L),lm}$ that is related to the model parameters $x$ and $p_{lm}$, forming the LEMONADE system (\ref{lemo-m}). 

\item
Importantly, in this ``natural" basis the problem diagonalizes: The $(l,m)$ components for the $M^{(L),lm}$ tensors only involve the corresponding $p_{lm}$ of the ODF. 
We can see that generally, $M^{(L),lm}$ are proportional to $p_{lm}$, and the highest order $M^{(l),lm}$ yields the highest order $p_{lm}$, Eq.~(\ref{lemo-odf}), --- a mathematical formulation for the term-by-term radial-angular connection of Sec.~\ref{sec:moments}. 

\item
Finally, we factor out the dependence on the choice of the basis, 
by writing system (\ref{lemo-m}) in rotationally invariant form, cf. Eq.~(\ref{rotinv}) above. 


\end{enumerate}

\subsection*{Taylor expansion of model \eq{S}--\eq{K} up to $\O(b^3)$}
\bigskip

To connect the moments to the model parameters, and to explore the low-energy landscape of the problem \eq{S=pK}--\eq{F}, let us expand the signal \eq{S}--\eq{K}. The $\O(b)$ term, $l=2$, yields the diffusion tensor 
\begin{subequations} \label{Msym}
\be \label{D}
M_{ij}^{(2)} \equiv D_{ij} = f \Da \langle n_i n_j \rangle + (1-f) \lp \Deperp \delta_{ij} + \dDe \langle n_i n_j \rangle \rp  
\ee
where $\langle n_i n_j \rangle = \int\! \d\n\, \P(\n) \, n_i n_j$ and 
\be \non
\dDe \equiv \Depar - \Deperp \,.
\ee
Expanding up to $\O(b^2)$ and $\O(b^3)$ yields the 4$^{\rm th}$ and 6$^{\rm th}$ order moments, correspondingly:
\bea \non
M_{ijkl}^{(4)} &=& f \Da^{2} \langle n_i n_j n_k n_l \rangle  
+ (1-f) \lb \Deperp^{2}\delta_{(ij}\delta_{kl)}  
\right. \\  &&  \left. 
+ 2\Deperp\dDe \langle n_{(i} n_{j}\rangle \delta_{kl)}   
+ \dDe^{2} \langle n_i n_j n_k n_l \rangle \vphantom{\Deperp^{2}}\rb; \quad
 \label{M}
\\
\non
M_{i_1\dots i_6}^{(6)} &=& f \Da^{3} \langle n_{i_1} \dots n_{i_6} \rangle  
+ (1-f) \lb \Deperp^{3}\delta_{(i_1i_2}\delta_{i_3 i_4}\delta_{i_5 i_6)}  
\right. \\   &&  \left. 
 \label{L}
+\, 3\Deperp^2\dDe \delta_{(i_1 i_2} \delta_{i_3 i_4}\langle n_{i_5} n_{i_6)}\rangle    
\right. \\  \non &&  \left.  
+\, 3\Deperp\dDe^2 \delta_{(i_1 i_2} \langle n_{i_3} \dots n_{i_6)}\rangle    
+ \dDe^{3} \langle n_{i_1} \dots n_{i_6} \rangle \vphantom{\Deperp^{3}}\rb. \quad
\eea
\end{subequations}
Here symmetrization \cite{thorne} over tensor indices between $_{(\dots)}$ is assumed, 
\new
such that, for instance, 
$S_{(ij)k} = \frac1{2!} (S_{ijk} + S_{jik})$ for any tensor $S_{ijk}$. \keep
\mpar{R1.11}
Therefore, 
\bea \non 
\delta_{(ij}\delta_{kl)} &=& \ts{\frac13} \lp \delta_{ij}\delta_{kl} + \delta_{ik}\delta_{jl} + \delta_{il}\delta_{jk}\rp , 
\\
\non \label{delta-nn}
\langle n_{(i} n_{j}\rangle \delta_{kl)} &=& \ts{\frac16} \lb 
   \langle n_{i} n_{j}\rangle \delta_{kl} 
+ \langle n_{i} n_{k}\rangle \delta_{jl}
+ \langle n_{i} n_{l}\rangle \delta_{jk} \right . 
\\ && \left. 
+ \langle n_{k} n_{l}\rangle \delta_{ij}
+ \langle n_{j} n_{l}\rangle \delta_{ik}
+ \langle n_{j} n_{k}\rangle \delta_{il}
\rb , 
\eea
such that
$\delta_{(ij}\delta_{kl)} g_i g_j g_k g_l = 1$, 
$\langle n_{(i} n_{j}\rangle\delta_{kl)} g_i g_j g_k g_l = \langle n_i n_j \rangle g_i g_j$.
Similarly,  symmetrized tensors in Eq.~\eq{L}, when convolved with $g_{i_1}\dots g_{i_6}$, yield the respective powers of the  product $\g\cdot \n$.  

In principle, one can proceed further, with the escalating complexity of relating the higher-order moments of the signal to the nonlinear combinations of the scalar model parameters $x = \{f, \Da, \Depar, \Deperp\}$ and of the ODF averages 
$\la n_{i_1}\dots n_{i_{l}}\ra \equiv \int\! \d\n\, \P(\n) \, n_{i_1}\dots n_{i_{l}}$. We would like to {\it invert} the above relations: to solve for the ODF expansion parameters $p_{lm}$ and the scalar  parameters $x$ in terms of the moments $M^{(l)}_{i_1\dots i_l}$, and to explore the properties of the solution.

The $N_c(6)=49$ equations \eq{Msym} provide an overdetermined nonlinear system  for $N_p(6)=31$ model parameters.  
To obtain an exact solution of this system we will utilize symmetry, by working in the irreducible representations of the SO(3) group, for which this challenging problem factorizes. We first remind of some useful facts on the SO(3) representations.

\subsection*{Isomorphic SO(3) representations: STF tensors and SH basis}
\bigskip

We will now outline how linear combinations of symmetric tensors $n_{i_1}\dots n_{i_l}$ form the SH set $Y_{lm}$, 
and how to form a basis in the space of symmetric moments $M^{(l)}_{i_1\dots i_l}$ using STF tensors. 

Consider any symmetric trace-free tensor $A_{i_1\dots i_l}$ of rank $l$ (STF-$l$ tensor), such that $A_{i_1\dots i_l}$ is symmetric with respect to permuting any pair of its indices, and any trace is zero, $A_{i_1\dots i_n \dots i_m \dots i_l}\delta_{i_n i_m} = 0$, $1\leq n,m\leq l$. It has $2l+1$ independent components; under any rotation, these components transform such that the STF properties are preserved. 
Viewed as a $2l+1$-dimensional vector, such components transform among themselves; technically, this means that STF-$l$ tensors 
generate an {\it irreducible representation} of the SO(3) group of rotations, of weight $l$ and dimension $2l+1$ \cite{thorne}. 

The $2l+1$ SH of order $l$ also generate an irreducible representation. Therefore, there is a 1-to-1 mapping between STF-$l$ tensors and SH set $Y_{lm}$, $-l\leq m \leq l$.  
This mapping is realized via the special, location-independent STF-$l$ tensors $\Y_{i_1\dots i_l}^{lm}$ [defined e.g. in 
Eq.~(2.11) of \cite{thorne}], that generate SH:
\be \label{Y=Yn}
Y_{lm}(\n) =  
\Y^{lm}_{k_1...k_l} \, n_{k_1}...n_{k_l} \,.
\ee
Generally, the $2l+1$ linearly independent tensors $\Y^{lm}_{k_1...k_l}$ form the basis for STF-$l$ tensors, such that 
$A_{k_1 \dots k_l} = \sum_{m=-l}^l A^{lm} \Y^{lm}_{k_1...k_l}$. 
Hence, any fully symmetric tensor $S_{k_1 \dots k_l}$ of rank $l$, with the number $n_c(l)$ of independent components [cf. main text, before Eq.~\eq{Nc}], 
can be projected on the STF-$l$, STF-$(l-2)$, STF-$(l-4)$, $\dots$ basis $\Y^{lm}_{k_1...k_l}$, $\Y^{l-2,m}_{k_1...k_{l-2}}$,  $\Y^{l-4,m}_{k_1...k_{l-4}},\dots$, by taking subsequent traces, each time reducing its rank by 2: $S_{k_1 \dots k_l}\delta_{k_{l-1}k_l}$, $S_{k_1 \dots k_l}\delta_{k_{l-3}k_{l-2}}\delta_{k_{l-1}k_l}$, and so on, so that the total number of its independent components $\sum_{l'=0,2,\dots}^l (2l'+1)$ is indeed given by $n_c(l)$.

\subsection*{$l=2$ example: Diffusion tensor invariants and FA}
\bigskip

Representing textbook-defined real-valued SH in the form $Y_{2m}(\n) = \Y_{ij}^{2m} \, n_i n_j$, 
we obtain the basis of 5 STF-$2$ tensors
\mpar{$\Y_{ij}^{2m}$ symmetrized}
\bea \non
\mkern-36mu
\Y_{ij}^{2,-2} &=& C_2 \cdot {\sqrt{3}\over2}\, (\delta_{i1}\delta_{j2} + \delta_{j1}\delta_{i2}) \,, 
\\ \non
\Y_{ij}^{2,-1} &=& C_2 \cdot {\sqrt{3} \over 2}\, (\delta_{i2}\delta_{j3} + \delta_{j2}\delta_{i3}) \,, 
\\ \non
\Y_{ij}^{2,0} &=& C_2 \cdot \lp \ts{\frac32} \delta_{i3}\delta_{j3} - \frac12 \delta_{ij} \rp \,,  
\\ \non
\Y_{ij}^{2,1} &=& C_2 \cdot {\sqrt{3}\over 2}\, (\delta_{i3}\delta_{j1} +  \delta_{j3}\delta_{i1})\,, 
\\ \non
\Y_{ij}^{2,2} &=& C_2 \cdot {\sqrt{3}\over 2} \lp\delta_{i1}\delta_{j1} -\delta_{i2}\delta_{j2}\rp 
\eea
where $C_l = \N_l/4\pi = \sqrt{(2l+1)/4\pi}$, that  
form the basis in the space of all symmetric trace-free $3\times 3$ matrices. 
The trace-free diffusion tensor components in the STF-2 basis are therefore the five $l=2$ components 
$D^{2m} = \Y^{2m}_{ij} D_{ij}$. 

Since the 2-norm in each SO(3) representation is rotation invariant (cf. {\it Theory}), the SH representation for the diffusion tensor yields the following {\it two} rotational invariants: the $l=0$ invariant 
$D^{00} = C_0 \,\delta_{ij}D_{ij} \equiv \frac1{\sqrt{4\pi}}\, \mathrm{tr}\, D $, and the $l=2$ invariant 
$\| D^{2m}\|^2 \equiv \sum_{m=-2}^2 (D^{2m})^2$, that can be expressed in terms of the STF part $D^{\rm STF}$ of $D_{ij}$,  
so that $D^{\rm STF}_{ii} = 0$:
\be \non
\mathrm{\, tr \,} (D^{\rm STF})^2 \equiv D^{\rm STF}_{ij} D^{\rm STF}_{ij} = \frac2{3C_2^2}\, \| D^{2,m}\|^2 \,, \quad 
D^{\rm STF}_{ij} \equiv D_{ij} - \frac{\mathrm{\, tr\,} D}3  \cdot \delta_{ij} \,.
\ee 
In the eigenbasis, where $D_{ij} = \mathrm{diag}\,(\lambda_1 \ \lambda_2 \ \lambda_3)$,
$\mathrm{\, tr \,} (D^{\rm STF})^2 =  \sum(\lambda_i - \bar\lambda)^2$, where mean diffusivity 
$\bar \lambda \equiv \frac13 \mathrm{tr}\,D$. This allows us to represent the familiar fractional anisotropy  
\be \label{FA}
\mathrm{FA} = 
\sqrt{{\frac32 \mathrm{\, tr\,} (D^{\rm STF})^2 \over \frac13 (\mathrm{tr\,}D)^2 + \mathrm{tr\,} (D^{\rm STF})^2}}
= 
\sqrt{3 \| D^{2m} \|^2 \over 5 |D^{00}|^2 + 2 \| D^{2m} \|^2}
\ee 
in terms of the $l=0$ and $l=2$ invariants, and calculate it {\it without diagonalizing} the diffusion tensor. 
Hence, using two basis-independent DTI metrics --- mean diffusivity and FA --- 
is equivalent to using the two SO(3) invariants of a rank-2 symmetric tensor:  its trace (i.e. the isotropic part), 
and the STF invariant $\mathrm{\, tr \,} (D^{\rm STF})^2$ quantifying its ``variance" relative to an isotropic tensor. 
(Note that, adding the third --- cubic in $D_{ij}$ --- invariant $\det D$, which would contribute to the eigenvalue ``skewness" $\sim \sum (\lambda_i - \bar\lambda)^3$, would define the three independent invariants of the matrix $D_{ij}$, i.e. the coefficients of its characteristic polynomial, or, equivalently, its three independent eigenvalues.)

\mpar{R1.11}
Likewise, for \del{kurtosis} \new the 4th-order cumulant tensor (from which kurtosis tensor is derived), \keep 
this approach yields 3 rotational invariants (out of total 12, cf. \cite{Ghosh2012}): $C^{(4),0} = C^{(4)}_{iijj}$ (mean); 
$C^{(4),2} =\| C^{(4),2m}\|$;
 and $C^{(4),4} = \| C^{(4),4m}\|$, where 
$C^{(4),2m} \propto \Y^{2m}_{ij} \delta_{kl} C^{(4)}_{ijkl}$ and $C^{(4),4m} \propto \Y^{4m}_{ijkl} C^{(4)}_{ijkl}$ 
are the STF components of the tensor $C^{(4)}_{ijkl}$. 
Generally, for the $l^{\rm th}$ order cumulant, this approach would yield $l/2 + 1$ SO(3) invariants, which can be calculated in any basis, without diagonalization.

\subsection*{LEMONADE derivation}
\bigskip

Based on the above general theory, the irreducible SO(3) representations of weight $l$ for the moment tensors in equations \eq{Msym} are selected by projecting them onto the special STF-$l$ tensors $\Y^{lm}_{k_1...k_l}$. 
Remarkably, the products $n_{i_1}\dots n_{i_l}$ yield the SH \eq{Y=Yn} after this projection. 
Since ODF is real, here we re-define 
$\Y^{lm}_{k_1...k_l} \to \sqrt{2}\,\Re \Y^{lm}_{k_1...k_l}$ for $m>0$ and 
$\Y^{lm}_{k_1...k_l} \to \sqrt{2}\,\Im \Y^{l|m|}_{k_1...k_l}$ for $m<0$, to work in real SH basis. 
Introducing the corresponding moments in the SH basis [$\N_l$ given before Eq.~(\ref{S=pK})]
\be \label{MLlm}
M^{(L),lm} = {4\pi \over \N_l}\ \Y^{lm}_{k_1...k_l} \delta_{k_{l+1} k_{l+2}} \dots \delta_{k_{L-1}k_{L}} \, M^{(L)}_{k_1...k_L}\,, 
\ee
we relate the SH moments $M^{(L),lm}$ to the model parameters 
by convolving equations \eq{Msym} with $\Y^{lm}_{k_1...k_l} \delta_{k_{l+1} k_{l+2}} \dots \delta_{k_{L-1}k_{L}}$, 
and by using the following identities, which can be proven by direct inspection, for $L=4$
\bea \non
\delta_{(ij} \delta_{kl)} \delta_{kl} &=& \ts{\frac53} \,\delta_{ij} \,,
\quad \delta_{(ij} \delta_{kl)} \delta_{ij}\delta_{kl} = 5 \,;
\\
\non
\langle n_{(i} n_{j}\rangle \delta_{kl)} \delta_{kl} &=& \ts{\frac16} \lb 7\langle n_i n_j \rangle + \delta_{ij} \rb , 
\quad
\langle n_{(i} n_{j}\rangle \delta_{kl)} \delta_{ij}\delta_{kl} =  \ts{\frac53} \,;
\eea
and for $L=6$:
\bea \non 
\delta_{(i_1i_2}\dots \delta_{i_5i_6)} \delta_{i_1i_2}\delta_{i_3i_4}\delta_{i_5i_6} = 7 \,, 
\\
\non
\delta_{(i_1i_2}\delta_{i_3i_4}\langle n_{i_5} n_{i_6)}\rangle \delta_{i_1i_2}\delta_{i_3i_4}\delta_{i_5i_6} = \ts{\frac73} \,, 
\\
\non
\delta_{(i_1i_2}\langle n_{i_3}\dots n_{i_6)}\rangle \delta_{i_1i_2}\delta_{i_3i_4}\delta_{i_5i_6} = \ts{\frac75} \,.
\eea
As a result, we obtain the minimal system for the overall $q$-space order $L\leq \lmax = 6$, and involving only angular orders $l=0,2$: 
\begin{subequations}\label{lemo-m}
\bea 
\label{D00-m}
\mkern-36mu
M^{(2),00} &=&   f \Da + (1-f) (3\Deperp + \dDe)   \\ 
\label{D20-m}
\mkern-36mu
\frac{M^{(2),2m}}{p_{2m}/\N_2} &=&   f\Da + (1-f) \dDe   \\ 
\label{M00-m}
\mkern-36mu
M^{(4),00} &=& f\Da^{2} \! + \! (1-f)\!\lb \! 5 \Deperp^{2} + \frac{10}3 \Deperp \dDe + \dDe^{2}\rb \qquad \\
\label{M20-m}
\mkern-36mu
\frac{M^{(4),2m}}{p_{2m}/\N_2} &=&   f\Da^{2} + (1-f)\!\lb \frac{7}3 \Deperp \dDe + \dDe^{2}\rb  \qquad \\
\label{L00-m} 
\mkern-36mu
M^{(6),00} &=& f\Da^{3} \\ \nonumber 
&  + & (1-f) \!\lb 7\Deperp^{2}(\Deperp + \dDe) + \frac{21}5 \Deperp \dDe^{2}  + \dDe^{3}\rb \qquad \\ 
\label{L20-m} 
\mkern-36mu
\frac{M^{(6),2m}}{p_{2m}/\N_2} &=&  f\Da^{3} + (1-f)\!\lb  \frac{21}5 \Deperp^{2}\dDe + \frac{18}5 \Deperp \dDe^{2}  + \dDe^{3}\rb 
\qquad \quad 
\eea
\new 
\mpar{R1.12}
The system \eq{lemo-m} involves minimal orders $L$ and $l$ enough to find all the 4 scalar kernel parameters $x$, as well as $p_{2m}$. 
Adding the orders $l=4,6$ would introduce respective extra parameters $p_{4m}$ and $p_{6m}$, and so it does not provide added practical benefit for determining the scalar parameters, especially given the generally lower accuracy and precision in determining the components $M^{(L),lm}$ with larger $l$ at a given $L$. 
\keep
Having found the parameters of the kernel \eq{K},  equation
\be \label{lemo-odf}
M^{(l),lm} = {p_{lm} \over \N_l} \cdot \lb f\Da^{l/2} + (1-f) \dDe^{l/2} \rb 
\ee
\end{subequations}
yields the ODF parameters $p_{lm}$ up to {\it arbitrary} order $l \leq l_{\rm max}$, as long as $M^{(l), lm}$ are linearly found from  series \eq{taylor} and Eq.~\eq{MLlm}. 
%
By defining the rotational invariants of the moments 
\be \label{MLl}
M^{(L),l} = \| M^{(L),lm} \| \equiv \lp{\sum_{m=-l}^l \left| M^{(L),lm}\right|^2 }\rp^{1/2} ,
\ee
$l=0,2, \dots, L$ (as we did for the signal in Sec.~\ref{sec:fact} and for the diffusion tensor, $\| D^{2,m}\|$ above), 
and using $p_2$ \new as defined in Eq.~(\ref{rotinv}), \keep we obtain the rotationally invariant system \eq{lemo} in the main text.  
Note that we do not use the factor $\N_l$ in the definition of the invariants (\ref{MLl}) because we intend to cancel it in going from Eqs.~(\ref{lemo-m}) to Eqs.~(\ref{lemo}).

\section{LEMONADE exact solutions: \\ Low-energy branches}
\setcounter{figure}{0}
\renewcommand\thefigure{D.\arabic{figure}} 
\renewcommand\theequation{D.\arabic{equation}}  
\label{app:lemo-sol}

To solve the system \eq{lemo}, we first focus on Eqs.~\eq{D00}--\eq{M20}, and eliminate $\Da$, $\Deperp$ and $\dDe$. 
Introducing the common scaling factor 
\be \label{Dbar}
\Dbar(p_2) \equiv \ts{\frac13}\left( M^{(2),0} - M^{(2),2}/p_2 \right) = (1-f) \Deperp \,, 
\ee
we make all quantities dimensionless functions of $p_2$ and $f$: 
\bea \label{dimensionless}
\begin{matrix}
d_a \equiv \ds{\Da \over \Dbar}\,, 
\quad d_2 \equiv {M^{(2),2} \over p_2 \Dbar} \,,  
\quad \dde \equiv {\dDe\over \Dbar} = {d_2 - f d_a \over 1- f} \,, \quad
\cr
m_0 \equiv \ds{M^{(4),0} \over \Dbar^2} \,, \quad m_2 \equiv {M^{(4),2} \over p_2 \Dbar^2}\,, 
\quad 
\deperp \equiv \ds{\Deperp\over \Dbar}  = \frac1{1-f}\,,  \quad
\end{matrix}
\eea
such that  moments $d_2$, $m_0$, $m_2$ are functions of $p_2$ explicitly, and via $f=f(p_2)$ to be found below. 
We will also need 
\be \label{dm}
\dm(p_2) \equiv m_0 - m_2 = 5\deperp + \dde =  {5 + d_2 -f d_a \over 1 - f} \,. 
\ee
Multiplying the dimensionless equation \eq{M20} by $f$, 
\[
fm_2 = (f d_a)^2 + f(d_2 - f d_a)\lb \frac73 \, \frac1{1-f} + \frac{d_2 - f d_a}{1-f} \rb
\]
and eliminating $d_a$ using equation \eq{dm}, the $f^3$ term fortuitously cancels, and we are left with a {\it quadratic} equation
\be \label{qeq}
a f^2 - \big(a+c-\ts{40 \over 3} \big) f + c = 0 \,, 
\ee
where the functions $a = a(p_2)$ and $c = c(p_2)$ are given by
\be \label{ac}
a = (\dm)^2 - \big(\ts{\frac73} + 2 d_2\big)\dm + m_2 \,, 
\quad
c = (\dm - 5 - d_2)^2 \,. \quad
\ee
We observe that, similar to the toy model \eq{toy} (cf. \ref{app:toy} below), the full LEMONADE system \eq{lemo} up to $\O(b^2)$ yields {\it two} possible solutions $f = f_\pm(p_2)$, corresponding to the two branches of $\sqrt{\D}$. Here, the discriminant of equation \eq{qeq}, expressed via the original parameters, using
$c = f^2/(1-f)^2 \cdot (5+d_2 - d_a)^2$ and $a = c/f + \frac{40}3/(1-f)$,
is  a {\it full square}
\mpar{R1.8: $\eta$ removed; results unchanged, formulated in terms of $\zeta$ only}
\new
\bea \non
\D = \big(a - c - \ts{40\over 3} \big)^2 - \ts{160\over 4}\,c = 
\lp \frac{40}3 {f\over 1-f} -  c\cdot {1-f\over f}\rp^2
\\ \equiv 
\lp {f\over 1-f}\rp^2 \lb  \frac{40}3 -  \big(5+d_2 -d_a\big)^2\rb^2 .
\label{Discr}
\eea
Analogously to our considerations in \ref{app:toy}, the solution
\be \label{fp2}
f_\zeta(p_2) = \big(a + c - \ts{\frac{40}3} + \zeta \sqrt{\D}\big)/(2a)  
\ee
corresponds to the correct branch $f_\zeta \equiv f$ when 
$\zeta \sqrt{\D} = \frac{40}3 {f\over 1-f} -  c\cdot {1-f\over f}$, equivalent to the branch index
\bea \non
\zeta &\equiv \sgn \lp \frac{40}3 {f\over 1-f} -  c\cdot {1-f\over f}\rp = 
\sgn \lb  \frac{40}3 -  \big(5+d_2 -d_a\big)^2\rb 
\\ \label{zeta}
&= \sgn \lp  \sqrt{\ts{40\over 3}} - \left| \beta - 4 \right| \rp  , 
\qquad \beta =  \frac{\Da - \Depar}{\Deperp} \,,
\eea
in terms of the original model parameters --- in this case, involving all three diffusivities, independent of $f$. 
(Here we used  $5+d_2 - d_a = 4 -\beta$.) The condition (\ref{zeta}) is 
equivalent to the branch selection \eq{branch} in the main text. Choosing the opposite branch will, roughly, swap the compartment diffusivity values, similar to the toy model (\ref{toy}) considered in detail in \ref{app:toy}. 
\keep


An important difference of the general solution \eq{fp2} from the toy model \eq{toy} is the remaining dependence on $p_2$, due to the arbitrary fiber ODF, leaving the model parameters {undetermined} at $\O(b^2)$: the branches $f_\pm(p_2)$ correspond to the {\it two 1-dimensional manifolds} of model parameters  
$\{f(p_2)$, $\Da(p_2)$, $\Depar(p_2)$, $\Deperp(p_2)$,  $p_2 \}_{\pm}$ which {exactly} satisfy the first 4 equations of the system \eq{lemo}. These manifolds correspond to the two trenches in the low-energy landscape of the full RotInv problem \eq{F}, Fig.~\ref{fig:landscape} and Supplementary Figs.~\ref{fig:landscape08}--\ref{fig:landscape04}, which are flat if our acquisition is only sensitive to $\O(b^2)$. It is the $\O(b^3)$ terms, corresponding to Eqs.~\eq{L00} and \eq{L20}, that in the noise-free case select the correct trench (elevating $F$ for the wrong one), and yield the value $p_2$ fixing the minimum of $F$ in the correct trench. 

\new
For each $\zeta$, substitution of Eq.~\eq{fp2} into Eqs.~\eq{L00} and \eq{L20} separately yields ${p_2}_\zeta^{(6),0}$ and ${p_2}_\zeta^{(6),2}$; these values should coincide, ${p_2}_\zeta^{(6),0}={p_2}_\zeta^{(6),2}$, for the correct $\zeta$ choice in the absence of noise and for the unbiased moments. Based on our experience with numerical simulations with different SNR levels, we suggest for now to choose $\zeta$ corresponding to the smallest difference: 
$\zeta = \arg\min_{\{\zeta = \pm\}} \big|{p_2}_\zeta^{(6),0} - {p_2}_\zeta^{(6),2}\big|$. 
We then further average the corresponding parameter values 
$x_\zeta \equiv (x_\zeta^{(6),0} + x_\zeta^{(6),2})/2$ to increase precision. 
The numerical solution is fastest ($\sim 1\,$ms/voxel on a desktop computer) by simply performing exhaustive search for the arg min of squares of each Eq.~\eq{L00} and \eq{L20}, on the discretized interval $0\leq p_2 \leq 1$.  
\keep

\section{From moments to cumulants}
\setcounter{figure}{0}
\renewcommand\thefigure{E.\arabic{figure}} 
\renewcommand\theequation{E.\arabic{equation}}  
\label{app:mom-cum}

Comparing term-by-term the Taylor expansion (\ref{taylor}) and the Taylor expansion of $\exp(\ln S)$ where $\ln S$ is given by Eq.~(\ref{cumexp}), we obtain the following relations:  
\bea
\non
M^{(2)}_{ij} &=& C^{(2)}_{ij} \,, 
\\ \non 
M^{(4)}_{ijkl} &=& 2C^{(4)}_{ijkl} + C^{(2)}_{(ij} C^{(2)}_{kl)} \,,
\\ \non
M^{(6)}_{i_1\dots i_6} &=& 6C^{(6)}_{i_1 \dots i_6} + 6 C^{(2)}_{(i_1 i_2} C^{(4)}_{i_3\dots i_6)} 
+  C^{(2)}_{(i_1 i_2} C^{(2)}_{i_3 i_4} C^{(2)}_{i_5 i_6)} \,,
\eea
and so on. Note that full symmetrization over all indices, as in Eq.~(\ref{delta-nn}), is essential. 
Here it was implemented as a Matlab function that symmetrizes any tensor by determining its rank $l$ and generating the appropriate code at runtime, employing  function {\tt evalc}.

\section{Multiple minima: A toy model}
\setcounter{figure}{0}
\renewcommand\thefigure{F.\arabic{figure}} 
\renewcommand\theequation{F.\arabic{equation}}  
\label{app:toy}

To develop  intuition about the problem \eq{S}--\eq{K}, we first consider its simple variant that already has the main signatures of the general solution. 
Suppose we were able to measure the kernel $\K(b,\xi)$ directly --- i.e. assume for a moment that an imaging voxel is small enough to contain only one fiber orientation $\n$ --- but our measurements were limited only to directions parallel ($||, \ \xi=1$) and transverse ($\perp, \ \xi=0$) to the fascicle. 
Since typical human dMRI has $bD_i\sim 1$, where $D_i = \{ \Da, \Depar, \Deperp\}$, fitting practically performs matching between the first few moments (Taylor coefficients) of the signal and of the series $\K \simeq 1 - bD + \frac{b^2}{2!} M - \dots$~. 
Matching up to $\O(b)$ and  up to $\O(b^2)$ respectively yields  
\bea \label{toy}
\mkern-36mu
\begin{matrix}
D^{\perp}  &=& (1-f) \Deperp \,,   \quad  D^{\parallel} &= &f \Da + (1-f) \Depar \,; \quad 
\cr
M^{\perp} &=& (1-f) \Deperp^2 \,,   \quad M^{\parallel} &= &f \Da^2 + (1-f)\Depar^2 \,. \quad
\end{matrix}
\eea
In $\perp$ direction, scalar parameters $\Deperp = M^\perp/D^\perp$ and $f = 1 - {D^\perp}^2/M^\perp$ are uniquely expressed via the moments. 
However, there are {\it two} possible solutions of the corresponding quadratic equation 
$\Depar^2 - 2D^\parallel \cdot \Depar + ({D^\parallel}^2 - f M^\parallel)/(1-f) = 0$ 
for $\Depar$ (and hence, $\Da$) in  $\parallel$ direction, cf. \cite{KM,wmdki}.  
The duality arises from choosing the $\zeta = \pm$ branch of the square root 
$\Depar_{,\zeta} = D^\parallel + \zeta \sqrt{\D}$, where $\D = \frac{f}{1-f} \big( M^\parallel - {D^\parallel}^2\big)$.
To understand which branch $\zeta$ to choose, let us express the discriminant $\D$ in terms of the original model parameters, using   equations \eq{toy}. Remarkably, $\D = f^2 \big( \Depar - \Da\big)^2$ is a {\it full square}.
\mpar{R1.8: $\eta$ removed; results unchanged}
\new
Assigning the branch index $\zeta = \sgn \big( \Depar - \Da \big)$ in terms of the ground truth parameters $\Depar$ and $\Da$
yields $\zeta \sqrt{\D} = \zeta \cdot f \big| \Depar - \Da\big| =  f  \big( \Depar - \Da\big)$, such that we 
get back the correct values $\Depar_{,\zeta} = \Depar$ and $\Da_{,\zeta} = \Da$.
%
However, with the wrong branch choice $\zeta = \sgn \big( \Da - \Depar\big)$, the apparent diffusivities differ from the true ones:
\keep
\bea \label{Dapp}
\begin{matrix}
\Da^{\rm app} = &  (2f-1) \Da + 2(1-f) \Depar \,,  \cr
{\Depar}^{\rm app} =&  2f \Da + (1-2f) \Depar  \,. 
\end{matrix}
\eea
Note that in this case, as expected, ${\Depar}^{\rm app} - \Da^{\rm app} = -\big( \Depar - \Da \big)$, i.e. the difference has the same absolute value and a wrong sign. 

To recap, there exist two solutions of equations \eq{toy} which, up to  $\O(b^2)$, exactly satisfy the toy parameter estimation problem. Hence, there will be two distinct minima in the toy  ``energy'' function (analog of Eq.~\eq{F}), because of the branch selection ambiguity.
This feature originates from the two - compartment nature of the model. 
It is the $\O(b^3)$ term that would elevate the wrong minimum above the true one. If  noise overwhelms the $\O(b^3)$ effect, there will be no way to select the correct branch $\zeta$ based on comparing the values of the energy function in both minima \cite{jelescu2016}. 
 Note that {\it wrong values \eq{Dapp} can be completely plausible}; in particular, for the symmetric case $f=1/2$, the diffusivities are {\it swapped} --- i.e. we mistake the intra-  for the extra-axonal. 
 
 For this toy example, branch choice is different from Eq.~\eq{branch} because in selecting $\parallel$ and $\perp$ directions we implicitly involved components of $M^{(4),4m}$, in addition to $M^{(4),2m}$, and constrained $p_2\equiv p_4 \equiv 1$. However, the qualitative origins of bi-modality are similar.

\section{Minimally rotating dMRI signal to obtain fiber directions without blurring or sharpening}
\setcounter{figure}{0}
\renewcommand\thefigure{G.\arabic{figure}} 
\renewcommand\theequation{G.\arabic{equation}}  
\label{app:K0}

When discussing Fig.~\ref{fig:odf}, we advocated a ``reference" dMRI deconvolution, with the measured dMRI signal represented as 
\be \label{S0}
S(\g) = \int\d\n\, \P^{(0)}(\n) \K^{(0)}(\n\!\cdot\!\g)  
\ee
such that the kernel $\K^{(0)}$ is {\it defined} by unit coefficients (\ref{Kl}),
$K_l^{(0)} = (-)^{l/2} \equiv e^{i\pi l/2}$, whose sign alternates as that in $K_l(b)$ for the model kernel (\ref{K}).  
To obtain the closed-form expression for $\K^{(0)}(\xi)$, we sum the series (\ref{K=KlPl})
\be \label{K0}
\K^{(0)}(\xi) =  \sum_{l=0}^\infty (2l+1)e^{i\pi l/2 -\epsilon l} P_l(\xi) \,,
\ee
where the  factor  $e^{-\epsilon l}$, $\epsilon \to +0$, is added for the regularization of the series.
(The imaginary terms are irrelevant, as they correspond to  odd $l$, absent in dMRI.)
Using the generating function 
\be \label{Fgen}
{\cal F}(\eta,\xi) = \sum_{l=0}^{\infty} \eta^l P_l(\xi) =  \frac1{\sqrt{1+\eta^2 - 2\eta\xi}} 
\ee 
of Legendre polynomials, we calculate the sum (\ref{K0}) as 
\be \label{Phi}
\K^{(0)}(\xi) = (1-2i \partial_z)|_{z=\pi/2+i\epsilon}\, {\cal F}(e^{iz}, \xi)  = \frac1{\sqrt{2}}\, {1\over (\epsilon - i\xi)^{3/2}} \,, \quad \epsilon \to +0 \,.
\ee
Conversely, we can check that the expansion coefficients (\ref{Kl}) of $\K^{(0)}(\xi)$ are equal to $(-)^{l/2}$. 
For that, instead of evaluating these integrals for each $l$, we  use  
the generating function (\ref{Fgen}): 
\[
\sum_{l=0}^\infty t^l K_l^{(0)} = \frac12 \int_{-1}^1 \!  {\d\xi\,  \K^{(0)}(\xi) \over \sqrt{1+t^2-2t\xi}}
= - \frac{\partial_\epsilon}{\sqrt{2}} \int_{-1}^1\! {\d\xi \over \sqrt{\epsilon-i\xi} \sqrt{1+t^2-2t\xi}} \,.
\]
The latter integration can be performed using an elementary substitution $u = \sqrt{\xi + i\epsilon}$; taking the derivative 
$\partial_\epsilon|_{\epsilon\to 0}$ leads to 
\be
\sum_{l=0}^\infty t^l K_l^{(0)} = \frac1{1-it} \,,
\ee
confirming $K_l^{(0)} = e^{i\pi l/2}$ upon expanding the geometric series. 

\begin{figure}[b!!]
\includegraphics[width=3.2in]{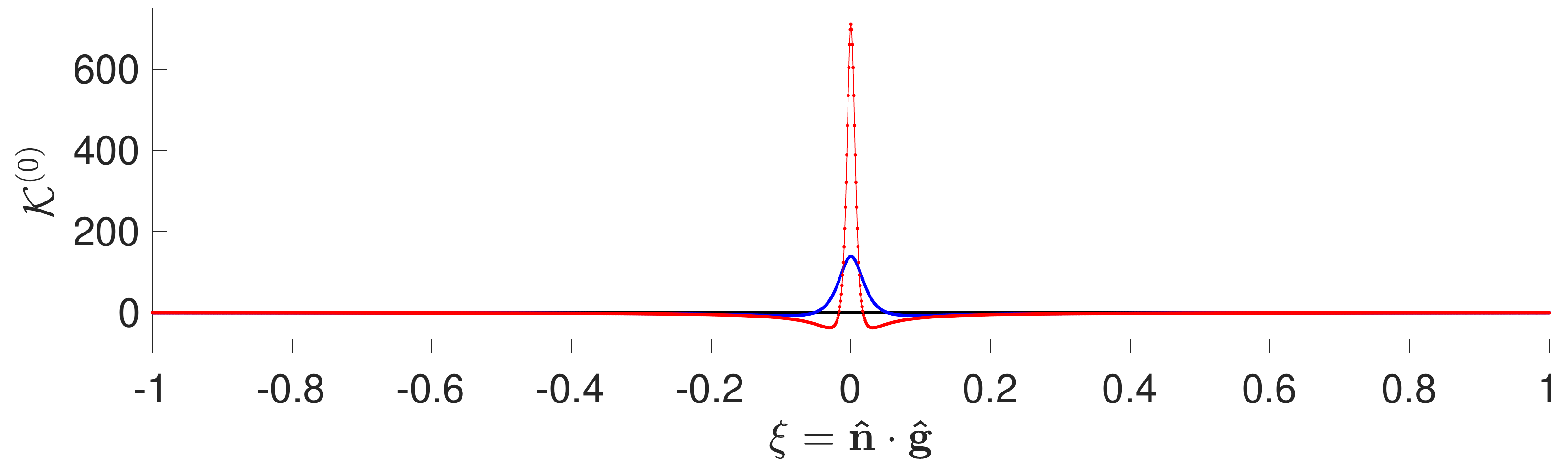}
\caption{The real part of kernel (\ref{Phi}) for $\epsilon = 0.03$ (blue) and $\epsilon = 0.01$ (red). Points are the result of summing even-order Legendre polynomials in Eq.~(\ref{K0}) up to $l=1000$; solid lines are given by the corresponding analytical formula (\ref{Phi}). 
Here, $\epsilon$ controls the peak width, i.e., the vicinity of the equator $\n\cdot\g = 0$ in Eq.~(\ref{S0}) (with $\g$ pointing at the North pole), 
where the fractional derivative (\ref{fracder}) is evaluated.
}
\label{fig:K0}
\end{figure}

The  equivalence of the integration kernel (\ref{Phi}) with a fractional derivative operator is best seen in the Fourier domain: 
\be \label{fracder}
\K^{(0)}(k)=\int_{-\infty}^\infty\! \d\xi\, e^{-ik\xi} \K^{(0)}(\xi) = 4\theta(k)\sqrt{\pi k} \quad\Rightarrow\quad 
\hat\K^{(0)} \propto \partial_\xi^{1/2} \,,
\ee 
since the $n$th order derivative in the Fourier domain corresponds to multiplication by $(ik)^n$. 
(The integration contour was deformed to pass alongside the branch cut, chosen along the negative $\Im$ axis of the complex variable $\xi$. Here, the positive $\epsilon=+0$ is essential, as it shifts the singularity below the $\Re \xi$ axis and makes the integral convergent, relying on the analytic continuation of Euler's $\Gamma$-function. The unit step function $\theta(k)$ reflects closing the contour in the upper- or lower-half-plane of $\xi$ according to Jordan's lemma.) 

The fractional derivative singularity is illustrated in Fig.~\ref{fig:K0}, where the peak at $\xi=0$ (corresponding to concentrating the weight on the equator $\n\cdot\g = 0$, Eq.~(\ref{S0})), expected from the Funk-Radon transform (FRT) relations \cite{Tuch2004,jensen2016}, also develops {\it negative lobes}. 
In an integral 
\[
\int\!\d\xi\,\K^{(0)}(\xi)\phi(\xi) \equiv  \lb \hat \K^{(0)}\phi \rb_{\xi = 0} \propto \partial^{1/2}_\xi \phi|_{\xi = 0}
\]
 with any function $\phi(\xi)$, these lobes compare the values of $\phi(\xi)$ in the vicinity of $\xi=0$. In contrast,  the asymptotic FRT kernel $\sim \sqrt{bD}\, e^{-bD\xi^2}$ in the fiber ball limit $b\to\infty$ \cite{jensen2016} acts as $\delta(\xi)$, yielding $\phi|_{\xi=0}$ when integrated with $\phi(\xi)$. Hence, the kernel (\ref{Phi}) is qualitatively more singular than the FRT. 
This singular behavior is reflected in the $l$-independent basis coefficients $|K_l^{(0)}|$, as compared to the FRT coefficients 
$P_l(0) = (-)^{l/2} (l-1)!!/l!!$, decaying with $l$ as $|P_l(0)|\sim (\pi l)^{-1/2}$ for $l\gg 1$ and de-emphasizing the higher-$l$ harmonics.

Hence, there exists a  hierarchy of the kernel coefficients in the SH basis. 
The coefficients $K_l^{(0)}$ exceed $P_l(0)$ for the FRT, which are in turn greater than 
the coefficients (\ref{Kl}) at {\it finite} $b$: 
\be \label{compare-FRT}
|K_l(b)| \leq |P_l(0)| < |K_l^{(0)}| \equiv 1 < {1\over |P_l(0)|}\,, 
\ee
since the corresponding functions $g_l \equiv K_l(b)/P_l(0) \leq 1$ for all $b$, at any given $l$ \cite{jensen2016}. 
Physically, this reflects that the kernel (\ref{K0}) in Eq.~(\ref{S0}) by design does not magnify or decrease the sharpness of the ``reference" ODF $\P^{(0)}(\n)$, while the fiber-ball kernel blurs the ODF, and the realistic (finite-$b$) kernel (\ref{K}) blurs it even more (that latter kernel is what, presumably, occurs in nature). 

Of course, it is the left-hand side of Eq.~(\ref{S0}), the dMRI signal, that is given, and we are trying to obtain the best approximation for the true ODF 
$\P(\n)$. Deconvolving the ODF from the signal (dividing $S_{\! lm}$ by $K_l$) progressively magnifies the signal's harmonics with larger $l$ for the FRT in the fiber-ball limit, and even more so for the finite-$b$ case, Fig.~\ref{fig:odf}, in order to compensate for the blurring of the ODF by the diffusion kernel $\K$. 
This results in the ODF sharpening (presumably, approaching the realistic ODF, at least in the absence of noise), relative to the minimally-rotated signal ODF $\P^{(0)}$ with $p_{lm}^{(0)} = S_{\! lm}/K^{(0)}_l$. 
We also note that the original Q-ball ODF reconstruction \cite{Tuch2004,Descoteaux2009} implies $p_{lm} \sim P_l(0) S_{\! lm}$, i.e. de-emphasizing the signal harmonics with the higher $l$, which puts Q-ball kernel $1/P_l(0)$ to the right of $|K_l^{(0)}|$ in the hierarchy (\ref{compare-FRT}). 

We can see that our ``minimal rotation" of the signal SH coefficients provides a natural reference for the ODF deconvolution methods. 
We also note an obvious property of the transform (\ref{S0}): since $\lb K_l^{(0)}\rb^2 \equiv 1$, the operator $\hat \K^{(0)}$ on a unit sphere applied twice yields identity; equivalently, it equals its inverse.

\newpage
\section*{References}
\bibliographystyle{elsarticle-harv}
\bibliography{rotinv-arXiv-v2.bbl}


\setcounter{figure}{0}%
\renewcommand\thefigure{S.\arabic{figure}}%
\newpage

{\bf \large Supplementary Information}%

\begin{figure*}[b!!]
\includegraphics[width=5.2in]{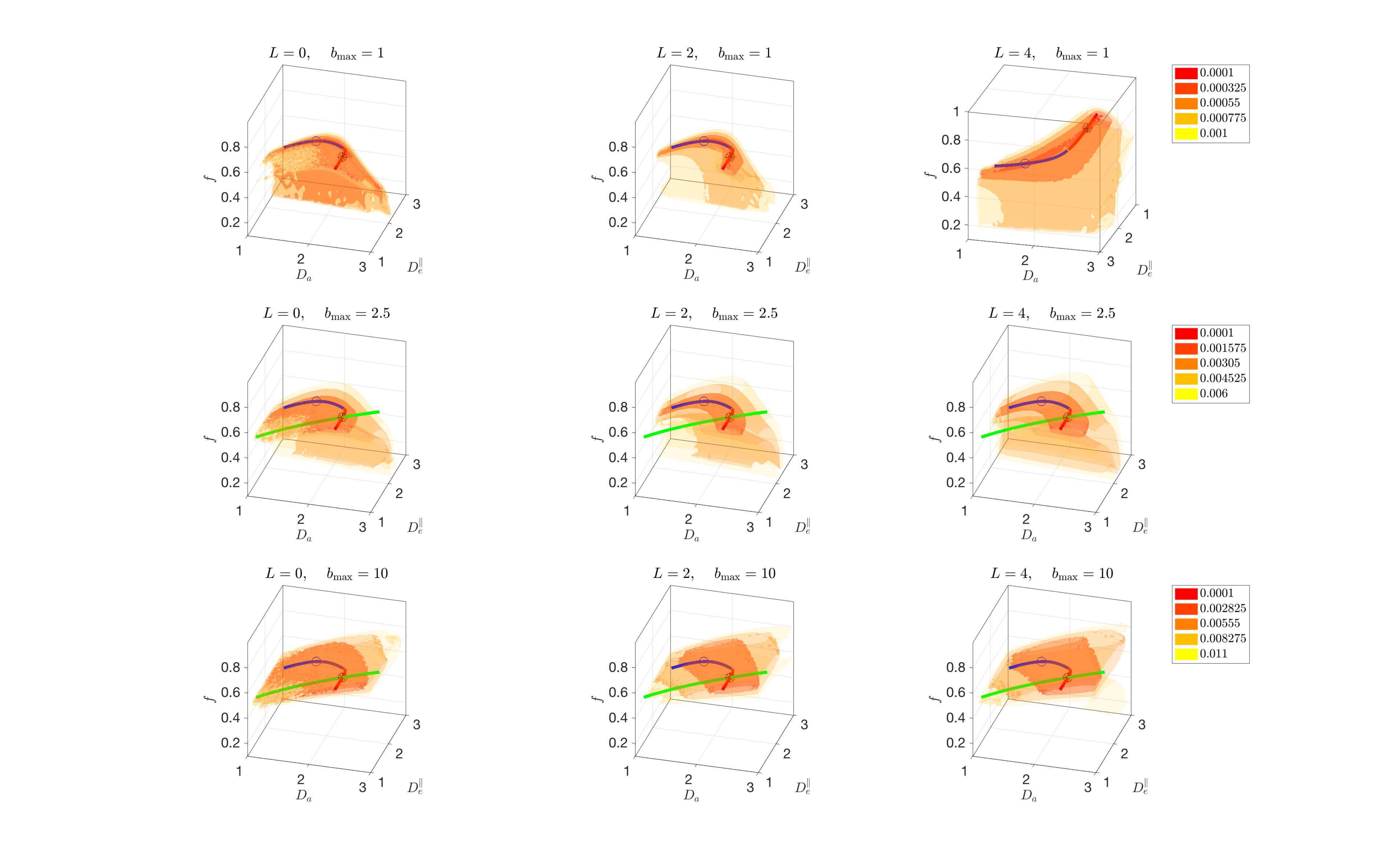}
\caption{{\bf Low-energy landscape of the problem \eq{F}}. The $F$-values are minimized with respect to $\Deperp$ and $p_l$, for the case when the two branches form a single trench within the feasible parameter range. Ground truth values 
$\{ f, \Da, \Depar, \Deperp, p_2 \} = \{0.7, 2.4, 1.5, 0.8, 0.7 \}$ correspond to three identical fiber segments crossing at an angle $\theta \approx 27^\circ$ to the tract axis. The simulated $b$-values correspond to those in our human experiments (see Sec.~\ref{sec:invivo}), with all the 21 $b$-shells uniformly rescaled to attain the maximal value $b_{\rm max}$, such that the bottom row corresponds to the actual acquisition. The two analytical LEMONADE branches (+ red, - blue) match the low-value manifolds, especially for low $b_{\rm max}$. 
Increasing $L$, the 2-dimensional surface ($L=0$, corresponding to the two constraints \eq{D00} and \eq{M00} for 4 scalar parameters) gradually turns into 1-dimensional trenches (the full system \eq{lemo}), while increasing $b_{\rm max}$ causes flattening of the landscape such that it eventually follows the surface $f/\sqrt{\Da} = \mbox{const}$ dominated by the intra-axonal water, with the extra-axonal water exponentially suppressed (green line). 
}
\label{fig:landscape08}
\end{figure*}

\begin{figure*}[t]
\includegraphics[width=5.2in]{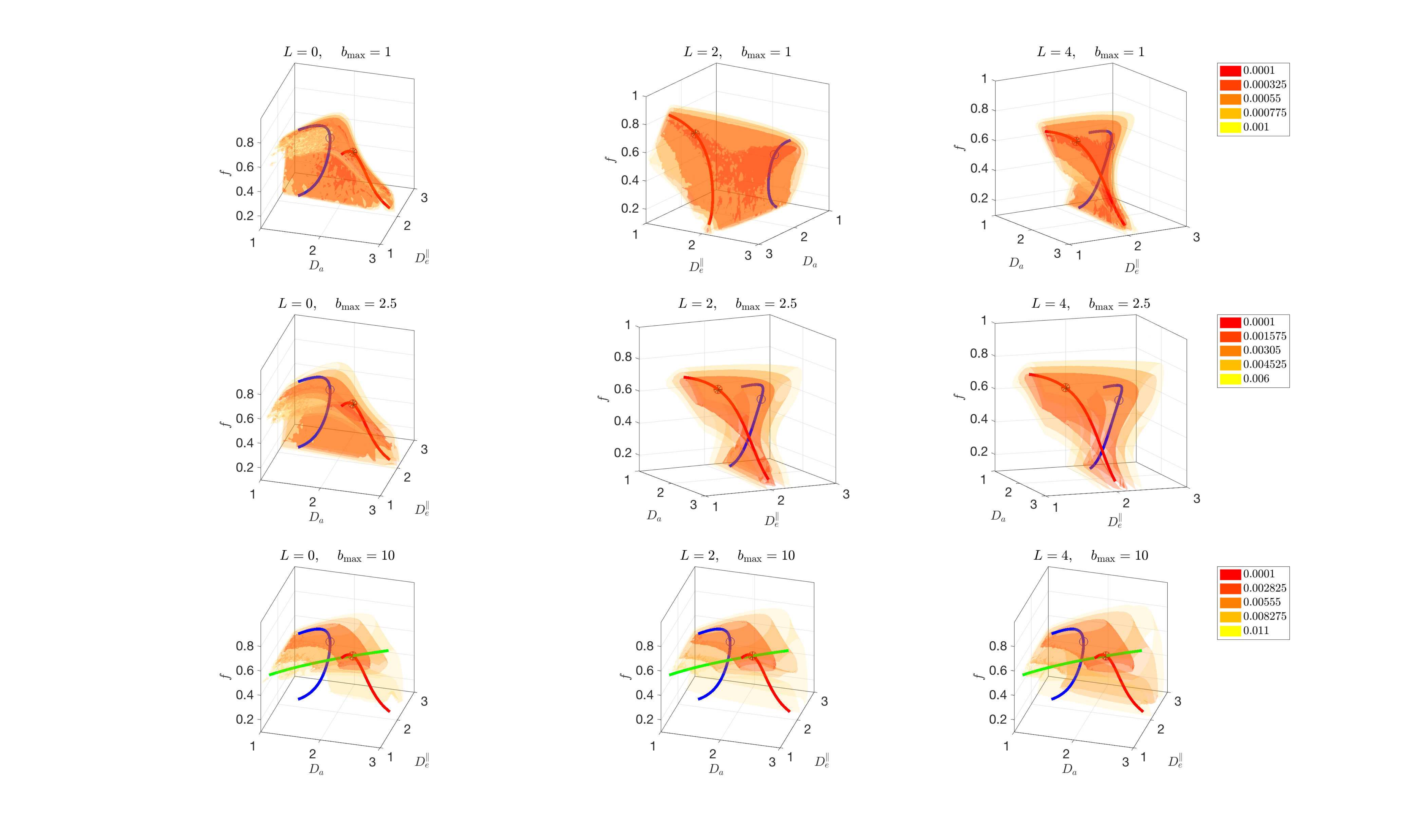}
\caption{
The same as in Fig.~\ref{fig:landscape08} but with $\Deperp = 0.4$. The landscape is highly sensitive to the ground truth values: merely altering one parameter, $\Deperp$, we now have two separate trenches passing through the physically feasible parameter range. They eventually connect (as in Fig.~\ref{fig:landscape08}), albeit outside this range.  In this case it is particularly easy for spurious minima (e.g. due to noise) to appear in-between the trenches. 
}
\label{fig:landscape04}
\end{figure*}

\begin{figure*}[t]
\includegraphics[width=2.5in]{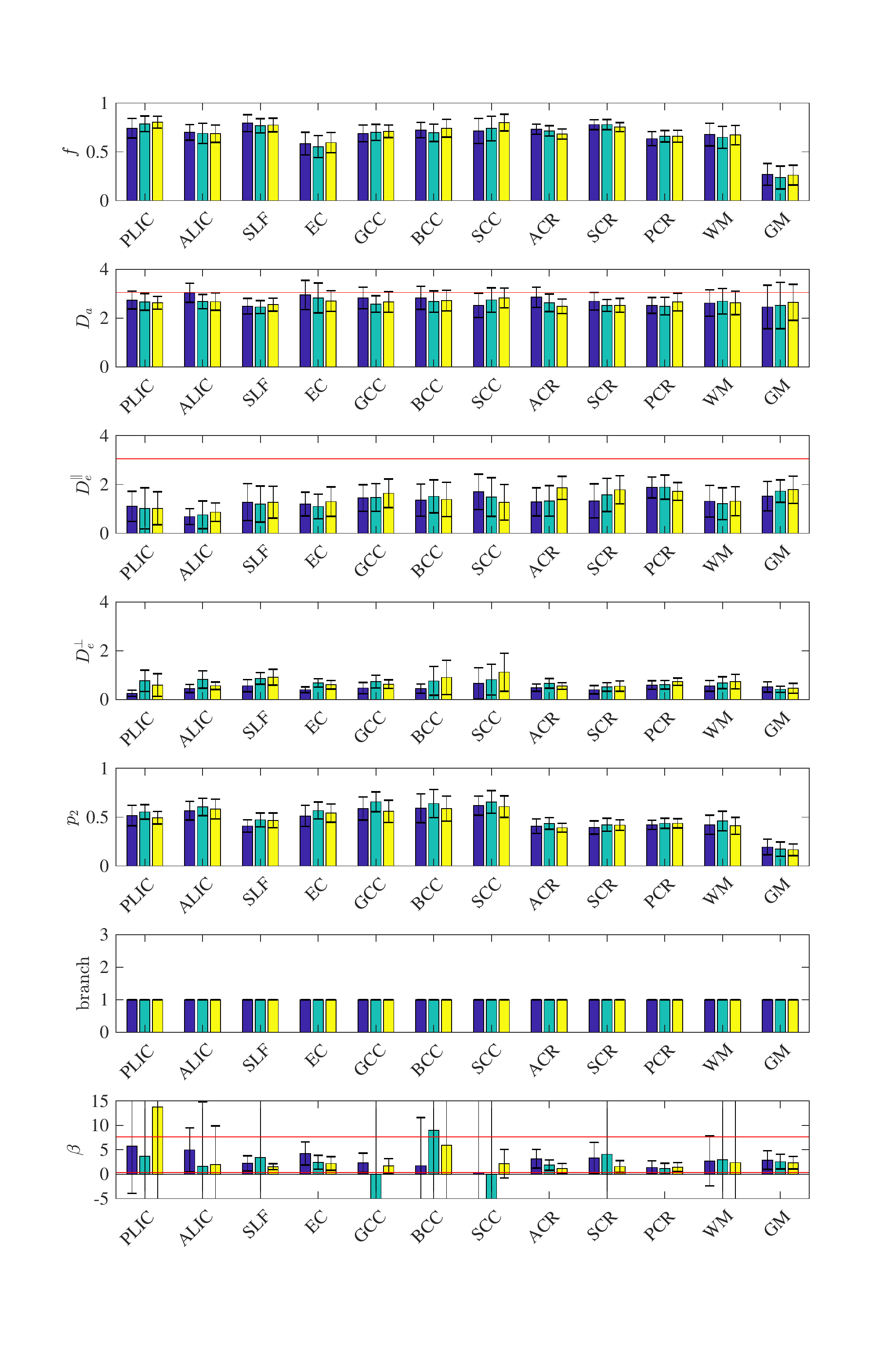}%
\includegraphics[width=2.5in]{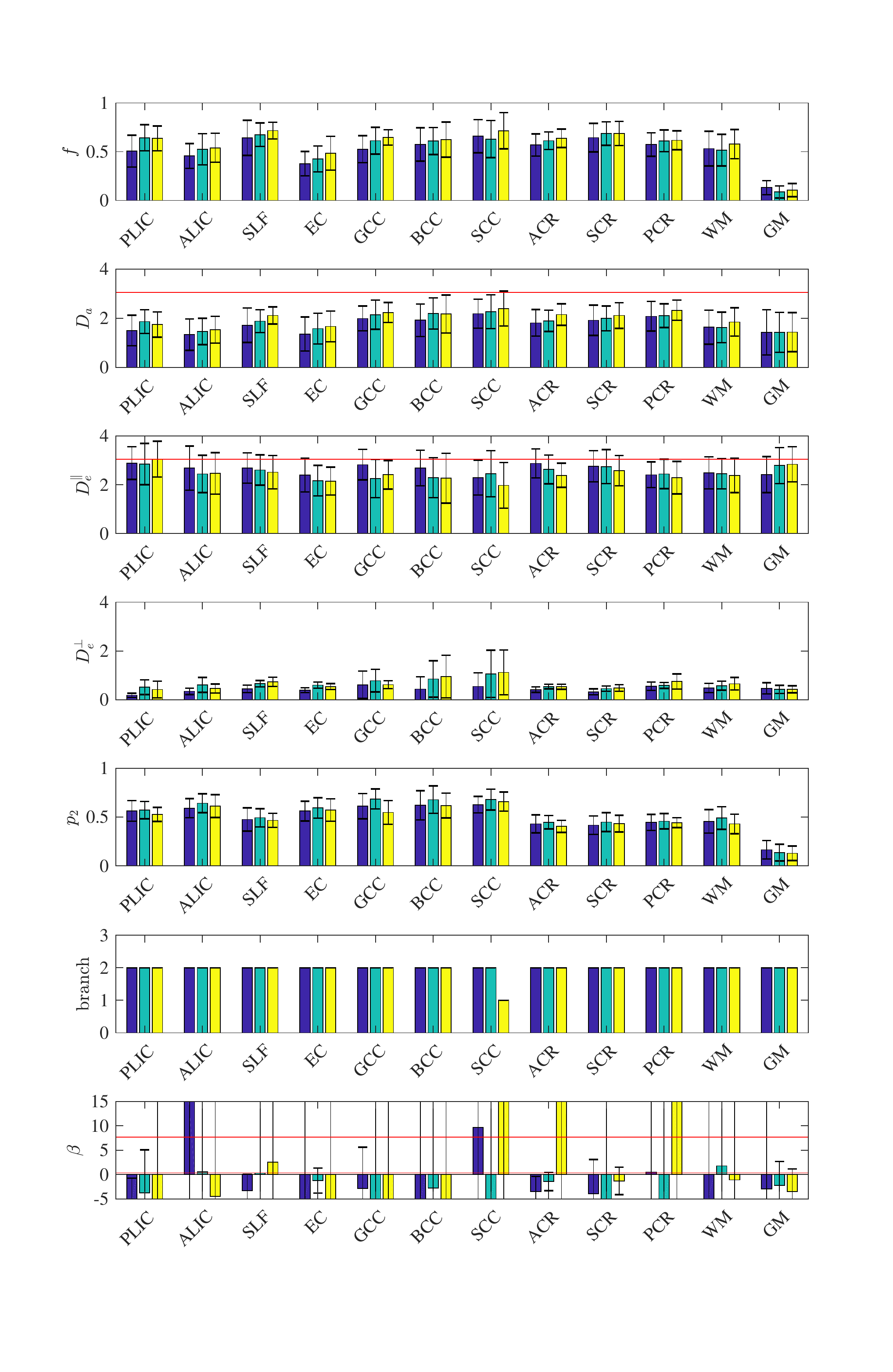}%
\includegraphics[width=2.5in]{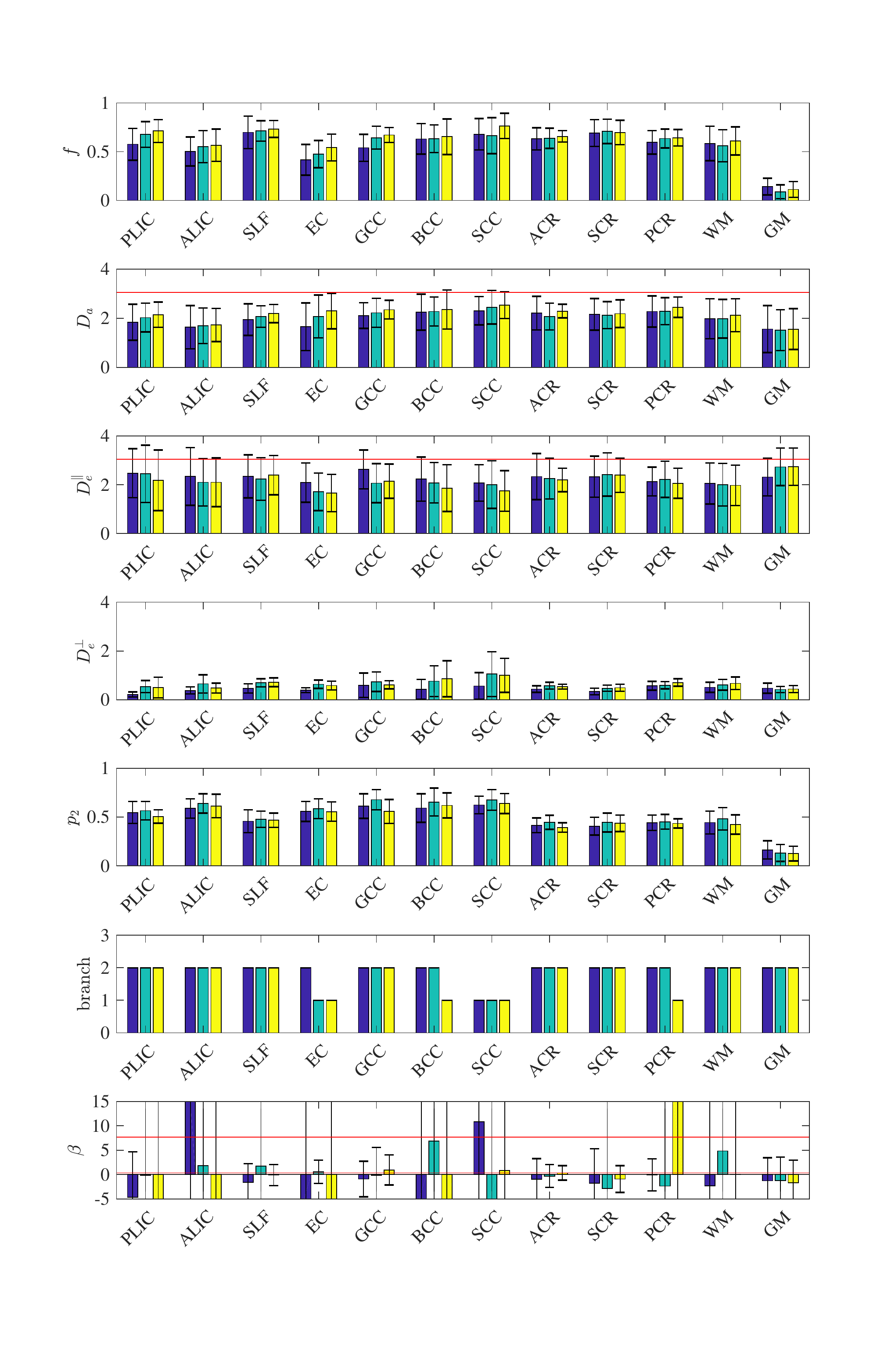}
\caption{
\new 
\rem{New Fig. }
{\bf ROI values} for the RotInv$\pm$ branches (left and center, correspondingly), as well as for the voxel-wise branch selection method RotInv $\zeta$, cf. Sec.~\ref{sec:branch}. Note diffusivity values $\Da > \Depar$ and $\Da < \Depar$ for the $\pm$ branches.  
Mean values for the RotInv $\zeta$ branch selection are quantitatively similar to those based on the prevalence method,  Fig.~\ref{fig:bars}, but the error bars are somewhat larger, presumably signifying the fact that our branch selection method is imperfect. 
\keep
}
\label{fig:bars-sel}
\end{figure*}

\setcounter{section}{0}
\renewcommand\thesection{S.\arabic{section}} 

\new
\section{Parameter landscape}

Figures \ref{fig:landscape08} and \ref{fig:landscape04} provide more examples of the topology of the minimization landscape, cf. Fig.~\ref{fig:landscape} from the main text.

\section{ROI values for the branch selection method }
\label{app:ROI}

Figure \ref{fig:bars-sel} provides ROI values for the RotInv$\pm$ estimations, as well as for the local branch selection method RotInv $\zeta$. Note that the RotInv $\zeta$ values look very similar to those obtained via the prevalence method, cf. Fig.~\ref{fig:bars} from the main text. 

\keep

\section{Noise propagation} \rem{Sec. moved from Appendix F to Suppl}
\label{app:noise}

\begin{figure*}[h!!!!!]
\vspace{-1cm}
{\bf a}\includegraphics[width=5.5in]{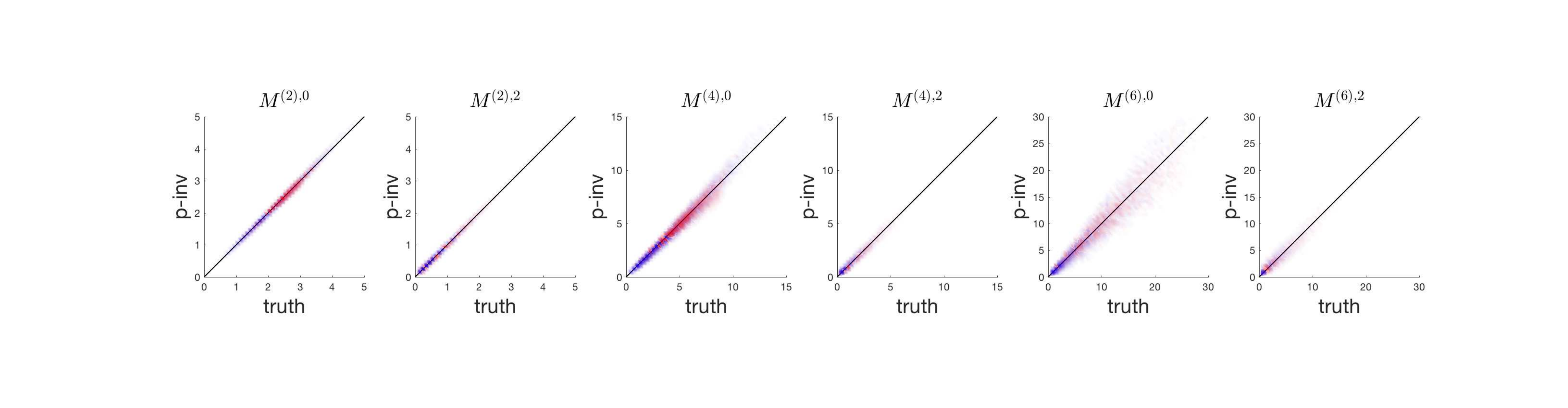}\\
\vspace{-5mm}
{\bf b}\includegraphics[width=5.5in]{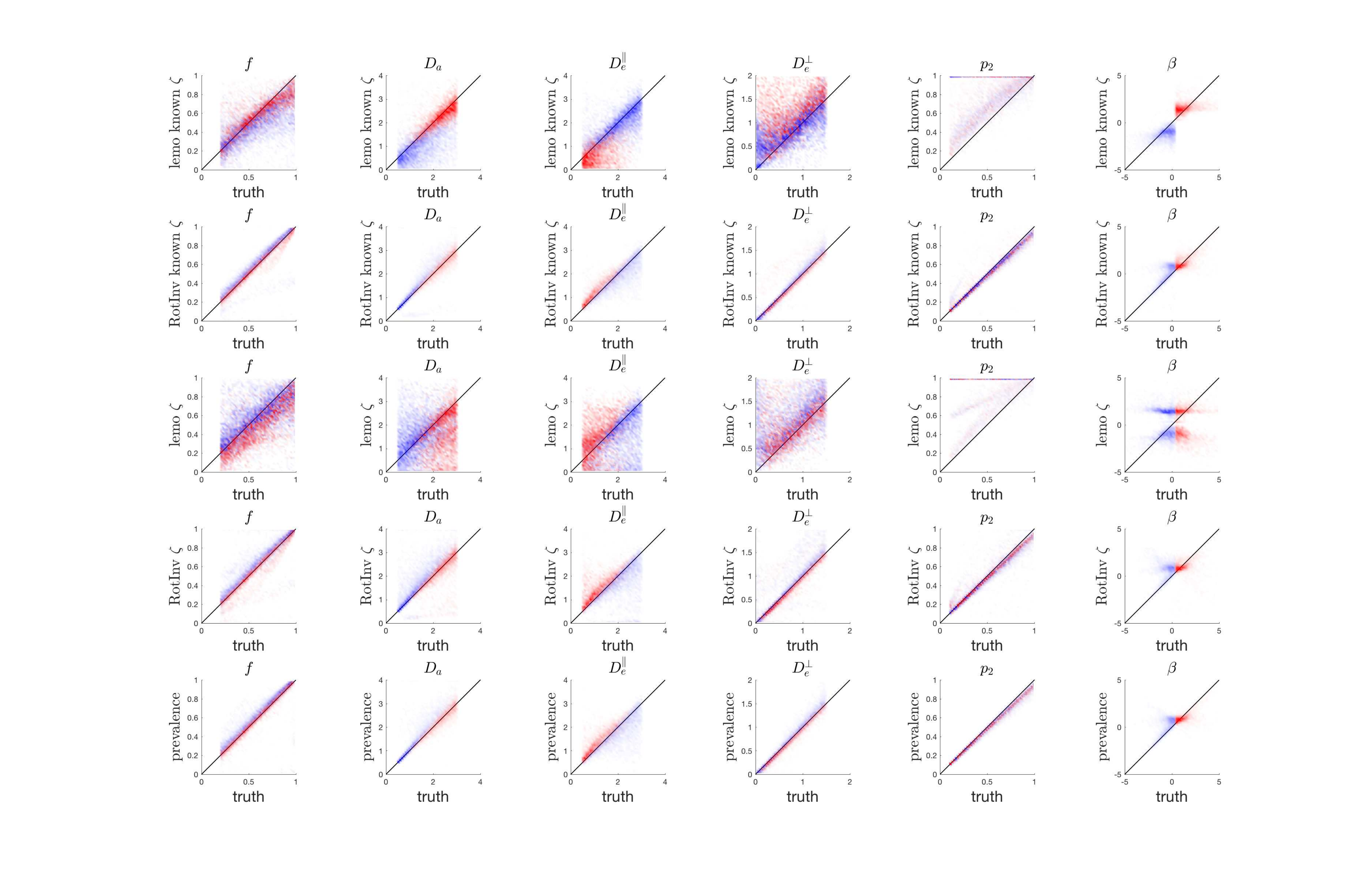}\\
\vspace{-5mm}
{\bf c}\includegraphics[width=5.5in]{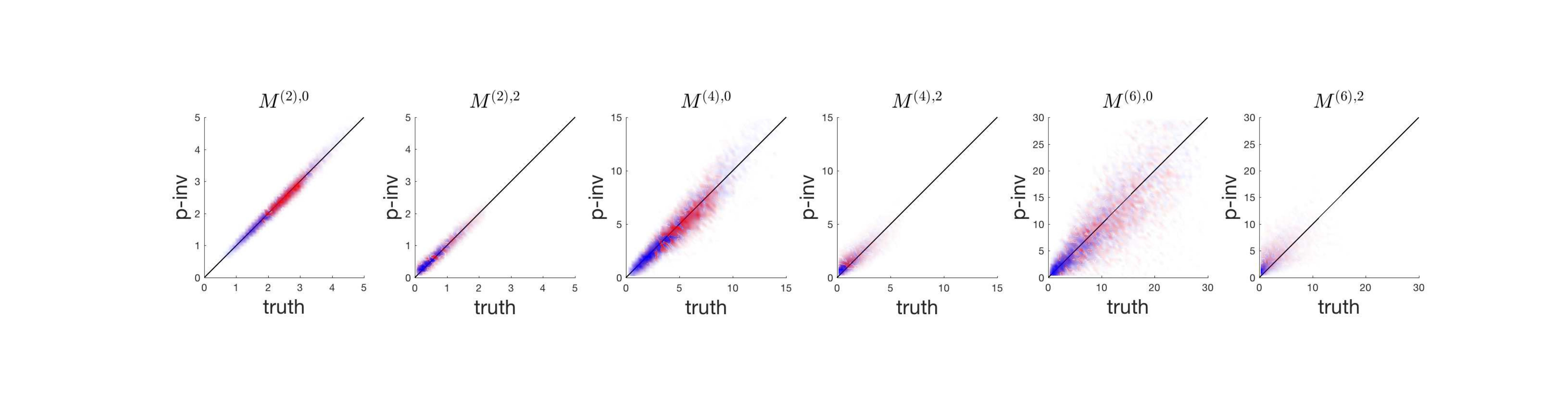}\\
{\bf d}\includegraphics[width=5.5in]{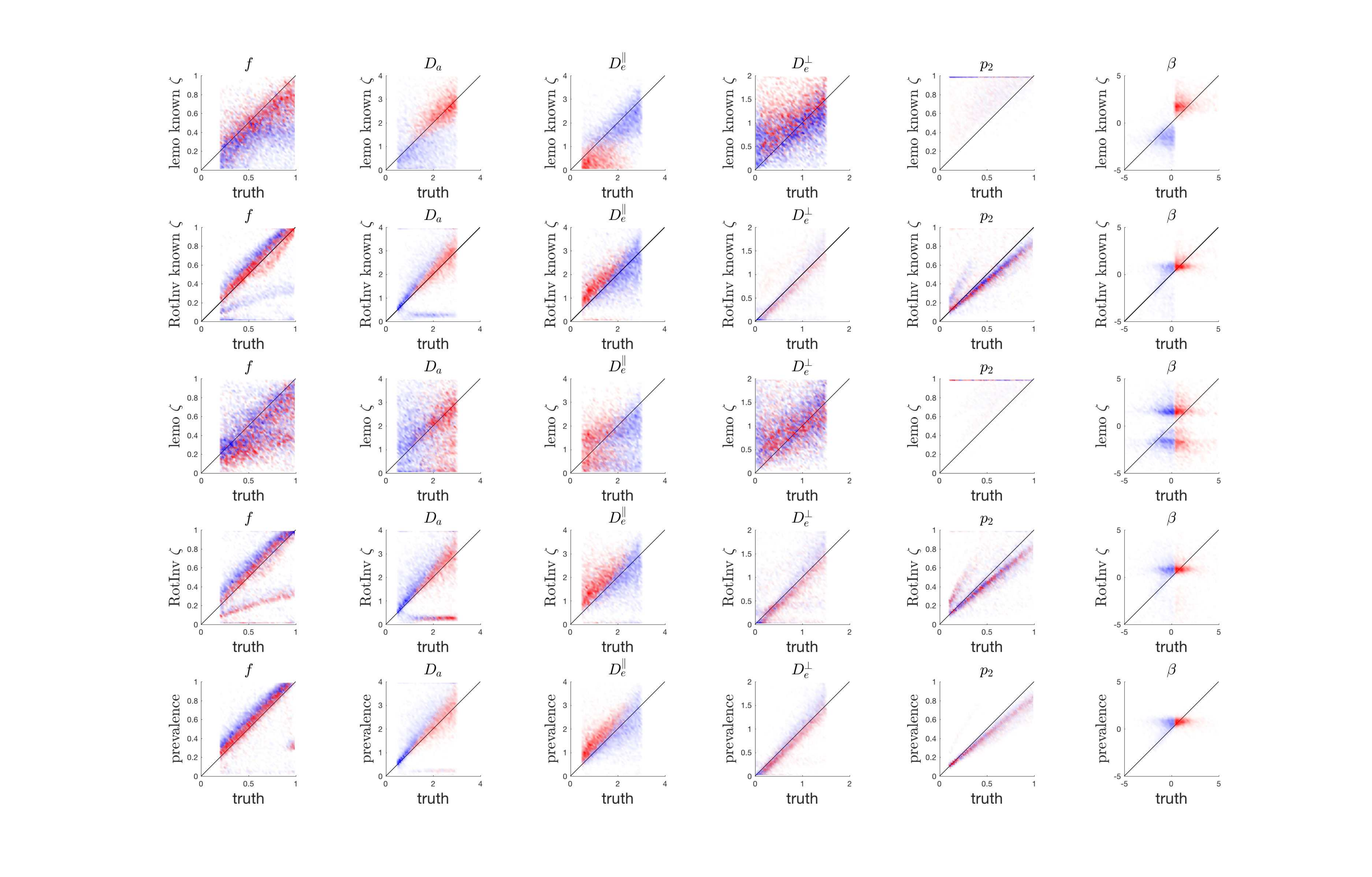}
\caption{
\rem{added lemo $\zeta$ and RotInv $\zeta$ rows}
{\bf Noise propagation} for {\bf a, b}: SNR = 100; {\bf c, d:} SNR = 33, at $b=0$, in estimating moments (panels {\bf a} and {\bf c}) and biophysical parameters  ({\bf b} and {\bf d}). 
Red/blue colors correspond to $\zeta=\pm$ branches assigned based on the ground truth values according to Eq.~\eq{branch}. 
}
\label{fig:noiseprop}
\end{figure*}

Figure \ref{fig:noiseprop} shows results  from Monte Carlo simulations of the full MRI protocol (see {\it Methods}) with 10,000 random combinations of ground truth values uniformly distributed within the biophysically relevant intervals ($x$-axis, ``truth"). The fiber geometry is three identical fiber segments with azimuthal angles $\phi = 0, \, \pm 2\pi/3$, crossing at an angle $\theta \approx 27^\circ$ with respect to the tract axis, as in Supplementary Figs.~\ref{fig:landscape08} and \ref{fig:landscape04}. Gaussian noise with variance $\sigma^2$ is addded to both real and imaginary parts of the signal, with absolute value at $b=0$ normalized to SNR$\, =1/\sigma$, such that the magnitude signal follows Rician distribution.

Branch degeneracy in Fig.~\ref{fig:noiseprop} manifests itself in that branch assignment is not apparent at the level of the rotational invariants --- in this case, the moments (panels {\bf a} and {\bf c}) ---  and becomes evident based on the parameter values 
(panels {\bf b} and {\bf d}). Practically, $\zeta = + $ corresponds to $\Da > \Depar$ and vice-versa, cf. Eq.~\eq{branch}. 
In panels {\bf b} and {\bf d}, top row corresponds to parameter estimation based on LEMONADE output, which subsequently  served as initialization for the nonlinear fitting of Eq.~\eq{F} (middle row), where the LEMONADE branch was pre-selected based on the ground truth values. 
We can see that noise results in decrease of precision, and that it can accidentally switch the branch. 
Addition of nonlinear fit \eq{F} notably improves both accuracy and precision relative to LEMONADE. 

\new
Note that knowing the branch index $\zeta$ beforehand (top two rows in {\bf b} and {\bf d}) makes estimation notably better. Unfortunately, in reality we do not know this index for any given voxel. The 3rd and 4th rows correspond to estimating parameters based on our local branch selection method of Sec.~\ref{sec:branch}. Note that the output RotInv $\zeta$ is almost as good as that of the (much lengthier) prevalence calculation (bottom row) for sufficiently large SNR (panel {\bf b}), whereas for the lower SNR, spurious parameter values appear (e.g. low $f$ and $\Da$).  
\keep

The prevalence method used 100 random initializations within the physically relevant domain of parameters. 
Generally, intra-axonal parameters $f$ and $\Da$ are more precise than extra-axonal $\Depar$ and $\Deperp$; 
unfortunately, the branch ratio $\beta$ is particularly imprecise, prompting the need for ``orthogonal" measurements to validate the branch index map in the whole brain.   

\end{document}